\documentclass[
letterpaper,
reprint,           
twocolumn,
superscriptaddress,
amsmath,           
amssymb,           
aps,               
prd,               
notitlepage,       
longbibliography,  
floatfix,          
nofootinbib,
]{revtex4-1}

\usepackage{amssymb,amsmath}
\usepackage{epsfig}
\usepackage{epstopdf}
\usepackage{tikz,tikz-3dplot}
\usepackage{aas_macros}
\usepackage[toc,page]{appendix}
\usepackage[mathscr]{euscript}
\usepackage{amsthm}
\usepackage[colorlinks, linkcolor=blue, citecolor=green, urlcolor=blue]{hyperref}
\usepackage{cleveref}
\usepackage{orcidlink}

\newcommand{\nobracket}{}

\newtheorem{corollary}{Corollary}
\newtheorem{theorem}{Theorem}

\crefname{figure}{Fig.}{Figs.}
\Crefname{figure}{Fig.}{Figs.}
\crefname{equation}{Eq.}{Eqs.}
\Crefname{equation}{Eq.}{Eqs.}

\begin{document}
\title{Full analytic expressions of overlap reduction functions for anisotropies of the stochastic gravitational-wave background with pulsar timing arrays}

\author{Kun Zhou~\orcidlink{0009-0001-1099-8623}}

\author{Jin Li~\orcidlink{0000-0001-8538-3714}}

\email{{cqujinli1983@cqu.edu.cn}}

\affiliation{College of Physics, Chongqing University, Chongqing 401331, China and Department of Physics and Chongqing Key Laboratory for Strongly Coupled Physics, Chongqing University, Chongqing 401331, China}

\begin{abstract}
	
Recent findings collectively suggest that pulsar timing arrays (PTAs) have detected the stochastic gravitational-wave background (SGWB) in the nanohertz frequency band, opening new avenues for testing theories of gravity and cosmology. PTAs serve as a unique tool for probing the nanohertz SGWB, enabling studies of supermassive black hole evolution and early-Universe physics. PTA data analysis fundamentally relies on cross-correlating timing residuals from pulsar pairs, for which the overlap reduction functions (ORFs) are critical in determining the sensitivity of such analyses. Conventional ORF calculations based on the ``short-wavelength approximation'' break down when handling the scalar longitudinal polarization mode and fail to accurately describe frequency-dependent effects and anisotropies. This work presents a significant advance by rigorously deriving the full response functions within an analytical cross-correlation framework. We systematically reveal, for the first time, the intrinsic symmetries and mathematical relationships among the anisotropic ORF integrals for different polarization modes. Through variable substitutions and coordinate rotations, we transform the complex integrals into analytically tractable forms, resolving the divergence difficulties encountered in the vector and scalar longitudinal modes. Building on these theoretical advances, we establish a universal framework capable of deriving fully analytical expressions for anisotropic ORFs to arbitrary order for all six polarization modes. As a direct application of this framework, we provide complete analytical expressions for anisotropic ORFs up to the order of $\ell \leq 5$. This universal framework is free from approximations; its $(0, 0)$ component naturally corresponds to the isotropic case and recovers the well-known Hellings--Downs curve. Compared to numerical integration, our results offer broader applicability, significantly faster computation, and higher precision. They provide a valuable theoretical and computational foundation for conducting high-precision anisotropic sky mapping, polarization-mode separation, and searches for new physics using PTA data.
	
\end{abstract}

\maketitle
\section{Introduction}\label{sec1}

The era of gravitational-wave astronomy was inaugurated by the landmark direct detection of the binary black hole merger GW150914 by the Laser Interferometer Gravitational-Wave Observatory (LIGO) on September 14, 2015 \cite{Abbott2016, Yunes2016}. Gravitational waves across different frequency bands reveal distinct physical origins. Ground-based interferometers like LIGO are sensitive to the $10$--$10^3$ Hz band, primarily detecting mergers of stellar-mass compact objects \cite{Abbott2021}. In contrast, the nanohertz frequency ($10^{-9}$ Hz) SGWB likely originates from the collective emission of supermassive black hole binaries (SMBHBs) throughout the Universe \cite{Rajagopal1995, Sesana2008} or from relic signals produced by early-Universe processes such as phase transitions or cosmic strings \cite{Caprini2010}. Detecting this background requires a galaxy-sized ultraprecise detector: the Pulsar Timing Array (PTA). By monitoring the timing residuals of a network of millisecond pulsars---nature's most stable clocks---distributed across our Galaxy, PTAs aim to extract a correlated signal characterized by a specific angular correlation pattern known as the Hellings--Downs (HD) curve \cite{Hellings1983}. This enables studies of SMBHB evolution and probes of early-Universe physics \cite{Burke-Spolaor2019}.

Recently, this field has witnessed exciting observational breakthroughs. Beginning in 2023, several international PTA collaborations---including the North American Nanohertz Observatory for Gravitational Waves (NANOGrav), the European Pulsar Timing Array (EPTA), the Parkes Pulsar Timing Array (PPTA), and the Chinese Pulsar Timing Array (CPTA)---have independently reported strong evidence for a common-spectrum stochastic process in their data, with spatial correlations consistent with the HD curve \cite{Agazie2023, Antoniadis2023a, Reardon2023, Xu2023, Caprini:2024lxj}. For instance, NANOGrav found the correlation with the HD curve in its 15-year dataset to have a significance of $3$--$4\sigma$ \cite{Agazie2023}. These nearly simultaneous announcements mark a transition from a long ``search'' phase into a new era of ``emerging evidence'' for nanohertz gravitational waves \cite{Laal2024}.

However, while the current evidence strongly suggests the presence of an SGWB, it also raises new scientific questions. First, the observed signal amplitude is generally higher than predictions for the standard astrophysical background from SMBHBs based on local galaxy relations \cite{Middleton2024}. Second, the precise agreement of the spectral characteristics with the HD curve requires further confirmation. Recent data from the MeerKAT PTA, for example, show that different assumptions about the noise model can significantly affect the correlation significance \cite{Miles2024}. These unresolved issues have sparked extensive theoretical discussions, including a reexamination of SMBHB population properties and in-depth studies of potential cosmological contributions to the signal, such as from domain walls or primordial black holes \cite{Gouttenoire2023, Ellis2024}.

Precisely determining the physical origin of the SGWB depends on the ability to map its potential anisotropy and to search for non-Einsteinian polarization modes. Anisotropy mapping can reveal an inhomogeneous distribution of the gravitational-wave energy density across the sky, helping to identify dominant source regions or distinguish a cosmological origin \cite{Mingarelli2013, Taylor2020}. Recent studies propose using frequency-dependent optimal statistics for anisotropic detection, suggesting that even low levels of anisotropy may be detectable against a more realistic SMBHB-generated background \cite{Bernardo2023}. Searches for additional polarization modes (e.g., scalar or vector modes) serve as direct probes for testing general relativity and exploring modified theories of gravity \cite{Lee2008, 2012prd_ORF_nonGR}.

The core theoretical tool enabling these precise analyses is the ORF, which quantifies the specific correlated signal induced by an SGWB between any pair of pulsars. The HD curve is precisely the ORF for an isotropic SGWB composed solely of the transverse-traceless tensor polarizations predicted by general relativity \cite{Hellings1983}. To handle the complex integrals involving pulsar distances and gravitational-wave vectors in ORF calculations, the traditional ``short-wavelength approximation'' is widely employed, neglecting oscillatory phase integrals related to the pulsar term \cite{Anholm2009}. However, this approximation leads to divergences for modes like the scalar longitudinal mode and fails to accurately describe finite-frequency effects and the general geometric configurations required for anisotropic analysis \cite{2012prd_ORF_nonGR}. 

The path toward an analytical calculation of the ORF in PTAs has been neither straightforward nor trivial, with its historical evolution intertwined with intrinsic difficulties. Since the derivation by \cite{Hellings1983} of the analytical special case---the HD curve for an isotropic, transverse-traceless tensor background---the quest for analytical expressions under more general conditions has been a persistent theoretical pursuit. When extending the formalism to include vector and scalar polarization modes allowed in modified gravity theories, researchers found that the traditional short-wavelength approximation leads to divergent integrals for the scalar longitudinal mode, yielding only limited expressions for modes like the vector modes \cite{Lee2008, 2012prd_ORF_nonGR}. This difficulty stems from the fact that when attempting to precisely account for finite pulsar distance effects (the oscillatory phase of the ``pulsar term'') and the finite frequency of the gravitational waves, the integrand becomes highly oscillatory and complex, rendering traditional integration techniques ineffective. Even numerical integration faces challenges in computational efficiency and accuracy at high frequencies. Furthermore, generalizing from an isotropic to an anisotropic background requires a spherical harmonic expansion of the sky distribution, which increases the dimensionality and analytical complexity of the integrals \cite{Gair2014}. Therefore, developing a precise analytical framework---free from specific approximations and capable of systematically handling arbitrary polarization modes and anisotropic sky distributions---has long been an outstanding core theoretical challenge in PTA data analysis.

Over the past decade, significant progress has been made in the analytical theory of ORFs. The work by \cite{Gair2014} systematically introduced the spherical harmonic expansion method into PTA analysis, establishing a complete framework for SGWB anisotropy studies. Earlier investigations by \cite{Lee2008, 2012prd_ORF_nonGR} examined ORFs for vector and scalar modes in the context of modified gravity. A recent major breakthrough was achieved by \cite{Hu2024}, who, for the first time and without relying on the short-wavelength approximation, derived complete analytical expressions for the ORFs applicable to all six polarization modes in general metric theories for an isotropic background, thereby resolving the computational problem for the scalar longitudinal mode. In contrast, by neglecting the pulsar term, the work by \cite{AnilKumar:2023yfw, Inomata:2024kzr} established a systematic framework based on spherical harmonic expansions to compute the ORFs induced by a gravitational-wave background of arbitrary anisotropy and polarization. This framework is applicable to both pulsar timing arrays and astrometry, with the results expressed in the form of numerically summable infinite series. However, it is limited by its reliance on the core approximation of neglecting the pulsar term, which prevents it from correctly handling the scalar longitudinal polarization mode and presents theoretical difficulties when dealing with pulsar pairs at small angular separations. Concurrently, studies on ORFs in modified gravity, particularly addressing the divergence of autocorrelation terms, have confirmed the regularizing role of finite-distance effects \cite{Cordes2024}. On the data analysis front, new techniques have been developed to address computational challenges in SGWB signal extraction, including fast spectral representation methods \cite{Lamb2023}, regularization approaches for Fourier-space likelihood functions \cite{Romano2017}, and global Gibbs schemes \cite{Laal2024}.

Despite these advances, current research has largely remained focused on specific polarization modes under an isotropic background. In anticipation of higher-precision data from next-generation PTAs (e.g., the Square Kilometre Array PTA) and the scientific goal of transitioning from ``detection'' to ``precision measurement'' \cite{Janssen2015}, a universal analytical ORF framework that seamlessly bridges isotropic and anisotropic analyses for arbitrary sky distributions (i.e., arbitrary spherical harmonic coefficients) is still lacking. The absence of such a tool severely limits our ability to extract the rich astrophysical and cosmological information encoded in the observed correlated signal, such as producing full-sky anisotropic power maps of the SGWB or systematically searching for faint non-Einsteinian polarization signals.

In this work, we aim to construct precisely such a universal analytical framework. We endeavor to derive general analytical expressions for the ORFs applicable to arbitrary anisotropic distributions, naturally encompassing the classical HD curve, existing isotropic solutions, and known analytical results for specific polarization modes as special cases. This work will provide an indispensable theoretical foundation for ultimately confirming the physical origin of the SGWB, enabling its precise cosmological measurement, and performing stringent tests of fundamental gravitational theories with PTAs.

The structure of this paper is as follows. In Sec.~\ref{sec2}, we briefly describe the correlation of PTA signals and outline the definition of the ORF integral. Sec.~\ref{sec3} presents the derivation of the analytical expansion for the ORFs of all polarization modes and analyzes the properties of the ORFs from different perspectives. Finally, a brief discussion is provided in Sec.~\ref{sec4}.

\section{Detecting Anisotropies of the SGWB with PTAs}\label{sec2}

\subsection{The SGWB and its anisotropy}

The metric perturbation corresponding to an SGWB can be expressed as
\begin{equation}
  h_{ab}  (\mathbf{x}, t) = \int_{- \infty}^{+ \infty} \mathrm{d} f \int
  \mathrm{d} \Omega_{\hat{k}} \, \mathrm{e}^{- 2 \pi i f (t -
  \hat{k} \cdot \mathbf{x})} \sum_A \tilde{h}_A  (f, \hat{k}) e_{ab}^A
  (\hat{k}), \label{h_ab_t}
\end{equation}
where $e^A_{ab} (\hat{n})$ are spin-2 polarization tensors and $\hat{k} = - \hat{n}$
is the unit wave vector of the gravitational wave. $A = \{+, \times, X, Y, B, L\}$
denotes the different polarization modes: $+, \times$
represent the tensor modes predicted by general relativity, while $X, Y$ and $B, L$
denote the vector and scalar modes allowed in general metric theories, respectively. Explicitly,
\begin{equation}
  \begin{aligned}
    e^+_{ab} (\hat{n}) & = \hat{\theta}_a  \hat{\theta}_b - \hat{\phi}_a
    \hat{\phi}_b, \quad e^{\times}_{ab} (\hat{n}) = \hat{\theta}_a
    \hat{\phi}_b + \hat{\phi}_a  \hat{\theta}_b,\\
    e^X_{ab} (\hat{n}) & = \hat{\theta}_a  \hat{n}_b + \hat{n}_a
    \hat{\theta}_b, \quad e^Y_{ab} (\hat{n}) = \hat{\phi}_a  \hat{n}_b +
    \hat{n}_a  \hat{\phi}_b,\\
    e^B_{ab} (\hat{n}) & = \hat{\theta}_a  \hat{\theta}_b + \hat{\phi}_a
    \hat{\phi}_b, \quad e^L_{ab} (\hat{n}) = \sqrt{2} \,
    \hat{n}_a  \hat{n}_b, \label{e_ab_n}
  \end{aligned}
\end{equation}
where
\begin{equation}
  \begin{aligned}
    \hat{n} & = (\sin \theta \cos \phi, \sin \theta \sin \phi, \cos \theta),\\
    \hat{\theta} & = (\cos \theta \cos \phi, \cos \theta \sin \phi, - \sin
    \theta),\\
    \hat{\phi} & = (- \sin \phi, \cos \phi, 0) . \label{n}
  \end{aligned}
\end{equation}
For an anisotropic, unpolarized, and stationary SGWB, the second-order expectation values satisfy
\begin{equation}
  \langle \tilde{h}_A (f, \hat{k}) \tilde{h}_{A'}^{\ast} (f', \hat{k}')
  \rangle = \frac{1}{2} \mathcal{P}_A (f, \hat{k}) \delta (f - f')
  \delta_{AA'} \delta^2  (\hat{k}, \hat{k}'), \label{S_h_old}
\end{equation}
where $\mathcal{P}_A (f, \hat{k})$ is the angular power spectral density, describing the power distribution of gravitational waves at frequency $f$ across the sky. The factor $\delta (f - f')$
arises from the assumption of stationarity. The factor $\delta_{AA'}$
indicates statistical independence between different polarization modes, assuming no preferred polarization component. Thus, we can write
$\mathcal{P}_+ (f, \hat{n}) =\mathcal{P}_{\times} (f, \hat{n}) =\mathcal{P}_T
(f, \hat{n}) / 2$ and $\mathcal{P}_X (f, \hat{n}) =\mathcal{P}_Y (f,
\hat{n}) =\mathcal{P}_V (f, \hat{n}) /
2$. Note that the two scalar modes should be considered as two independent polarization modes. The factor
$\delta^2  (\hat{k}, \hat{k}')$
stems from the assumption of spatial homogeneity and isotropy.

If the angular power spectral density $\mathcal{P} (f, \hat{n})$
is expanded in spherical harmonics, for an unpolarized and anisotropic SGWB with intensity
$I$, we can define
\begin{eqnarray}
  \langle \tilde{h}_A (f, \hat{k}) \tilde{h}_{A'} (f', \hat{k}') \rangle & = &
  \delta (f - f')  \frac{\delta^{(2)}  (\hat{k} - \hat{k}')}{4 \pi}
  \delta_{AA'} \nonumber\\
  &  & \sum_{\ell m} \tilde{I}_{A,\ell m} (f)  \tilde{Y}^{\ell m} (\hat{k}),
  \label{eq:aver-hh-aa}
\end{eqnarray}
where $\tilde{Y}^{\ell m} (\hat{k}) \equiv
\sqrt{4 \pi} Y^{\ell m} (\hat{k})$
are the normalized spherical harmonics such that $\tilde{Y}^{00} (\hat{k}) =
1$. Therefore, measuring the anisotropy translates to measuring $\tilde{I}_{\ell m} (f)$.

For an isotropic background, the $(0, 0)$ term in Eq.~\eqref{eq:aver-hh-aa} corresponds to the second-order moment, which is entirely determined by the power spectral density $S_h (f)$:
\begin{equation}
  S_{A,h} (f) = \int \mathrm{d}^2 \Omega_{\hat{n}} \,
  \mathcal{P}_A (f, \hat{n}) = \frac{3 H_0^2}{2 \pi^2}
  \frac{\Omega_{\mathrm{GW}} (f)}{f^3}, \label{e:Sh-Omega_g}
\end{equation}
a relation that can be derived in detail from Ref.~\cite{Allen:1997ad}, where
\begin{equation}
  \Omega_{\mathrm{GW}} (f) = \frac{1}{\rho_c}  \frac{\mathrm{d}
  \rho_{\mathrm{GW}}}{\mathrm{d} \ln f}
\end{equation}
denotes the gravitational-wave energy density contained in the frequency interval from $f$ to $f + \mathrm{d} f$.
Here, $\rho_c \equiv \frac{3 c^2
H_0^2}{8 \pi G}$ is the critical energy density of the present Universe, with $H_0$ and $G$
being the Hubble constant and gravitational constant, respectively. If anisotropies induced by specific sources are considered, it is necessary to relate
$\tilde{I}_{\ell m} (f)$
to the angular power spectrum of the sources. In general, this relationship can be expressed as
\cite{LISACosmologyWorkingGroup:2022kbp}
\begin{equation}
  \tilde{I}_{\ell m} (f) = \frac{1}{\sqrt{4 \pi}}  \frac{3 H_0^2}{4 \pi^2}
  \frac{\Omega_{\mathrm{GW}} (f)}{f^3} \delta_{\mathrm{GW}, \ell m},
  \label{I-to-delta}
\end{equation}
where $\delta_{\mathrm{GW}, \ell m}$ denotes the spherical harmonic components of the density contrast:
\begin{equation}
  \delta_{\mathrm{GW}}  (f, \hat{k}) = \sum_{\ell} \sum_{m = - \ell}^{\ell}
  \delta_{\mathrm{GW}, \ell m} (f) Y^{\ell m} (\hat{k}) . \label{e:dec}
\end{equation}
In general, we have:
\begin{equation}
  \langle \delta_{\mathrm{GW},\ell m} \delta_{\mathrm{GW},\ell' m'}^{\ast}
  \rangle = C_{\mathrm{GW},\ell m} (f) \delta_{\ell \ell'} \delta_{mm'} .
  \label{Cellm-def}
\end{equation}
Furthermore, assuming that $C_{\mathrm{GW},\ell m} (f)$ is the same for the same $\ell$
but different $m$, we have:
\begin{equation}
\langle \delta_{\mathrm{GW},\ell m} \delta_{\mathrm{GW},\ell' m'}^{\ast}
\rangle = C_{\mathrm{GW},\ell} (f) \delta_{\ell \ell'} \delta_{mm'} .
\label{Cell-def}
\end{equation}
$C_{\mathrm{GW},\ell m} (f)$ and $C_{\mathrm{GW},\ell} (f)$
are referred to as the angular power spectra of the sources. A detailed discussion can be found in Ref.~
\cite{LISACosmologyWorkingGroup:2022kbp}.

\subsection{Cross-correlation of PTA signals}

Since the metric perturbation is weak, the timing signal for a single pulsar can be expressed as
\begin{equation}
  \label{h_t} h (t) = \int_{- \infty}^{+ \infty} \mathrm{d} f \int
  \mathrm{d}^2 \Omega_{\hat{n}} \, R^{ab} (f, \hat{n}) h_{ab}
  (f, \hat{n}) e^{i 2 \pi f t} .
\end{equation}
Here, the response function for Doppler frequency measurements is given by \cite{Romano2017}
\begin{eqnarray}
  \label{e:R_ab} R^{ab} (f, \hat{n}, \vec{r}_2) & = & \frac{1}{2}  \frac{u^a
  u^b}{1 + \hat{n} \cdot \hat{u}}  \left( 1 - e^{- \frac{i 2 \pi f L}{c} (1 +
  \hat{n} \cdot \hat{u})} \right) \nonumber\\
  &  & e^{i 2 \pi f \hat{n} \cdot \vec{r}_2 / c},\label{e:pulsarresponse-earthonly}
\end{eqnarray}
where $\hat{u}$ is the unit vector pointing from Earth to the pulsar, $L$
is the distance from the pulsar to Earth, and $\vec{r}_2 = L_2  \hat{r}_2$
is the position vector of Earth, with $L_2$ being the distance from Earth to the pulsar.

In the frequency domain, the signal expressed in the polarization basis is
\begin{equation}
  \label{h_f} \tilde{h} (f) = \int \mathrm{d}^2 \Omega_{\hat{n}}  \sum_A R^A
  (f, \hat{n})  \tilde{h}_A  (f, \hat{n}),
\end{equation}
where $R^A (f, \hat{n}) = R^{ab} (f, \hat{n}) e^A_{ab} (\hat{n})$.

The cross-correlation of signals from two pulsars $i$ and $j$ is:
\begin{equation}
  \label{h_I_h_J} \langle \tilde{h}_{A,i} (f) \tilde{h}_{A,j}^{\ast} (f') \rangle =
  \frac{1}{2} \delta (f - f')  \sum_{\ell m} \int_0^{\infty} \mathrm{d} f
  \,  \Gamma_{ij}^{A, \ell m} (f)  \tilde{I}_{A,\ell m} (f),
\end{equation}
where
\begin{equation}
  \Gamma_{ ij}^{A, \ell m} (f) \equiv \frac{1}{4 \pi}  \int \mathrm{d}^2
  \Omega_{\hat{n}} \, \tilde{Y}^{\ell m} (\hat{k}) R^A_{ij} (f,
  \hat{k}), \label{eq:response}
\end{equation}
with $A = \{T, V, B, L\}$. Here, $\tilde{Y}^{\ell m} (\hat{k}) =
\sqrt{4 \pi} Y^{\ell m} (\hat{k})$, so the $(0, 0)$ term of $\Gamma_{ij}^{A, \ell m}
(f)$ corresponds to the isotropic case. $R^A_{ij} (f, \hat{k})$
is written as
\begin{equation}
  R^A_{ij} (f, \hat{k}) \equiv \alpha \sum_A R_i^A (f, \hat{k}, r_i) R_j^{A
  \ast} (f, \hat{k}, r_j), \label{e:GammaIJ-New}
\end{equation}
where $\alpha = \frac{1}{2}$ for tensor and vector modes and $\alpha = 1$
for scalar breathing and scalar longitudinal modes. The $(\ell, m)$
order ORFs are then
\begin{equation}
  \Gamma^{A, \ell m}_{ij} (a_u, a_v, \gamma) = \frac{1}{4 \pi}  \int
  \mathrm{d} \Omega_{\hat{n}} \, R^A_{ij} (a_u, a_v, \theta,
  \phi)  \tilde{Y}^{\ell m} (\theta, \phi) .
\end{equation}

In practical data analysis, for $N_{\mathrm{p}}$
pulsars, the number of independent pulsar pairs is $N_{\mathrm{pairs}} =
N_{\mathrm{p}}  (N_{\mathrm{p}} - 1) /
2$. According to information-theoretic criteria, the maximum number of modes that can be stably reconstructed satisfies the constraint
$\sum_{\ell = 0}^{\ell_{\mathrm{max}}} (2 \ell + 1) \lesssim
N_{\mathrm{pairs}}$. For current PTAs, $N_{\mathrm{p}} \approx
70$, so $N_{\mathrm{pairs}} \approx 2415$, and theoretically $\ell_{\mathrm{max}} \sim
50$ can be detected. However, in practice, due to limitations in signal-to-noise ratio and uneven pulsar distribution, the actual detectable
$\ell_{\mathrm{max}}$
is much smaller. In such cases, having an analytical form of the ORFs would be very convenient.

\subsection{Symmetries of the ORFs}\label{sec:ORFssymmetry}

As expected, similar to space-based gravitational-wave detectors \cite{Zhou:2023rop, LISACosmologyWorkingGroup:2022kbp}, the anisotropic overlap reduction functions $\Gamma_{ij}^{\ell m}(f)$ for PTAs should also possess corresponding symmetry relations. This is indeed the case. However, these symmetries are governed by the intrinsic properties of PTAs themselves and differ significantly from those applicable to space-based detectors, necessitating a separate investigation. In the following, we derive several important symmetry properties of $\Gamma_{ij}^{\ell m}(f)$. As will be shown, these symmetry relations not only carry physical implications and serve to simplify calculations, but also provide a useful means of verifying computational results. The corresponding proofs are given in Appendix \ref{sec:ORFsymmetryproof}.

\begin{theorem}
  For any rigid rotation $\mathscr{R}$, we have
  \begin{equation}
    \Gamma_{\mathscr{R} i \, \mathscr{R} j}^{\ell m} (a_u, a_v,
    \gamma) = \sum_{m' = - \ell}^{\ell} [D_{mm'}^{(\ell)} (R)]^{\ast}
    \Gamma_{ij}^{\ell m'} (a_u, a_v, \gamma), \label{rot-R}
  \end{equation}
  where $D_{mm'}^{(\ell)}$ are the matrix elements of the Wigner D-matrix.
\end{theorem}

\begin{corollary}
  For a rotation about the $z$ axis by an angle $\alpha$, we have
  \begin{equation}
    \Gamma_{\mathscr{R}_z (\alpha) i, \, \mathscr{R}_z (\alpha)
    j}^{\ell m} (a_u, a_v, \gamma) = \mathrm{e}^{i m \alpha} \Gamma_{ij}^{\ell
    m} (a_u, a_v, \gamma) . \label{rotz-R}
  \end{equation}
\end{corollary}

\begin{theorem}
  If $\ell + m = 2 n + 1$, $n \in \mathbb{N}$, then
  \begin{equation}
    \Gamma_{ij}^{\ell m} (a_u, a_v, \gamma) = 0. \label{eq:R-prop-lm}
  \end{equation}
\end{theorem}

\begin{theorem}
  Regarding $a_u$ and $a_v$, we have
  \begin{equation}
    \Gamma_{ij}^{\ell m} (a_u, a_v, \gamma) = \Gamma_{ji}^{\ell m \ast} (a_v,
    a_u, \gamma) . \label{eq:R-rop-lm4}
  \end{equation}
\end{theorem}

\begin{theorem}
  When $\gamma = 0$ or $a_u, a_v \rightarrow + \infty$, for any
  $(\ell, m)$, we have
  \begin{equation}
    \Gamma_{ij}^{\ell, - m} = (- 1)^m \Gamma_{ij}^{\ell m} .
    \label{eq:R-prop-menom-app-old}
  \end{equation}
\end{theorem}

Because of the chosen reference frame and the symmetries of the spherical harmonics, Eq.~\eqref{eq:R-prop-lm} holds when $\ell + m$ is odd, which means the corresponding ORFs vanish. Therefore, we need to compute only the ORFs for $\ell + m$ even. With respect to $a_u$ and $a_v$, the ORFs satisfy the exchange symmetry relation \eqref{eq:R-rop-lm4}, so the final expressions can be further simplified. In practical situations, when considering multiple pulsar pairs, the reference frame may be arbitrary, and the ORFs can then be obtained via Eq.~\eqref{rot-R}. For the special cases $\gamma = 0$ and the short-wavelength approximation $a_u, a_v \rightarrow +\infty$, all modes except the longitudinal mode (since terms containing $a_u$ and $a_v$ cannot be eliminated) satisfy Eq.~\eqref{eq:R-prop-menom-app-old}.

\section{Anisotropic Overlap Reduction Functions}\label{sec3}
\begin{figure}[!t]
	\begin{center}
		\resizebox{240pt}{220pt}{\includegraphics{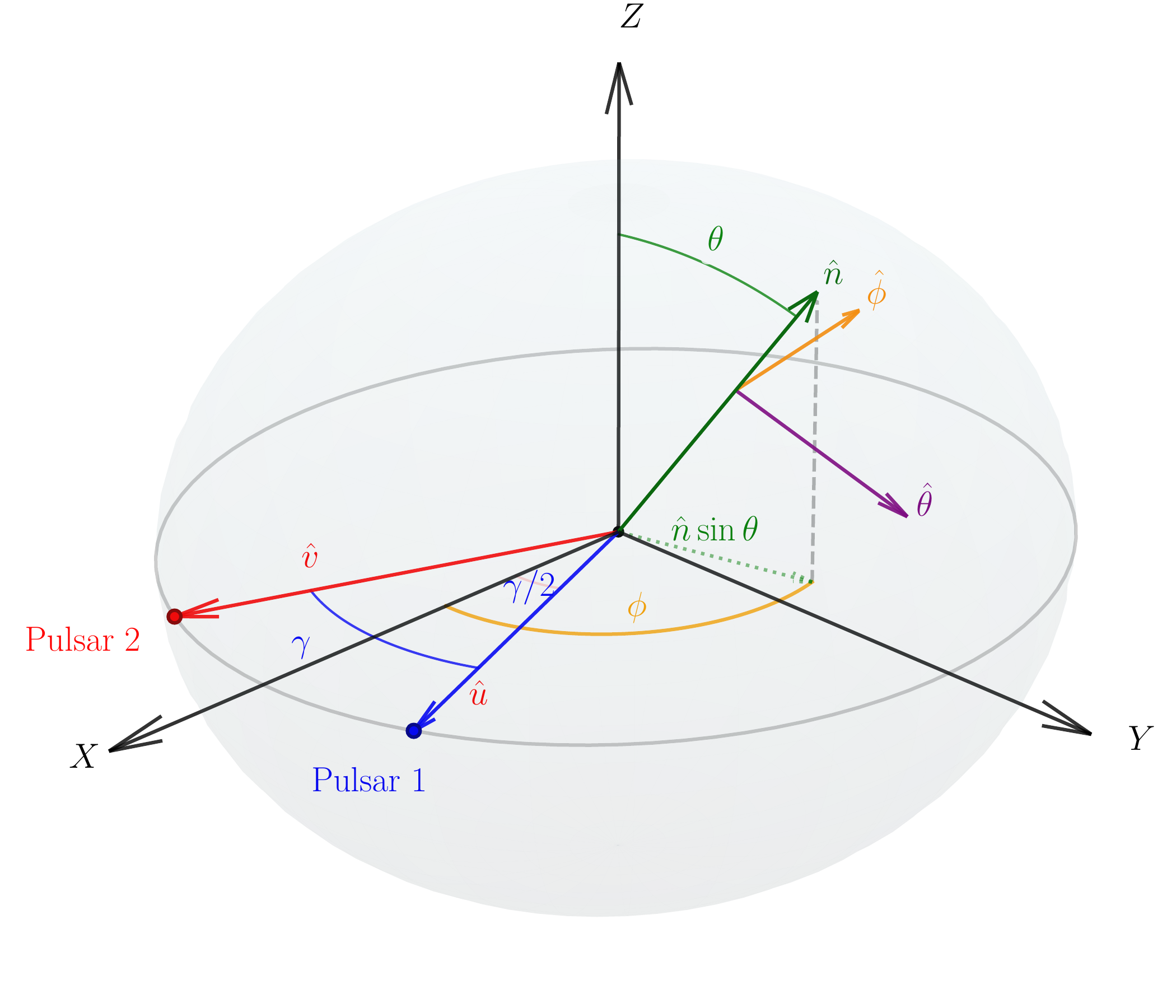}}
		\caption{\label{fig:coord} Our convention for the coordinates.}
	\end{center}
\end{figure}

The ORF depends only on the relative directions of the two pulsars and their distances from Earth. Therefore, we can choose a suitable coordinate system to simplify the integration. We choose Earth as the origin, such that $\vec{r}_2 = 0$. The direction vectors of the two pulsars are $\hat{u} = (\cos
\frac{\gamma}{2}, \sin \frac{\gamma}{2}, 0)$ and $\hat{v} = (\cos
\frac{\gamma}{2}, - \sin \frac{\gamma}{2}, 0)$, respectively, where $\gamma$
is the angular separation between the two pulsars. For convenience, we adopt the coordinate system shown in Fig.~\ref{fig:coord} and define
\begin{equation}
  \begin{aligned}
  \tilde{x} = \hat{n} \cdot \hat{u} = \sin \theta \cos \tilde{\phi},
  \tilde{y} = \hat{n} \cdot \hat{v} = \sin \theta \cos \underset{\sim}{\phi},
  \end{aligned}
\end{equation}
where $\underset{\sim}{\phi} = \phi - \frac{\gamma}{2}$ and $\tilde{\phi} =
\phi + \frac{\gamma}{2}$. Substituting into
\eqref{e:GammaIJ-New}, for the tensor modes we have
\begin{eqnarray}
  R^T_{ij} (a_u, a_v, \tilde{x}, \tilde{y}) & = & \frac{1}{8} (f^{T_1}  (a_u,
  a_v, \tilde{x}, \tilde{y}) + \nonumber\\
  &  & f^{T_2}  (a_u, a_v, \tilde{x}, \tilde{y})),  \label{eq:RvuTsm}
\end{eqnarray}
\begin{eqnarray}
  R^V_{ij} (a_u, a_v, \tilde{x}, \tilde{y}) & = & \frac{1}{8} f^V  (a_u, a_v,
  \tilde{x}, \tilde{y}) .  \label{eq:RvuVsm}
\end{eqnarray}
For the breathing and longitudinal modes,
\begin{equation}
  R^A_{ij} (a_u, a_v, \tilde{x}, \tilde{y}) = \frac{1}{4} f^A  (a_u, a_v,
  \tilde{x}, \tilde{y}) \label{eq:RvuBLsm},
\end{equation}
where $a_v = 2 \pi f L_v / c$, $a_u = 2 \pi f L_u / c$, and $L_v$ and $L_u$
are the distances from the two pulsars to Earth, respectively. Let
\begin{equation}
  P (a_u, a_v, \tilde{x}, \tilde{y}) = (e^{i a_v (\tilde{x} + 1)} - 1)
  (e^{- i a_u (\tilde{y} + 1)} - 1),
\end{equation}
\begin{equation}
  Q (a_u, a_v, \tilde{x}, \tilde{y}) = \frac{P (a_u, a_v, \tilde{x},
  \tilde{y})}{(\tilde{x} + 1)  (\tilde{y} + 1)},
\end{equation}
and then
\begin{eqnarray}
  f^B  (a_u, a_v, \tilde{x}, \tilde{y}) & = & f^{T_1}  (a_u, a_v, \tilde{x},
  \tilde{y}) \nonumber\\
  & = & (\tilde{x} - 1)  (\tilde{y} - 1) P (a_u, a_v, \tilde{x}, \tilde{y}),
  \label{eq:funcBT1}
\end{eqnarray}
\begin{equation}
  f^{T_2}  (a_u, a_v, \tilde{x}, \tilde{y}) = - 2 Q (a_u, a_v, \tilde{x},
  \tilde{y}) \cos^2 (\theta) \sin^2 (\gamma) \label{eq:funcTT2},
\end{equation}
\begin{equation}
  f^V  (a_u, a_v, \tilde{x}, \tilde{y}) = Q (a_u, a_v, \tilde{x}, \tilde{y}) 4
  \tilde{x}  \tilde{y}  (\cos \gamma - \tilde{x}  \tilde{y}),\label{eq:funcV}
\end{equation}
\begin{equation}
  f^L  (a_u, a_v, \tilde{x}, \tilde{y}) = Q (a_u, a_v, \tilde{x}, \tilde{y}) 2
  \tilde{x}^2  \tilde{y}^2 .\label{eq:funcL}
\end{equation}
By transforming the integration variables $\{\theta, \phi\}$ to $\{\tilde{x},
\tilde{y}\}$, the integration domain changes from the unit sphere to an elliptical region in the $xy$
plane. Because of Eq.~\eqref{eq:R-prop-lm}, we only need to consider where $\ell + m$ is even.
The $\ell m$-order ORF can then be expressed in the $(\tilde{x}, \tilde{y})$
representation as the following integral
\begin{eqnarray}
  \Gamma^{A, \ell m}_{ij} (a_u, a_v, \gamma) & = & R_{ij}  (a_u, a_v,
  \tilde{x}, \tilde{y}) \odot \tilde{Y}^{\ell m} (\tilde{x}, \tilde{y})
  \nonumber\\
  & = & \frac{1}{4 \pi}  \int_E R^A_{i j} (a_u, a_v, \tilde{x}, \tilde{y})
  \tilde{Y}^{\ell m} (\tilde{x}, \tilde{y}) \nonumber\\
  &  & \frac{1}{\sin (\gamma) \cos \theta} d \tilde{x} d \tilde{y} .
  \label{eq:orf-def-xy-old}
\end{eqnarray}
This integral can be evaluated analytically.
Based on the above integration algorithm and using the symmetry relations described in Sec.~\ref{sec:ORFssymmetry}, one can, in principle, compute the ORFs to any order. These formulas may involve special functions.
\begin{equation}
  E_1 (z) = \int_z^{\infty} \frac{e^{- t}}{t} dt, \quad | \arg (z) | < \pi,
\end{equation}
\begin{equation}
  \mathrm{Si} (x) = \int_0^x \frac{\sin t}{t} dt,
\end{equation}
\begin{equation}
  \mathrm{Ci} (x) = - \int_x^{\infty} \frac{\cos t}{t} dt = \gamma_E + \ln x +
  \int_0^x \frac{\cos t - 1}{t} dt, \quad x > 0,
\end{equation}
where $\gamma_E$ is the Euler constant ($\gamma_E \approx 0.5772156649$).

Here we present the analytical forms of all ORFs for $\ell \leqslant 5$.
For the detailed expressions of these formulas, 
please see Ref.~\cite{zhou2026:git}
The following subsections will analyze these results in detail.

\subsection{Tensor modes}
\begin{figure*}[!t]
	\resizebox{230pt}{142.5pt}{\includegraphics{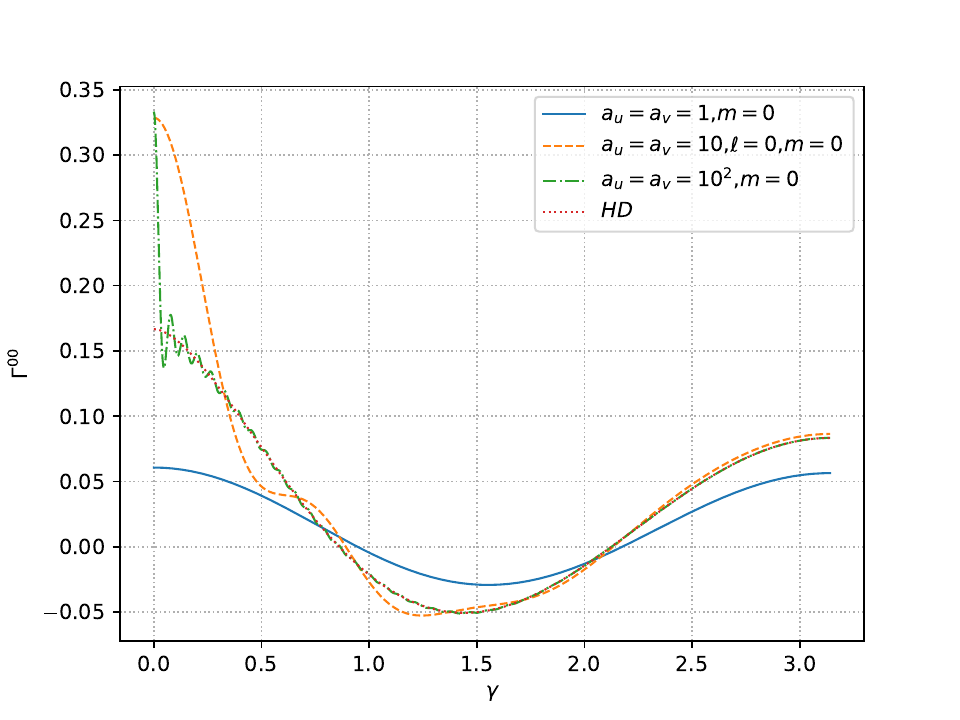}}\resizebox{230pt}{142.5pt}{\includegraphics{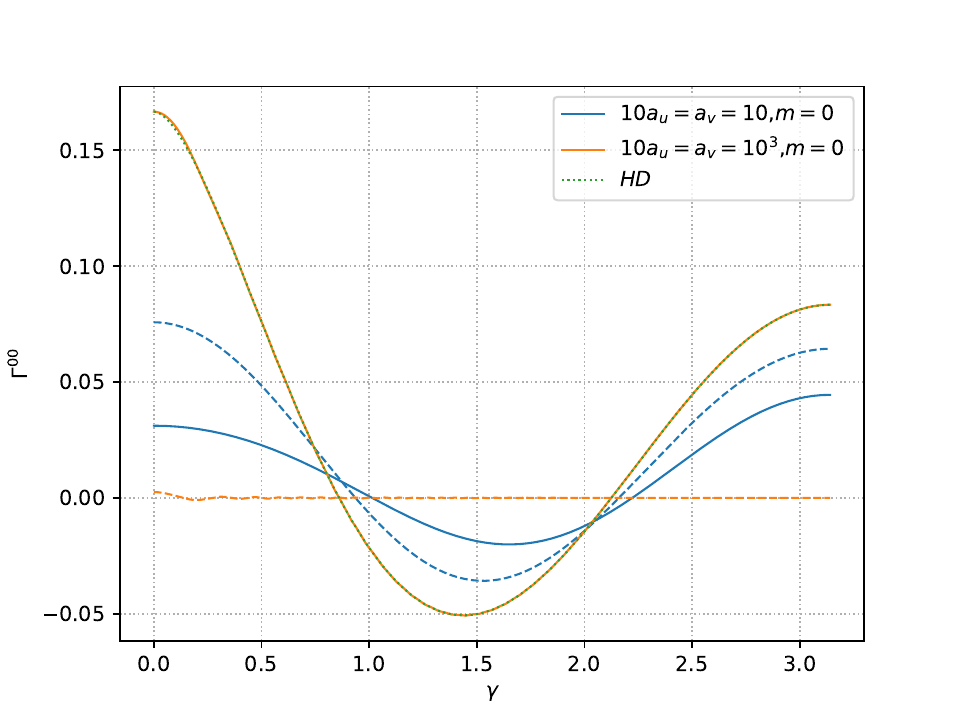}}
	
	\resizebox{230pt}{142.5pt}{\includegraphics{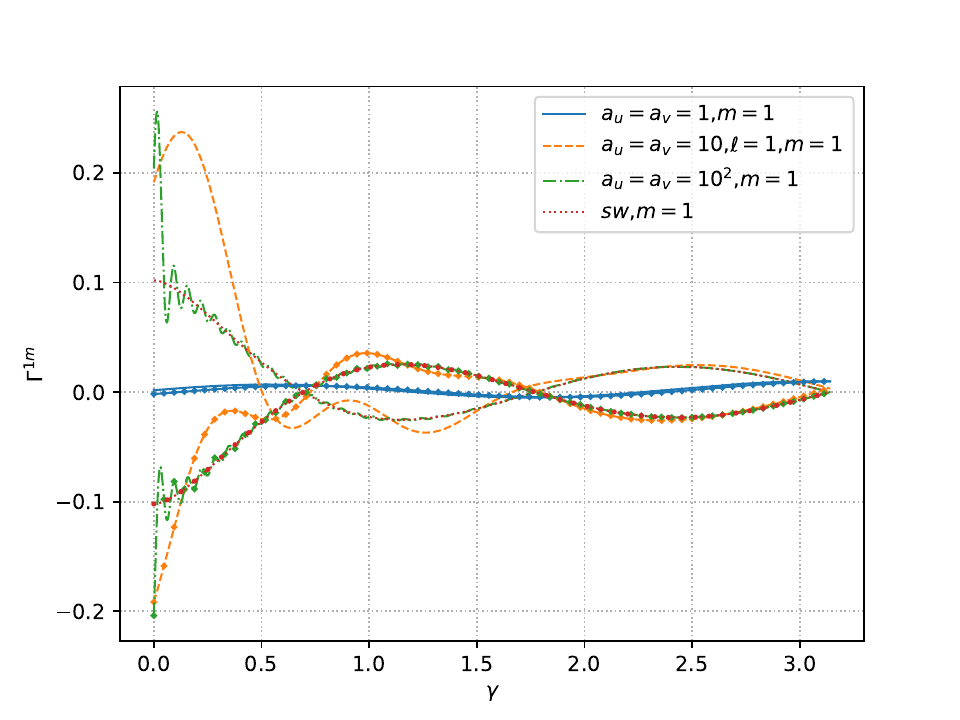}}\resizebox{230pt}{142.5pt}{\includegraphics{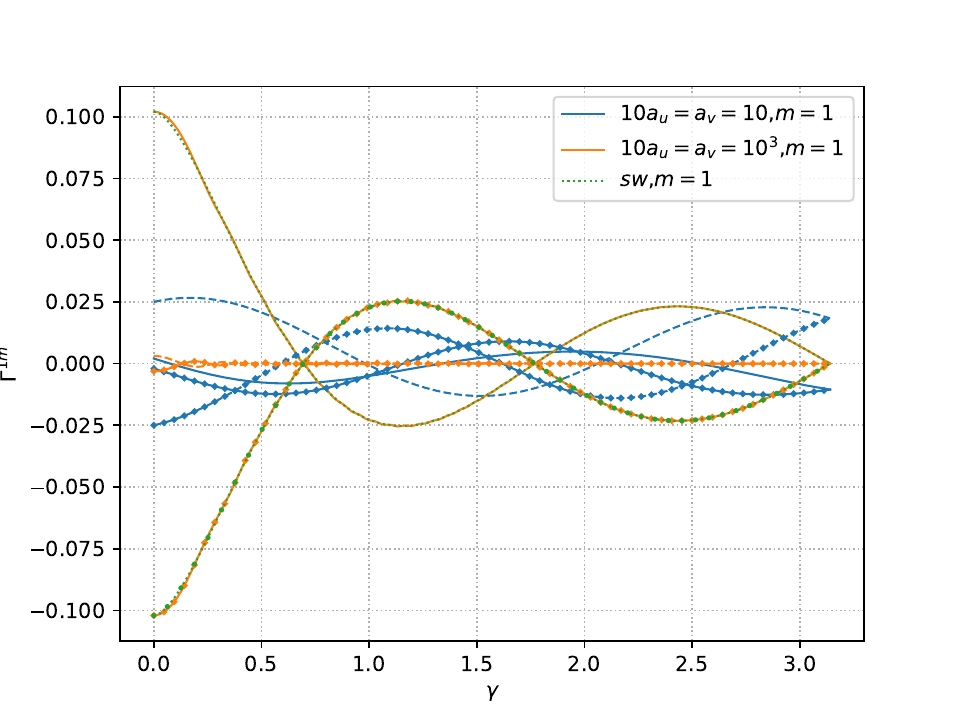}}
	
	\resizebox{230pt}{142.5pt}{\includegraphics{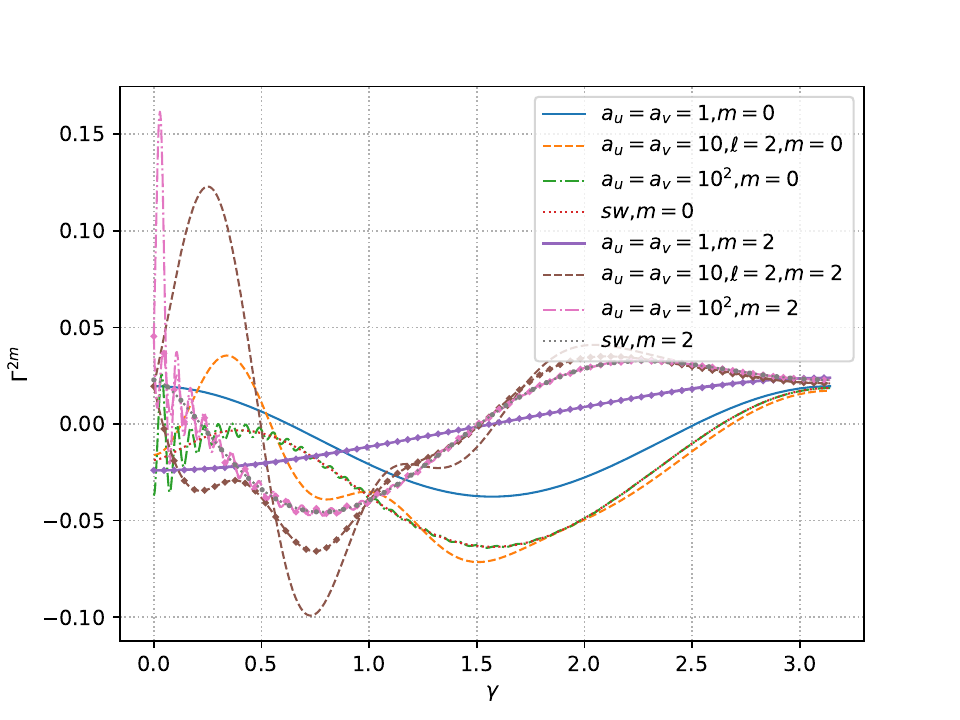}}\resizebox{230pt}{142.5pt}{\includegraphics{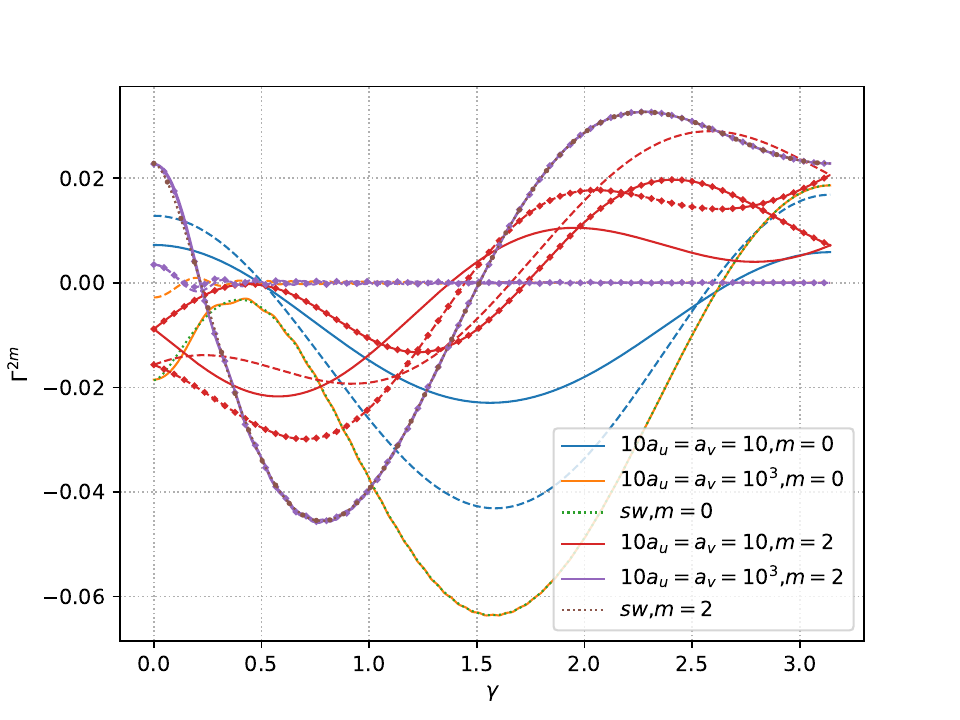}}
	
	\caption{\label{fig:orfT-lm-gamma} ORFs for tensor modes with $\ell = 0, 1,
		2$. Left: $L_u = L_v$. Right: $L_u \neq
		L_v$. Solid lines: real part; dashed lines: imaginary part. Markers correspond to $m <
		0$ cases: diamonds for the general form and circles for the short-wavelength approximation.}
\end{figure*}
\begin{figure*}[!t] 
	\resizebox{230pt}{142.5pt}{\includegraphics{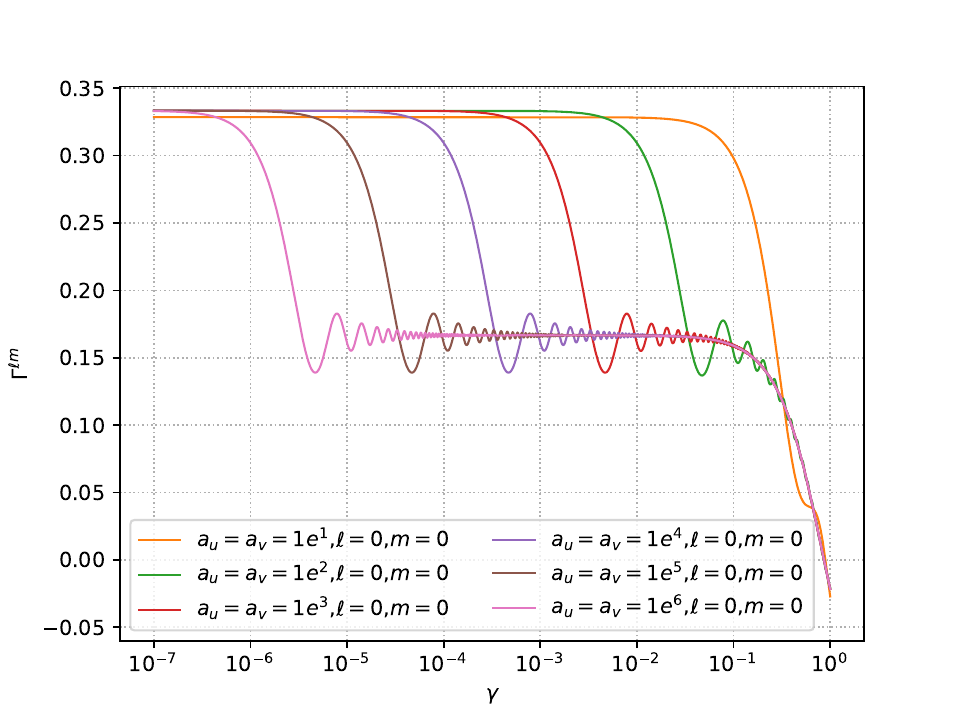}}\
	\resizebox{230pt}{142.5pt}{\includegraphics{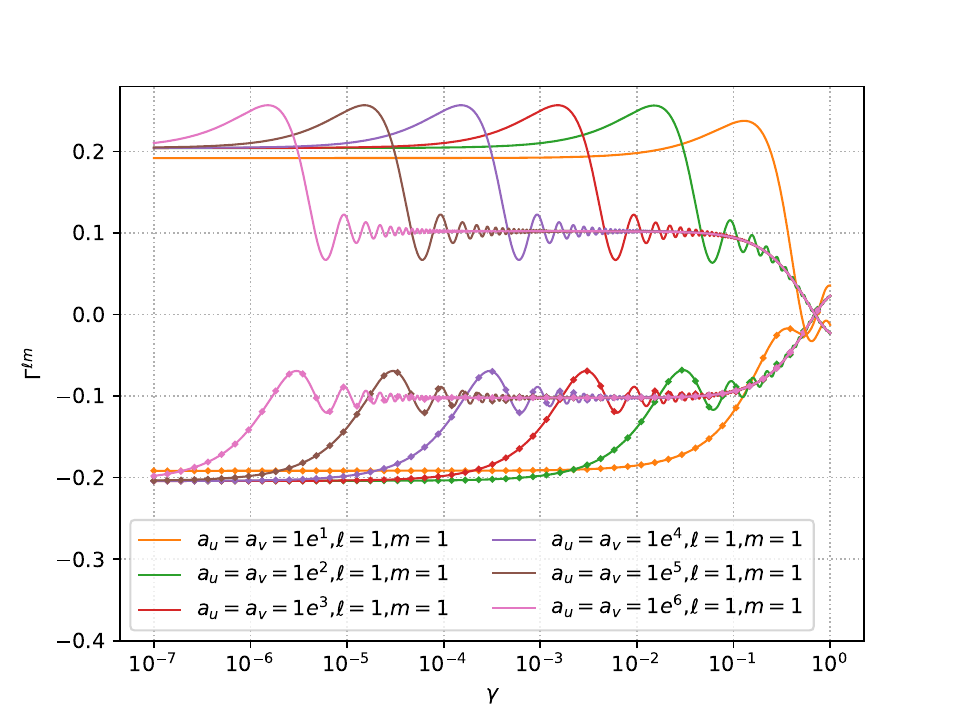}}\
	\resizebox{230pt}{142.5pt}{\includegraphics{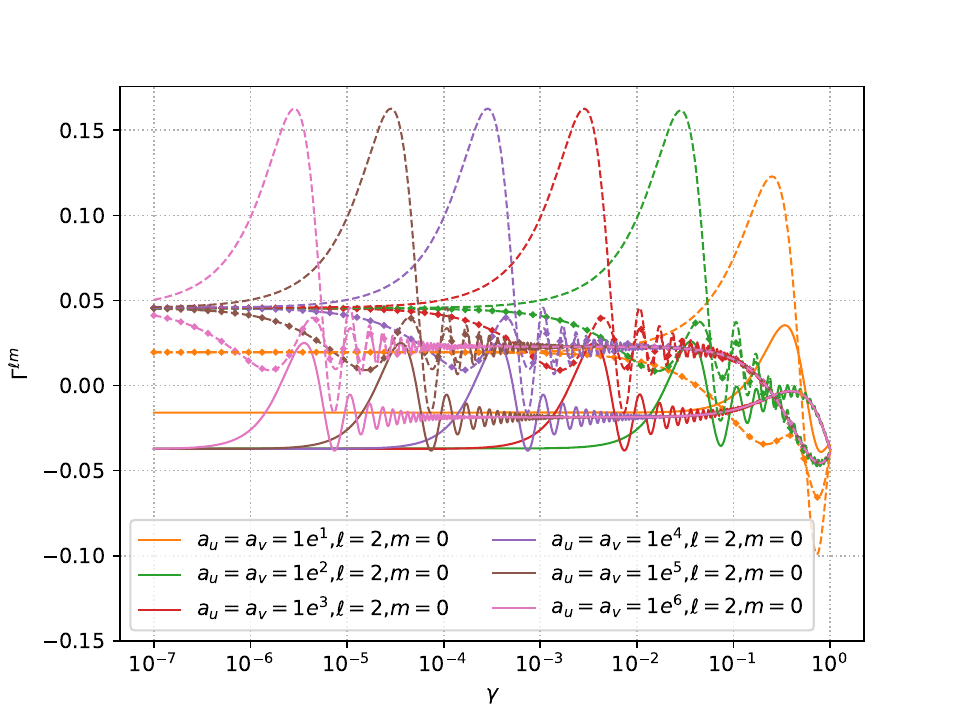}}\
	\resizebox{230pt}{142.5pt}{\includegraphics{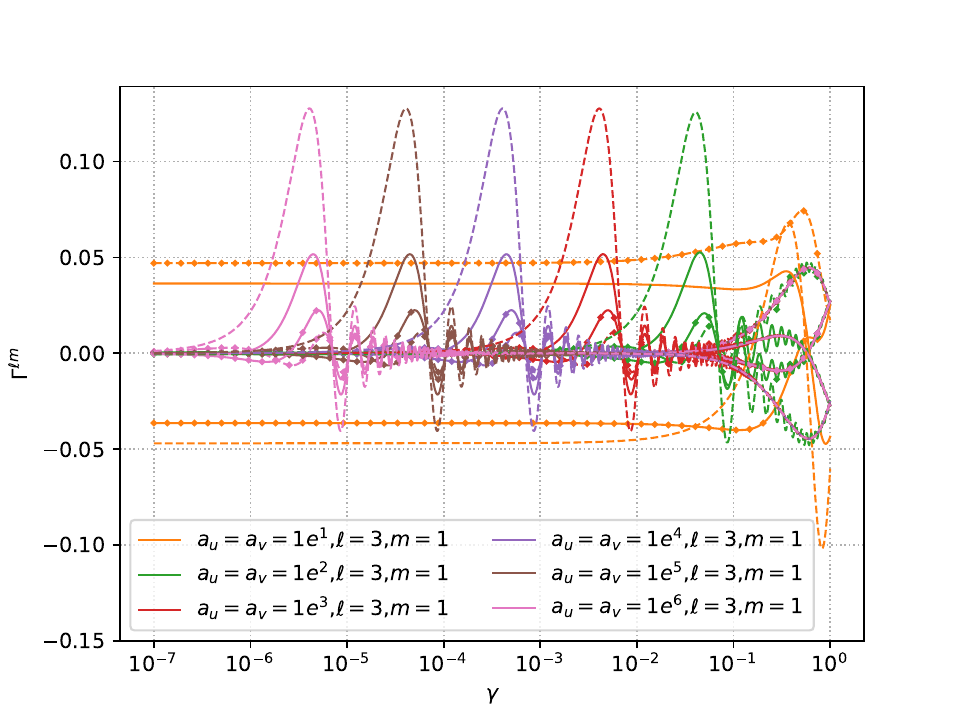}}\
	\caption{\label{fig:orfT-lm-litgamma} Behavior of the tensor mode at small angular separations for $L_u = L_v$. For $\ell=2$: solid line, $(2,0)$; dashed line, $(2,2)$. For $\ell=3$: solid line, $(3,1)$; dashed line, $(3,3)$. Markers correspond to $m<0$ cases: diamonds for the general form and circles for the short-wavelength approximation.}
\end{figure*}
For tensor modes, with $\tilde{\gamma} = \gamma/2$, the analytical results in the short-wavelength approximation $a_v,
a_u \gg 1$ simplify to the following forms.

\begin{equation}
\Gamma_{sw}^{T,0, 0} = \log (\sin (\tilde{\gamma})) \sin (\tilde{\gamma})^2 +
\frac{1}{24}   \cos (\gamma) + \frac{1}{8},
\end{equation}
\begin{eqnarray}
\Gamma_{sw}^{T,1, 1} & = & \frac{1}{48}   \sqrt{6}  ( 24
\log (\sin (\tilde{\gamma})) \sin (\tilde{\gamma}) \tan
(\tilde{\gamma}) - 3  \cos \left( 3 
\tilde{\gamma} \right) \nonumber\\
&  & + 5  \cos (\tilde{\gamma}) ),
\end{eqnarray}
\begin{eqnarray}
\Gamma_{sw}^{T,2, 0} & = & - \frac{1}{120}   \sqrt{5} ( 60
\left( 3  \sec (\tilde{\gamma})^2 + \cos(\gamma) - 4 \right) \log (\sin (\tilde{\gamma})) \nonumber\\
&  &- 44  \cos(\gamma) + 45 ),
\end{eqnarray}
\begin{eqnarray}
\Gamma_{sw}^{T,2, 2} & = & \frac{1}{480}   \sqrt{30}  ( 120
\log (\sin (\tilde{\gamma})) \tan (\tilde{\gamma})^2 - 5
\cos \left( 4  \tilde{\gamma} \right)  \nonumber\\
&  &- 30
\cos (\gamma) + 37 ),
\end{eqnarray}

The terms with $m <
0$ can be derived from the above expressions using the symmetry relation \eqref{eq:R-prop-menom-app-old}. The $(0,
0)$ term, when normalized, corresponds to the HD function \cite{Hellings1983}
\begin{equation}
\label{HD} \gamma_{HD} (\gamma) = \frac{3}{2} \left( \frac{1 - \cos
	\gamma}{2} \right) \ln \frac{1 - \cos \gamma}{2} - \frac{1 - \cos \gamma}{8}
+ \frac{1}{2} .
\end{equation}

Moreover, as $\gamma \rightarrow 0$ (i.e., for pulsars with vanishing angular separation), both the analytical expressions and the short-wavelength approximation converge to the same value. Since no approximation is made, the analytical expressions are in excellent agreement with numerical integration. When the distances of the two pulsars are equal, i.e., $a_v = a_u$, the ORFs are functions of a real variable. As shown in Fig.~\ref{fig:orfT-lm-gamma}, as $a_u$ increases, the analytical expressions agree with the short-wavelength approximation. For long wavelengths, $\lim_{\gamma \rightarrow 0} \Gamma^{T,\ell m}$ is a fixed value that depends only on $a_u$. In realistic situations, the distances of the two pulsars are generally different, making the ORFs a set of functions of a complex variable that satisfy the symmetries described in Sec.~\ref{sec:ORFssymmetry}. As shown in Fig.~\ref{fig:orfT-lm-gamma}, when $a_v \neq a_u$, i.e., when the two pulsars are at unequal distances, the imaginary part of the ORFs depends on the ratio of $a_v$ to $a_u$. The larger this ratio, the larger the imaginary part. However, when both $a_v$ and $a_u$ are large, the imaginary part approaches zero.

As shown in Fig.~\ref{fig:orfT-lm-litgamma}, for $a_v = a_u$, the behavior when the angular separation is small is illustrated. As $a_u$ increases, the ORFs decay faster; as $\ell$ increases, the ORF values gradually decrease. For fixed $\ell$, the ORFs for $m > 0$ are larger than those for $m < 0$, and as $|m|$ increases, the ORF values increase. From these highly similar curves, we see that the ORFs for two nearby pulsars depend only on their distance. For pulsars with identical directions, i.e., $\gamma = 0$, let $a_0 = a_u - a_v$. In this case, the ORF can be expressed as $K^{A,lm}_{ij}$ plus the sum of its limits for $a_u \to 0$, $a_v \to 0$, and $(a_u,a_v) \to (0,0)$, where
\begin{equation}
K_{sw}^{T,0, 0} = - \frac{i}{4  a_0} + \frac{1}{4
	a_0^2} - \frac{i  e^{\left( - 2 i
		a_0 \right)}}{8  a_0^3} + \frac{i}{8
	a_0^3},
\end{equation}
\begin{eqnarray}
K_{sw}^{T,1, 1} & = & - \frac{1}{16}   \sqrt{6}  \biggl( \left( -
\frac{i}{a_0^3} - \frac{3}{a_0^4} \right) e^{\left( - 2 i 
	a_0 \right)} + \frac{2 i}{a_0}\nonumber\\
&  & - \frac{4}{a_0^2} - \frac{5 i}{a_0^3} +
\frac{3}{a_0^4} \biggl),
\end{eqnarray}
\begin{eqnarray}
K_{sw}^{T,2, 0} & = & - \frac{1}{16}   \sqrt{5}  \biggl( \left( -
\frac{i}{a_0^3} - \frac{9}{a_0^4} + \frac{18 i}{a_0^5} \right) e^{\left( - 2
	i  a_0 \right)} - \nonumber\\
&  & \frac{2 i}{a_0} + \frac{8}{a_0^2} +
\frac{19 i}{a_0^3} - \frac{27}{a_0^4} - \frac{18 i}{a_0^5} \biggl),
\end{eqnarray}
other terms are determined by the following relations
\begin{equation}
K_{i j}^{A, 2, 2} = - \frac{1}{2}   \sqrt{6} K_{i j}^{A, 2,
	0} \label{eq:KAIJ-lm-rela22},
\end{equation}
\begin{equation}
K_{i j}^{A, 3, 3} = - \frac{\sqrt{15}}{3} K_{i j}^{A, 3, 1}
\label{eq:KAIJ-lm-rela31},
\end{equation}
where $A = \{T, V, B, L\}$. It is also easy to verify that, when $a_u$ is finite, $\lim_{\gamma
	\rightarrow 0}  \Gamma^{T,\ell m}  = F(a_u)$.
Specifically, the autocorrelation responses for a single detector for tensor modes are
\begin{equation}
\Gamma_{i i}^{T,0, 0} = - \frac{1}{2  a_v^2} +
\frac{1}{3},
\end{equation}
\begin{equation}
\Gamma_{i i}^{T,1, 1} = - \frac{1}{24}   \sqrt{6}  \left(
\frac{12}{a_v^2} + \frac{9  \left( \cos \left( 2
	a_v \right) - 1 \right)}{a_v^4} - 2 \right),
\end{equation}
\begin{equation}
\Gamma_{i i}^{T,2, 0} = \frac{1}{120}   \sqrt{5}  \left(
\frac{120}{a_v^2} - \frac{135  \left( \cos \left( 2
	a_v \right) + 3 \right)}{a_v^4} - 2 \right),
\end{equation}
other terms satisfy Eqs.~\eqref{eq:KAIJ-lm-rela22} to \eqref{eq:KAIJ-lm-rela31}, with the letter $K$ replaced by $\Gamma$.

For $a_u \gg 1$, the ORFs converge. But as $a_u$ increases, when $\gamma \rightarrow 0$, the ORFs tend to a constant. In general, for fixed $(\ell, m)$, the real part of the ORF increases with growing $a_v$ and $a_u$, while the imaginary part decreases and approaches zero. As $\ell$ increases, the ORF values gradually decrease. For fixed $\ell$, the ORFs for $m > 0$ are larger than those for $m < 0$, and they increase with $|m|$.

\subsection{Vector modes}
\begin{figure*}[!t]
	\resizebox{230pt}{142.5pt}{\includegraphics{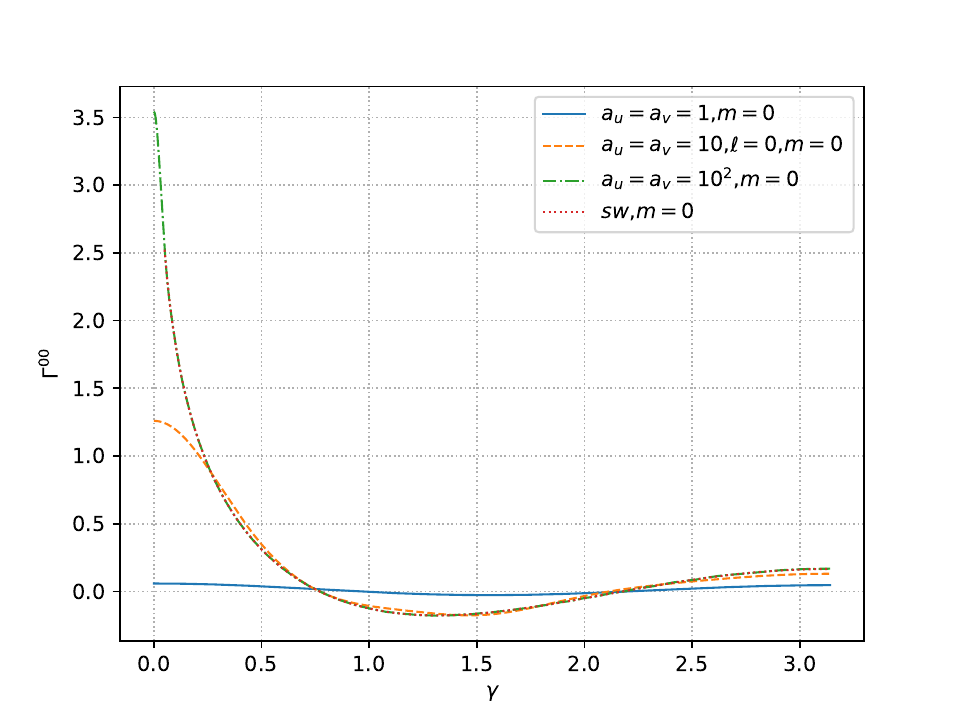}}\resizebox{230pt}{142.5pt}{\includegraphics{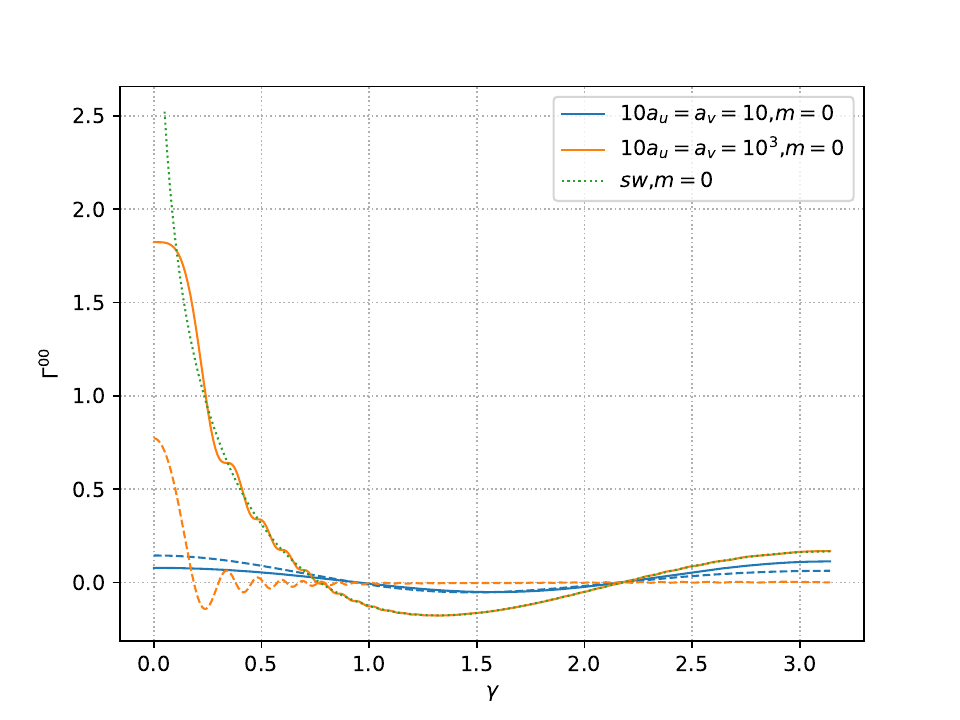}}
	
	\resizebox{230pt}{142.5pt}{\includegraphics{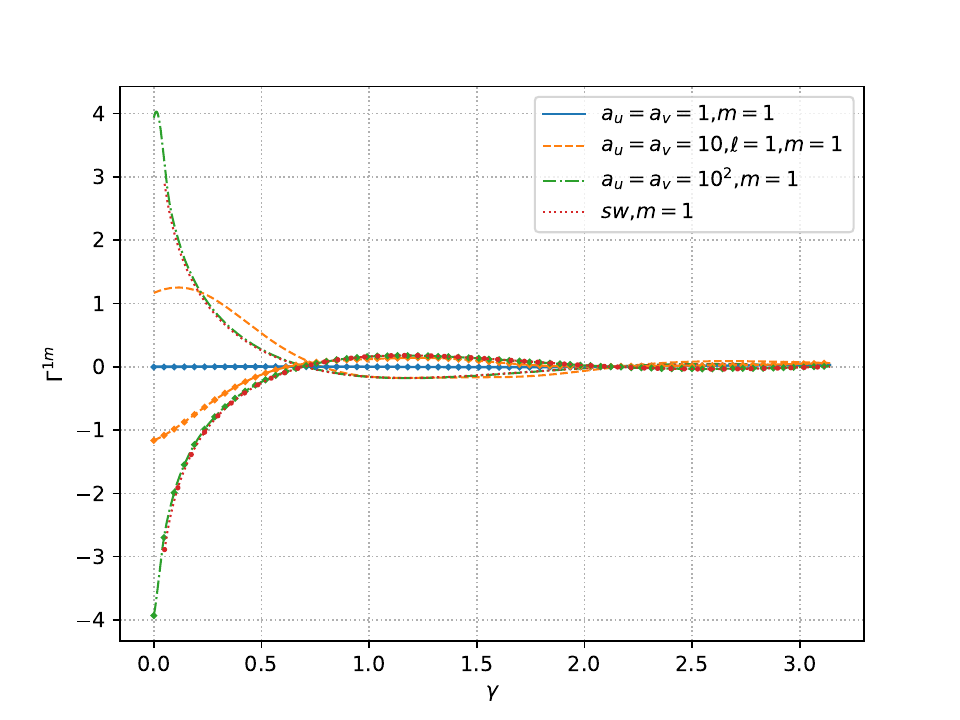}}\resizebox{230pt}{142.5pt}{\includegraphics{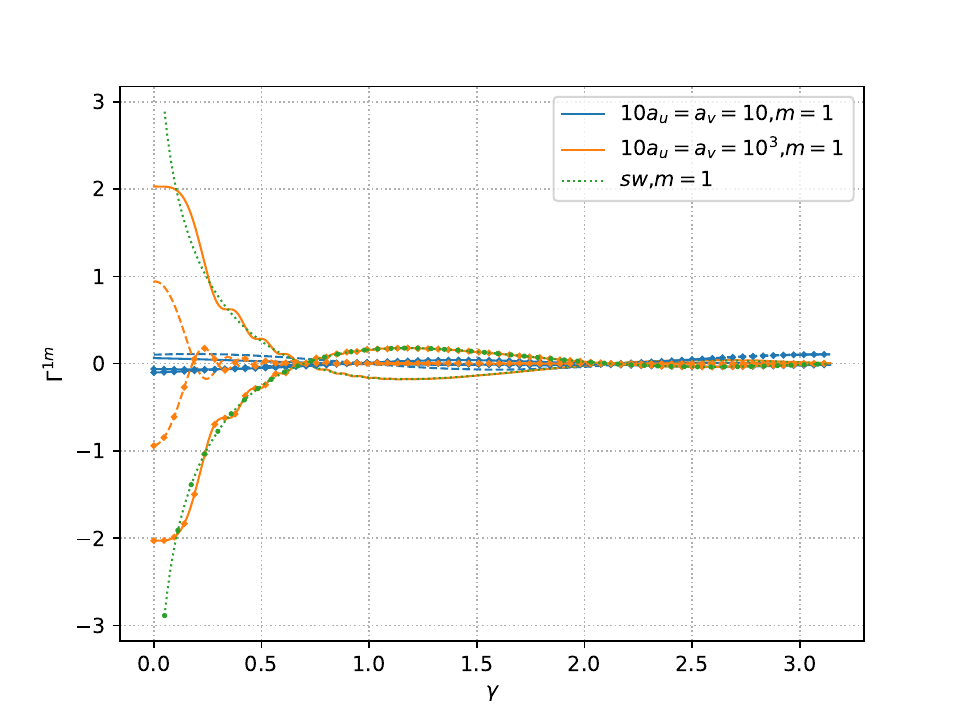}}
	
	\resizebox{230pt}{142.5pt}{\includegraphics{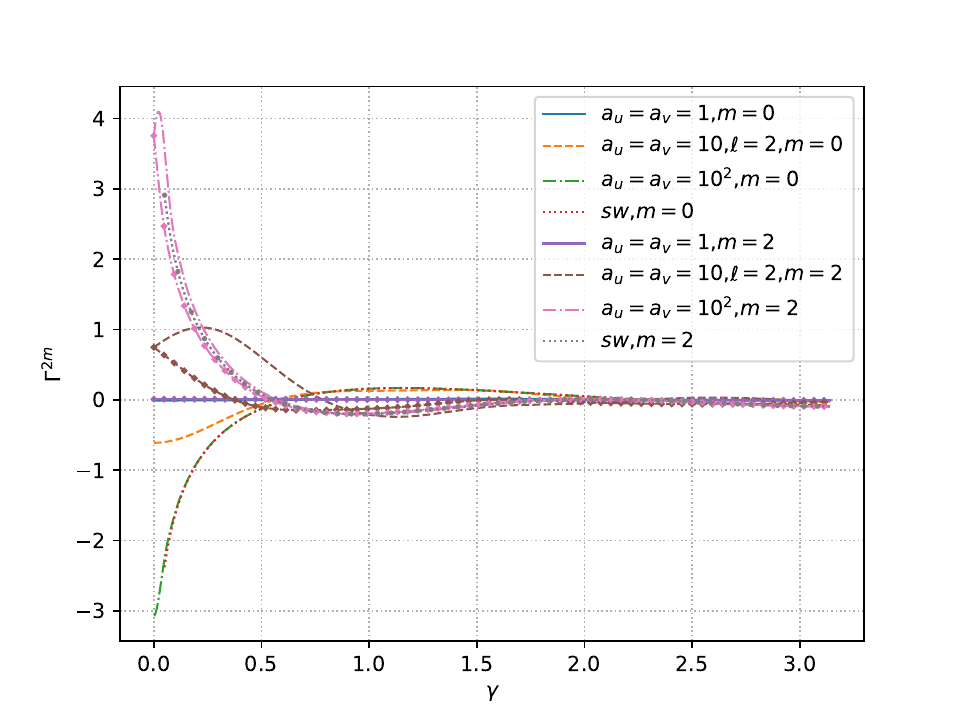}}\resizebox{230pt}{142.5pt}{\includegraphics{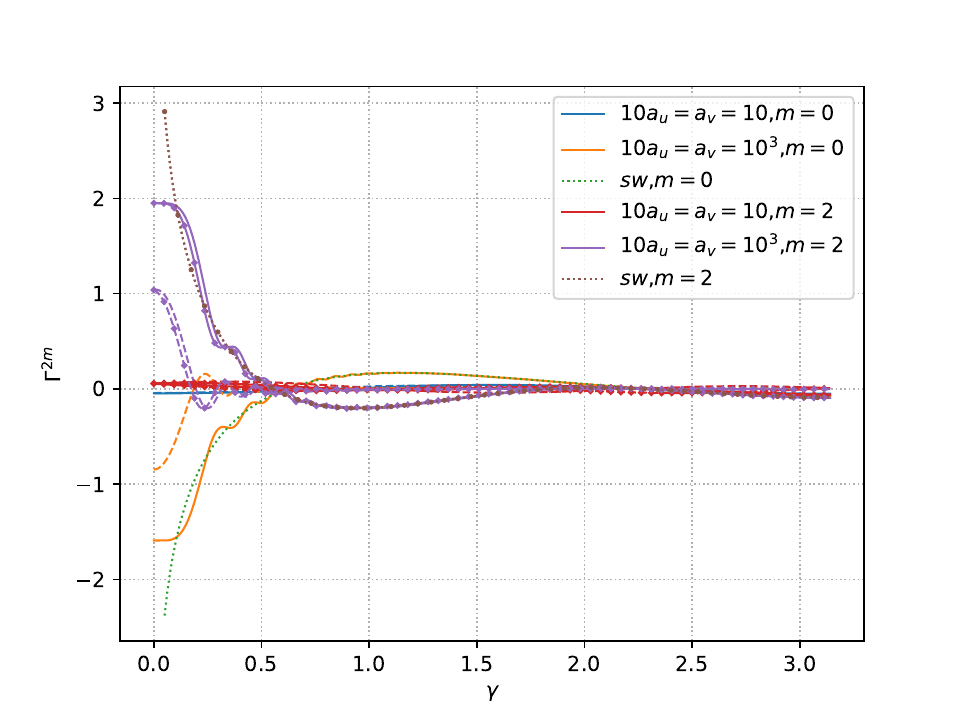}}
	
	\caption{\label{fig:orfV-lm-gamma} ORFs for vector modes with $\ell = 0, 1, 2$. The panel layout and graphical conventions follow Fig.~\ref{fig:orfT-lm-gamma}.}
\end{figure*}
\begin{figure*}[!t] 
	\resizebox{230pt}{142.5pt}{\includegraphics{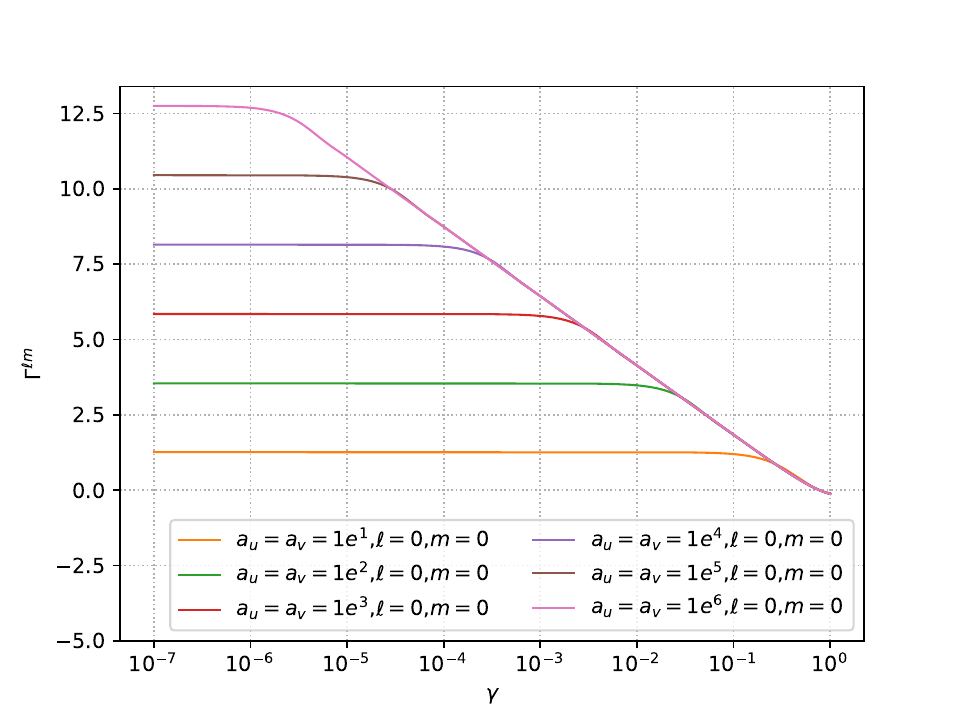}}\resizebox{230pt}{142.5pt}{\includegraphics{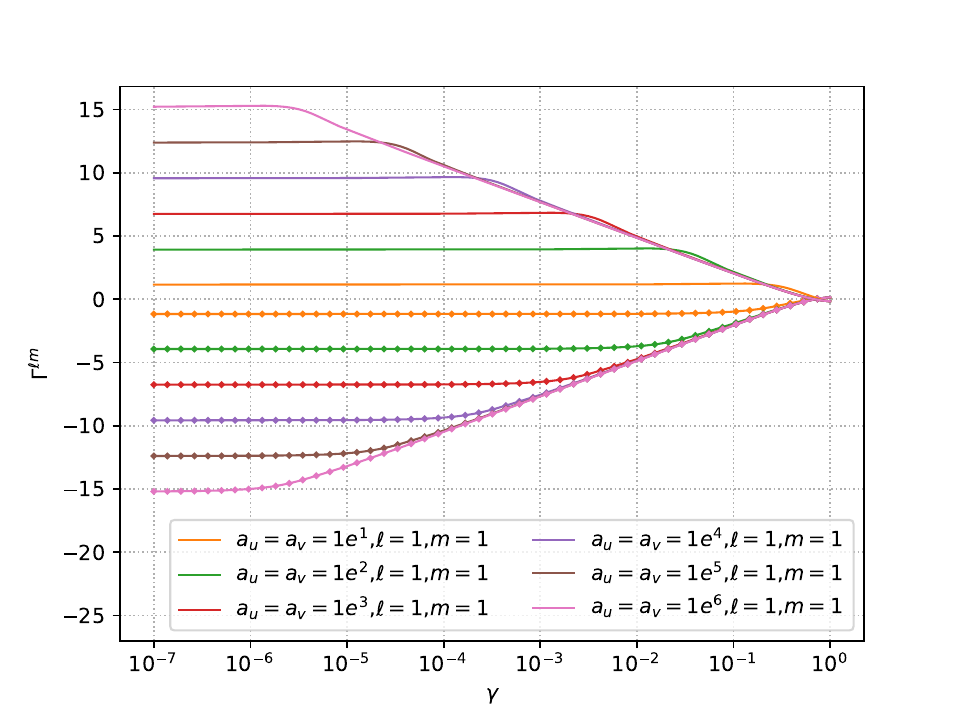}}
	\resizebox{230pt}{142.5pt}{\includegraphics{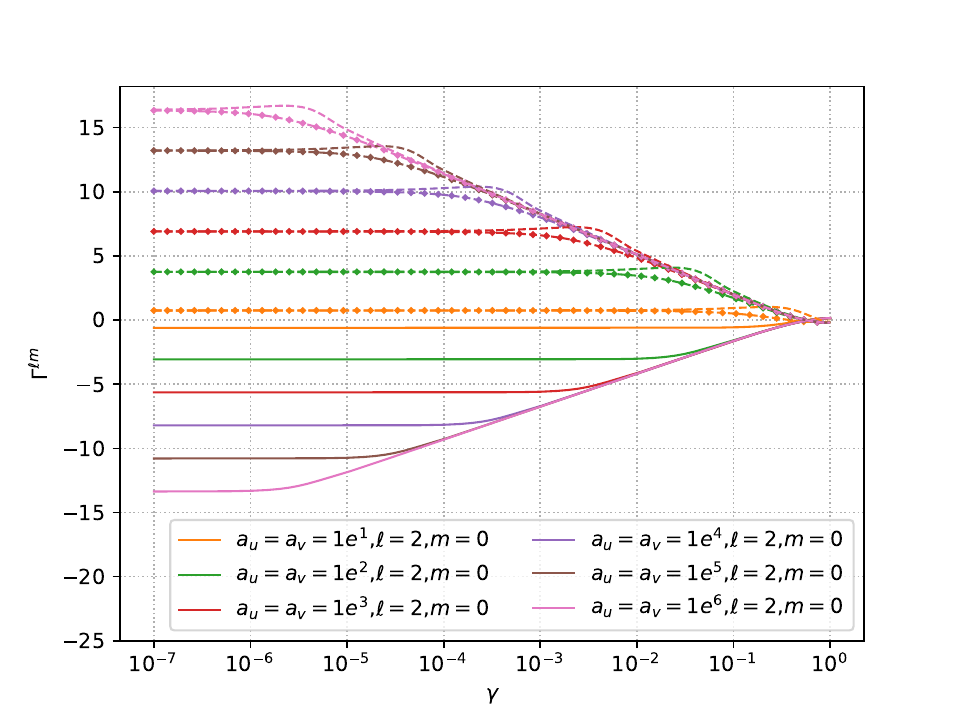}}\resizebox{230pt}{142.5pt}{\includegraphics{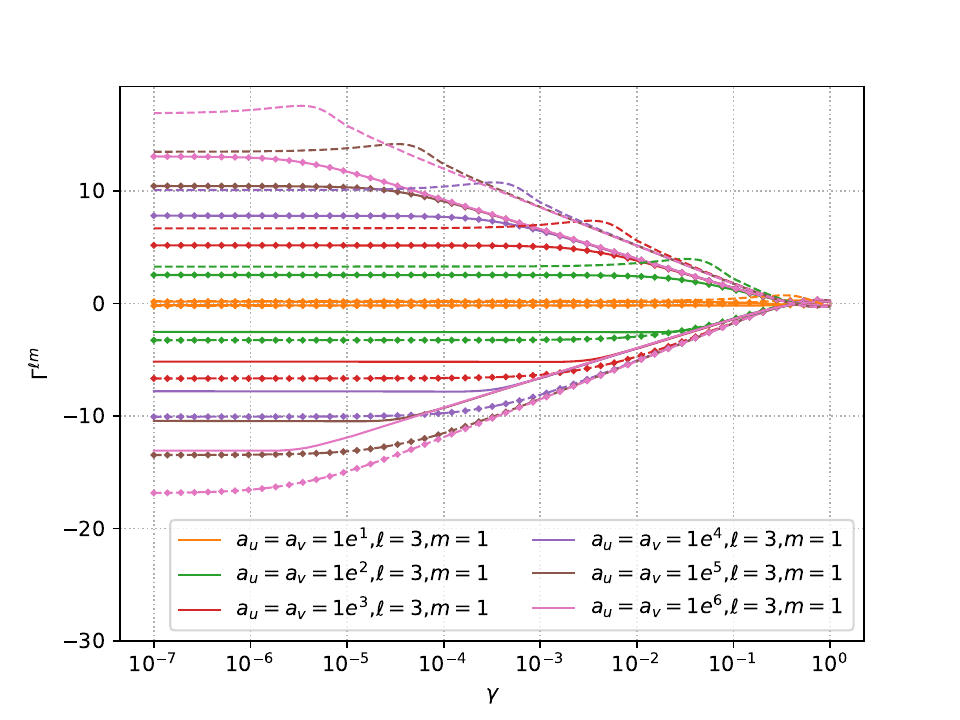}}
	\caption{\label{fig:orfV-lm-litgamma} Behavior of the vector modes at small angular separations for $L_u = L_v$. Line styles and markers as in Fig.~\ref{fig:orfT-lm-litgamma}.}
\end{figure*}

For vector modes, the analytical results in the short-wavelength approximation $a_u, a_v \gg
1$ simplify to the following forms.

\begin{equation}
\Gamma_{sw}^{V,0, 0} = - \frac{2}{3}   \cos (\gamma) - \log (\sin
(\tilde{\gamma})) - \frac{1}{2},
\end{equation}
\begin{eqnarray}
\Gamma_{sw}^{V,1, 1} &= &- \frac{1}{24}   \sqrt{6}  ( 12
\log (\sin (\tilde{\gamma})) \sec (\tilde{\gamma}) + 3
\cos \left( 3  \tilde{\gamma} \right)\nonumber\\
&  & + 13
\cos (\tilde{\gamma}) ),
\end{eqnarray}
\begin{eqnarray}
\Gamma_{sw}^{V,2, 0} &=& \frac{1}{60}   \sqrt{5}  ( 30
\left( 3  \sec (\tilde{\gamma})^2 - 2 \right)
\log (\sin (\tilde{\gamma})) + \nonumber\\
&  & 2  \cos (\gamma) + 45 ),
\end{eqnarray}
\begin{eqnarray}
\Gamma_{sw}^{V,2, 2} & = & - \frac{1}{120}   \sqrt{30}  ( 30
\log (\sin (\tilde{\gamma})) \sec (\tilde{\gamma})^2 + 5
\cos \left( 4  \tilde{\gamma} \right)\nonumber\\
&  & + 15
\cos (\gamma) + 27 ).
\end{eqnarray}

The terms with $m <
0$ can be derived from the above expressions using the symmetry relation \eqref{eq:R-prop-menom-app-old}. The $(0,
0)$ term cannot be normalized, as it grows with $a_u$.
The short-wavelength approximation for the isotropic case corresponds to \cite{Lee2008}
\begin{equation}
\Gamma_{V,sw} (\gamma) = - \frac{1}{2} - \frac{2}{3} \cos \gamma - \ln (\sin
\frac{\gamma}{2}).
\end{equation}

Moreover, as $\gamma \rightarrow 0$ (i.e., for pulsars with vanishing angular separation), the full analytical form of the ORFs converges, while the short-wavelength approximation diverges. Since no approximation is made, the analytical expressions are in excellent agreement with numerical integration. When the distances of the two pulsars are equal, i.e., $a_v =
a_u$, the ORFs are functions of a real variable. As shown in Fig.~\ref{fig:orfV-lm-gamma}, as $a_u$ increases, the analytical expressions agree with the short-wavelength approximation for sufficiently large $\gamma$. For the full expression, larger $a_v$ is needed for better agreement with the short-wavelength approximation. For long wavelengths, $\lim_{\gamma \rightarrow 0} \Gamma^{V,\ell m}$ is a fixed value that depends only on $a_u$. In practice, similarly to the tensor modes, the ORFs in this case form a set of complex-valued functions that satisfy the symmetries described in Sec.~\ref{sec:ORFssymmetry}. As shown in Fig.~\ref{fig:orfV-lm-gamma}, when $a_v \neq
a_u$, i.e., when the two pulsars are at unequal distances, the imaginary part of the ORFs depends on the ratio of $a_v$ to $a_u$. The larger this ratio, the larger the imaginary part. However, when both $a_v$ and $a_u$ are large, the real part agrees with the short-wavelength approximation, while the imaginary part does not approach zero for small $\gamma$ and cannot be ignored.

As shown in Fig.~\ref{fig:orfV-lm-litgamma}, for $a_v = a_u$, the behavior when the angular separation is small is illustrated. As $a_u$ increases, the ORFs increase, and their evolution is smooth without significant decay, indicating that the exponential term plays a dominant role; as $\ell$ increases, the ORF values do not show a clear decreasing trend as in the tensor modes;
for fixed $\ell$, there is no significant difference in the ORF values between $m > 0$ and $m < 0$ cases, and as $|m|$ increases, the ORF values increase. From these highly similar curves, we see that the ORFs for two nearby pulsars depend only on their distance. For pulsars in the same direction, i.e., $\gamma = 0$, the full analytical expression will give the exact value, while the short-wavelength expression $\Gamma_{sw}^{V,\ell, m}$ diverges. For these analytical ORF formulas, direct substitution of $\gamma = 0$ may lead to divergence. Similar to the tensor modes, we have
\begin{eqnarray}
K_{sw}^{V,0, 0} & = & - \frac{1}{4}   \left( \frac{i}{a_0} -
\frac{2 i}{a_0^3} \right) e^{\left( - 2 i  a_0 \right)} +
\frac{5 i}{4  a_0} - \frac{1}{a_0^2} - \frac{i}{2
	a_0^3} - \nonumber\\
&  & \frac{1}{2}   E_1  \left( 2 i
a_0 \right) - \frac{1}{2}   \log (- a_0),
\end{eqnarray}
\begin{eqnarray}
K_{sw}^{V,1, 1} & = & - \frac{1}{8}   \sqrt{6} \biggl[\nobracket
\left( \frac{i}{a_0} - \frac{1}{a_0^2} + \frac{2 i}{a_0^3} + \frac{6}{a_0^4}
\right) e^{\left( - 2 i  a_0 \right)} + 2  E_1
\left( 2 i  a_0 \right) \nonumber\\
&  &  + 2  \log (- a_0) - \frac{7 i}{a_0} + \frac{9}{a_0^2} + \frac{10 i}{a_0^3} -
\frac{6}{a_0^4}\biggl] ,
\end{eqnarray}
\begin{eqnarray}
K_{sw}^{V,2, 0} & = & - \frac{1}{8}   \sqrt{5}  \biggl[\nobracket
\left( - \frac{i}{a_0} + \frac{5 i}{a_0^3} + \frac{18}{a_0^4} - \frac{36
	i}{a_0^5} \right) e^{\left( - 2 i  a_0 \right)} -\nonumber\\
&  & 2
E_1  \left( 2 i  a_0 \right) - 2
\log (- a_0) + \frac{11 i}{a_0} - \frac{22}{a_0^2} - \frac{41 i}{a_0^3} \nonumber\\
&  & +
\frac{54}{a_0^4} + \frac{36 i}{a_0^5} \biggl] ,
\end{eqnarray}
other terms can be obtained from \eqref{eq:KAIJ-lm-rela22} to \eqref{eq:KAIJ-lm-rela31}.

Furthermore, when $a_u$ is finite, $\lim_{\gamma \rightarrow 0} 
\Gamma^{V,\ell m}  = F(a_u)$.
Specifically, the autocorrelation responses for a single detector for vector modes are
\begin{equation}
\Gamma_{i i}^{V,0, 0} = \gamma_E + \frac{2}{a_v^2} - \mathrm{Ci} \left(
2  a_v \right) + \log \left( 2  a_v \right) -
\frac{7}{3},
\end{equation}
\begin{eqnarray}
\Gamma_{i i}^{V,1, 1} & = & \frac{1}{12}   \sqrt{6}
\biggl[\nobracket 6  \gamma_E - \frac{3  \left( \cos
	\left( 2  a_v \right) - 9 \right)}{a_v^2} + \frac{18
	\left( \cos \left( 2  a_v \right) - 1
	\right)}{a_v^4} \nonumber\\
&  & - 6  \mathrm{Ci} \left( 2  a_v \right) +
6  \log \left( 2  a_v \right) - 16 \biggl] \nobracket,
\end{eqnarray}
\begin{eqnarray}
\Gamma_{i i}^{V,2, 0} & = & - \frac{1}{30}   \sqrt{5} 
\biggl[\nobracket  15  \gamma_E + \frac{165}{a_v^2} - \frac{135
	\left( \cos \left( 2  a_v \right) + 3
	\right)}{a_v^4} - \nonumber\\
&  & 15  \mathrm{Ci} \left( 2  a_v \right) +
15  \log \left( 2  a_v \right) - 47 \biggl]\nobracket  , 
\end{eqnarray}
other terms satisfy Eqs.~\eqref{eq:KAIJ-lm-rela22} to \eqref{eq:KAIJ-lm-rela31}, with the letter $K$ replaced by $\Gamma$.

In the limit $\gamma \rightarrow 0$, $\Gamma_{i i}^{V,\ell, m}$ grows logarithmically with $2 a_v$. The $(0, 0)$ order corresponds to the isotropic case and agrees with Eq.~(A42) in \cite{Lee2008}.

In general, for fixed $(\ell, m)$, the real part of the ORF increases with growing $a_v$ and $a_u$, while the imaginary part decreases and approaches zero. As $\ell$ increases, the ORF values do not show a clear change in magnitude. For fixed $\ell$, the ORFs for $m > 0$ are larger than those for $m < 0$, and they increase with $|m|$.

\subsection{Scalar breathing mode}
\begin{figure*}[!t]
	\resizebox{230pt}{142.5pt}{\includegraphics{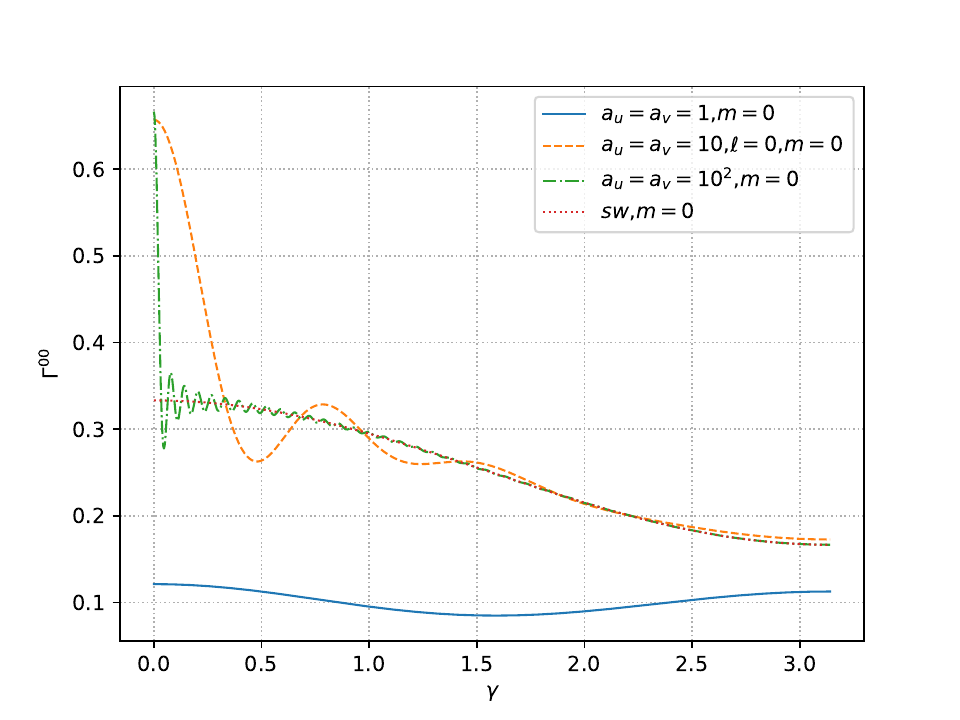}}\resizebox{230pt}{142.5pt}{\includegraphics{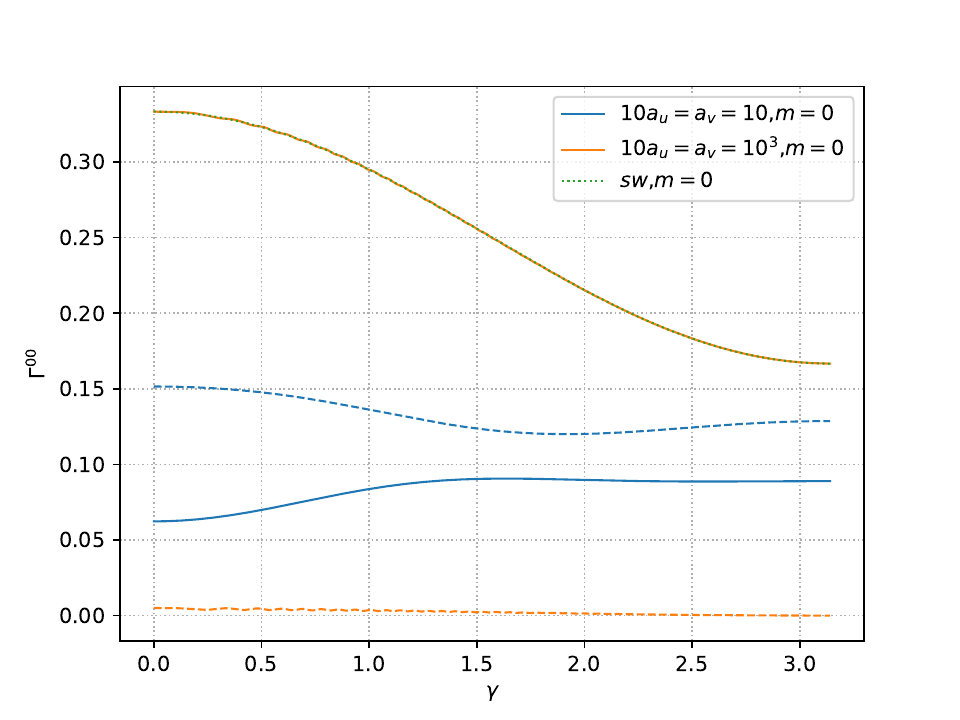}}
	
	\resizebox{230pt}{142.5pt}{\includegraphics{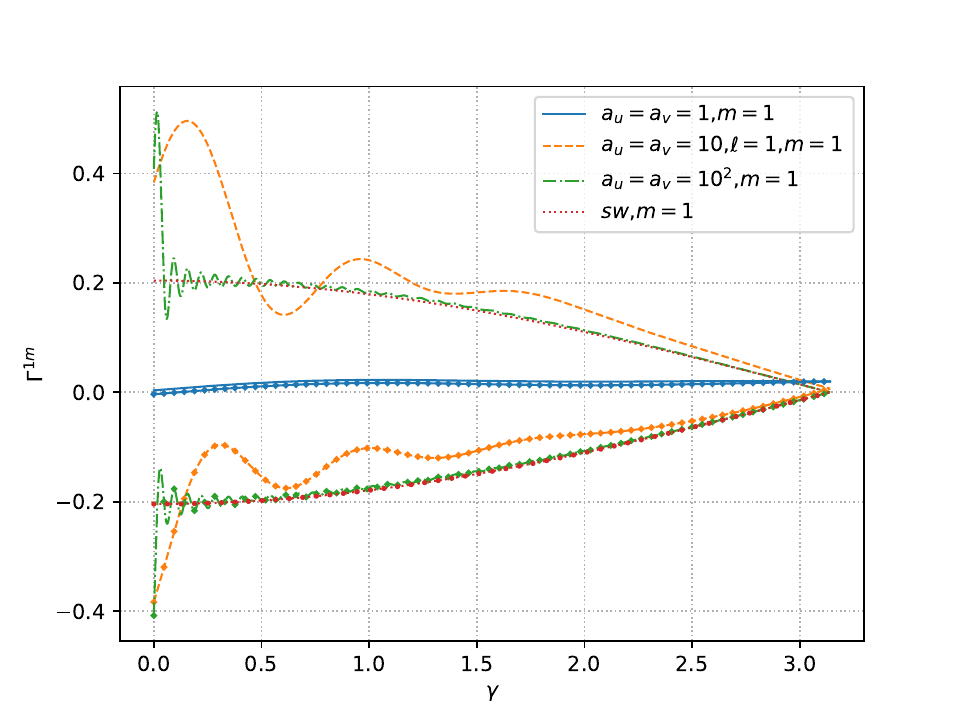}}\resizebox{230pt}{142.5pt}{\includegraphics{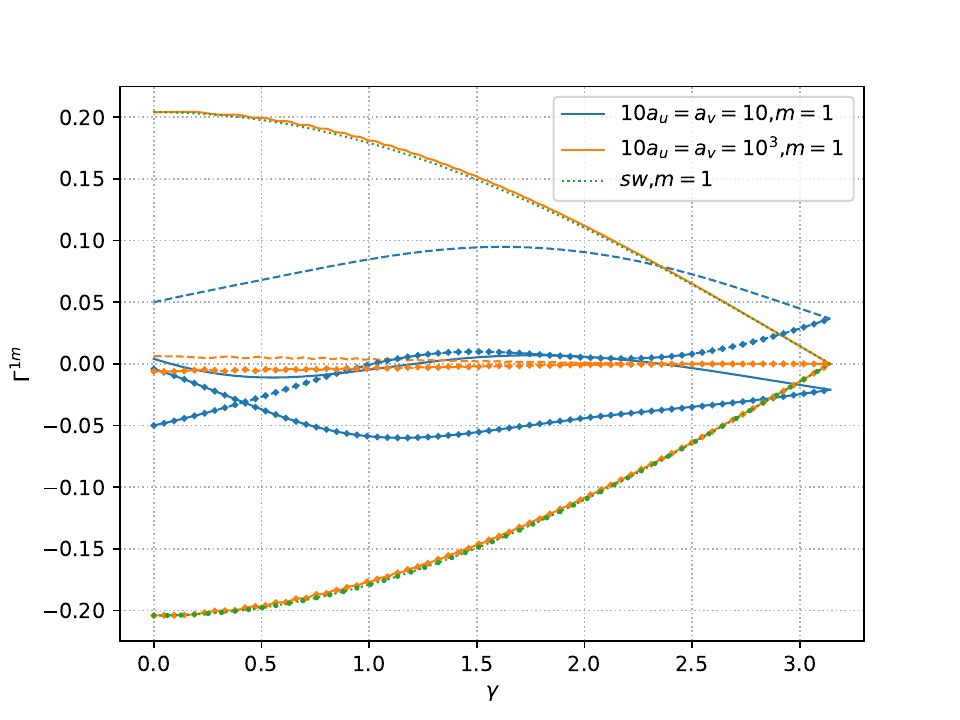}}
	
	\resizebox{230pt}{142.5pt}{\includegraphics{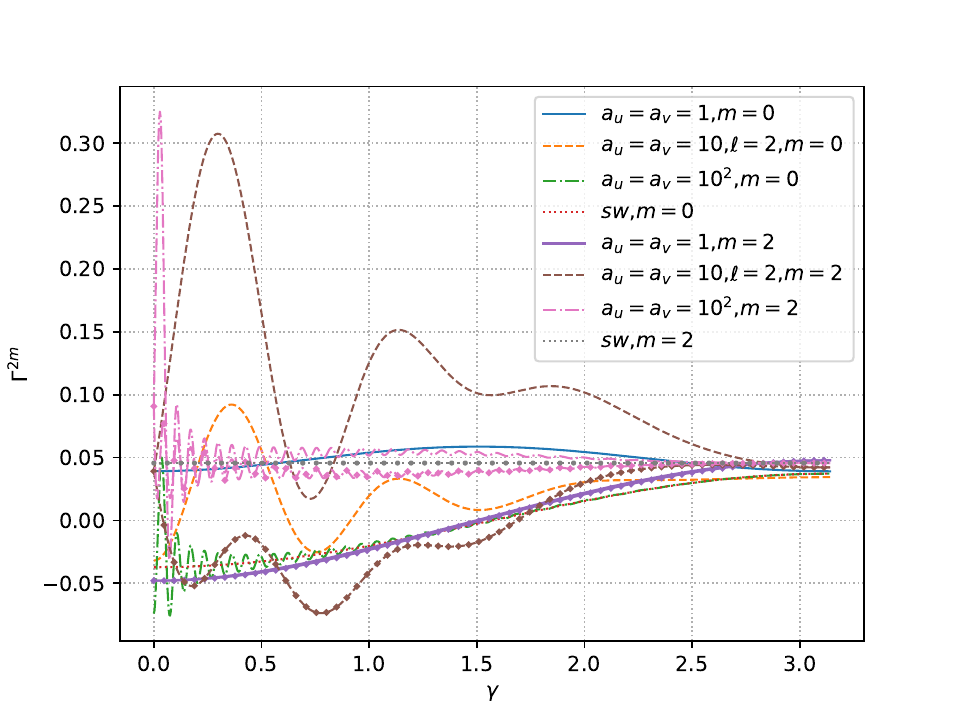}}\resizebox{230pt}{142.5pt}{\includegraphics{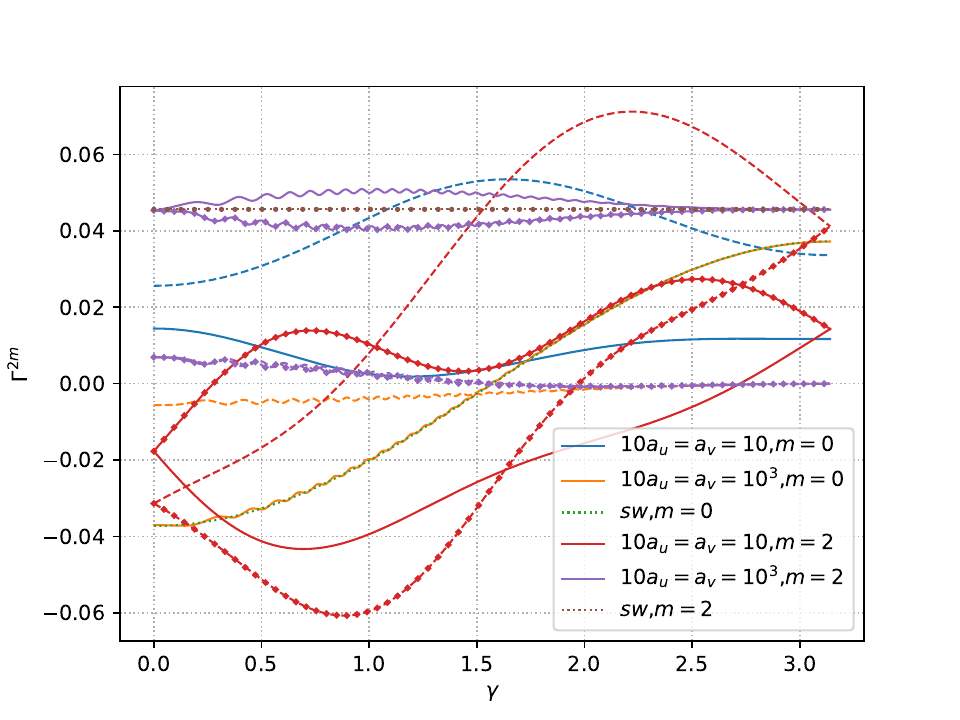}}
	
	\caption{\label{fig:orfB-lm-gamma} ORFs for scalar breathing mode with $\ell = 0, 1, 2$. The panel layout and graphical conventions follow Fig.~\ref{fig:orfT-lm-gamma}.}
\end{figure*}
\begin{figure*}[!t] 
	\resizebox{230pt}{142.5pt}{\includegraphics{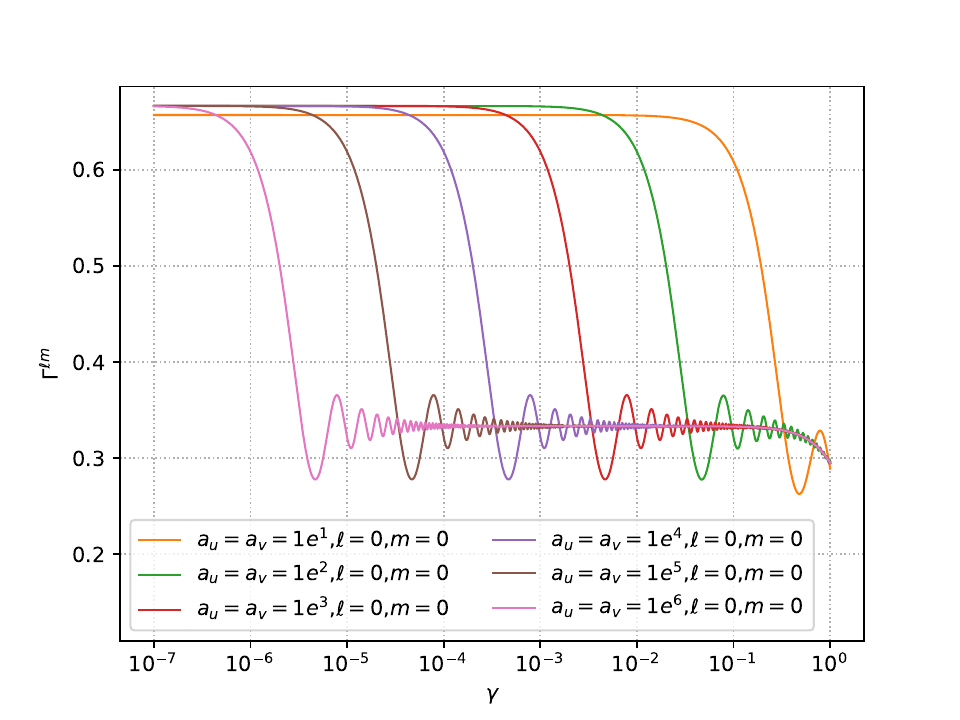}}\resizebox{230pt}{142.5pt}{\includegraphics{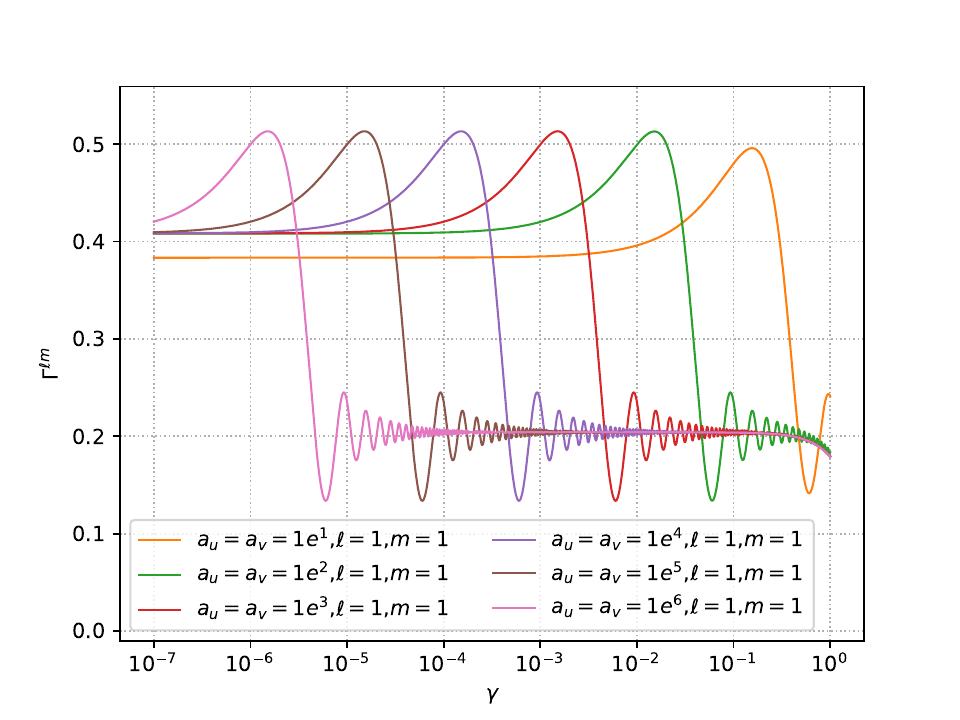}}
	\resizebox{230pt}{142.5pt}{\includegraphics{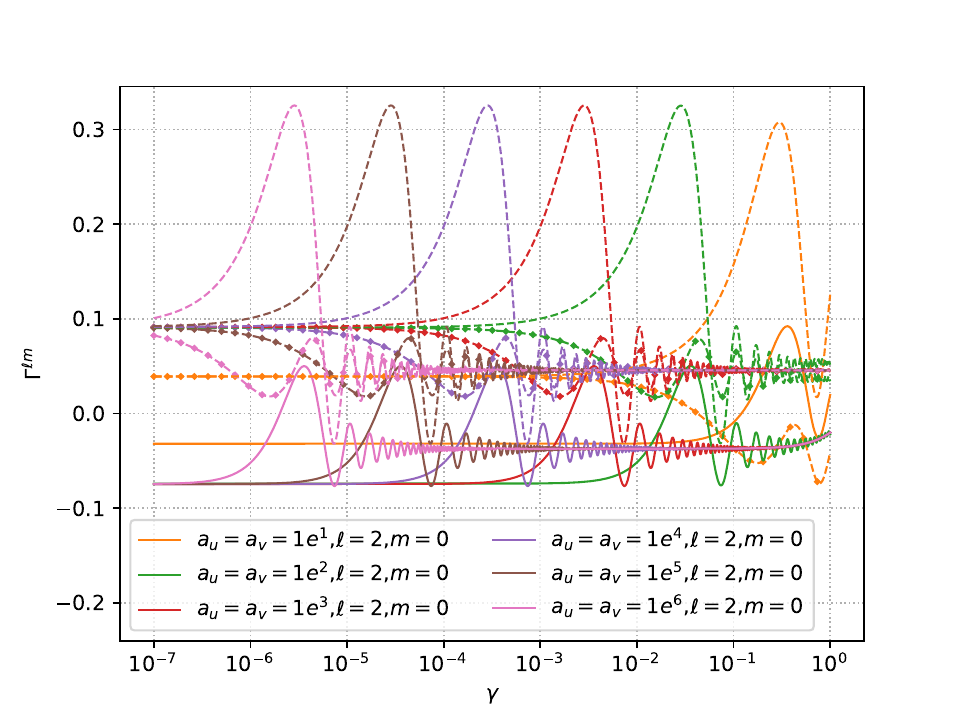}}\resizebox{230pt}{142.5pt}{\includegraphics{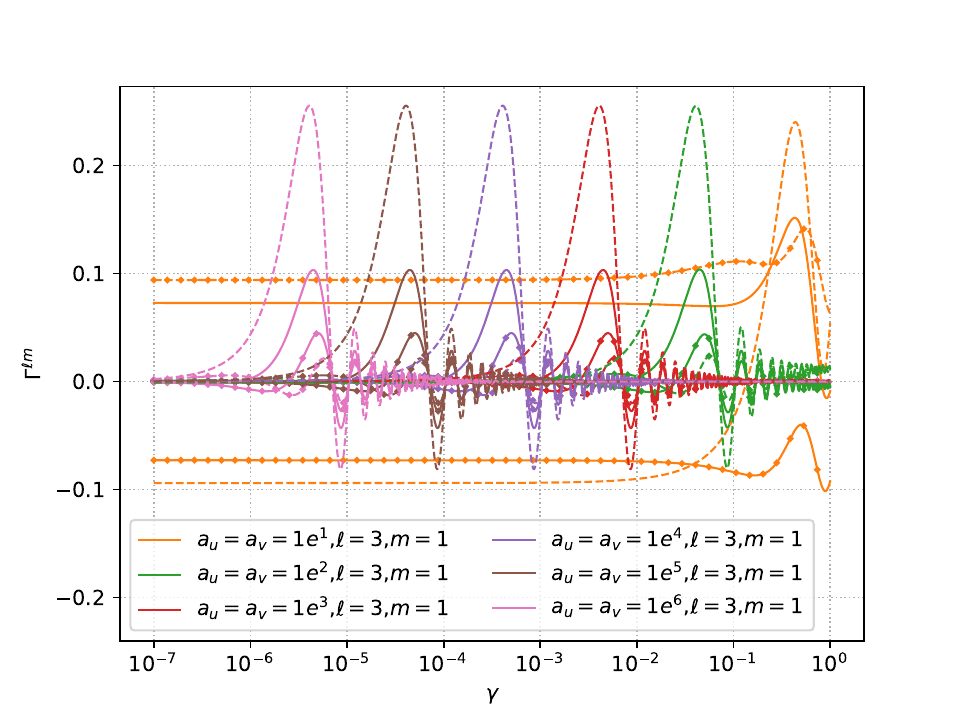}}
	\caption{\label{fig:orfB-lm-litgamma} Behavior of the scalar breathing mode at small angular separations for $L_u = L_v$. Line styles and markers as in Fig.~\ref{fig:orfT-lm-litgamma}.}
\end{figure*}

For the scalar breathing mode, the analytical results in the short-wavelength approximation $a_u, a_v \gg
1$ simplify to the following forms.
\begin{equation}
\Gamma_{sw}^{B,0, 0} = \frac{1}{12}   \cos (\gamma) +
\frac{1}{4},
\end{equation}
\begin{equation}
\Gamma_{sw}^{B,1, 1} = \frac{1}{12}   \sqrt{6} \cos
\left( \frac{1}{2}   \gamma \right),
\end{equation}
\begin{equation}
\Gamma_{sw}^{B,2, 0} = - \frac{1}{60}   \sqrt{5} \cos
(\gamma),
\end{equation}
\begin{equation}
\Gamma_{sw}^{B,2, 2} = \frac{1}{120}   \sqrt{30}.
\end{equation}
For $\ell \geqslant 3$, $\Gamma_{sw}^{B,\ell, m} = 0$.
The terms with $m <
0$ can be derived from the above expressions using the symmetry relation \eqref{eq:R-prop-menom-app-old}. The $(0,
0)$ term, when normalized, corresponds to the isotropic short-wavelength expression \cite{Lee2008}
\begin{equation}
\gamma_{sw}^B (\gamma) = \frac{3}{8} + \frac{1}{8} \cos \gamma .
\end{equation}

We observe that as $\gamma \rightarrow 0$ (i.e., for a pulsar pair with vanishing angular separation), both the analytical form and the short-wavelength approximation converge to the same value. We also find that for $\ell \geqslant 3$, the short-wavelength approximation of the ORFs is identically zero. This is a major difference between the scalar breathing mode and other modes. Comparison with the tensor modes indicates that in the short-wavelength approximation, the dominant term in the high-order ORF integrals is Eq.~\eqref{eq:funcTT2}. Since no approximation is made, the analytical expressions are in excellent agreement with numerical integration. When the distances of the two pulsars are equal, i.e., $a_v = a_u$, the ORFs are functions of a real variable. As shown in Fig.~\ref{fig:orfB-lm-gamma}, as $a_u$ increases, the analytical expressions agree with the short-wavelength approximation. For long wavelengths, $\lim_{\gamma \rightarrow 0} \Gamma_{ij}^{B, \ell m}$ is a fixed value that depends only on $a_u$. As with the other modes, the ORFs are therefore given by a set of complex-valued functions that adhere to the symmetries presented in Sec.~\ref{sec:ORFssymmetry}. As shown in Fig.~\ref{fig:orfB-lm-gamma}, when $a_v \neq a_u$, i.e., when the two pulsars are at unequal distances, the imaginary part of the ORFs depends on the ratio of $a_v$ to $a_u$. The larger this ratio, the larger the imaginary part. However, when both $a_v$ and $a_u$ are large, the imaginary part approaches zero.

As shown in Fig.~\ref{fig:orfB-lm-litgamma}, for $a_v = a_u$, the behavior when the angular separation is small is illustrated. As $a_u$ increases, the ORFs decay faster; as $\ell$ increases, the ORF values gradually decrease. For fixed $\ell$, the ORF for $m > 0$ is larger than that for $m < 0$, and as $|m|$ increases, the ORF values increase. From these highly similar curves, we see that the ORFs for two nearby pulsars depend only on their distance. For pulsars in the same direction, i.e., $\gamma = 0$, the ORFs have the form
\begin{equation}
\Gamma_{i j}^{B, \ell m} = 2 \Gamma_{i j}^{T, \ell m}.
\end{equation}
When $a_u$ is finite, $\lim_{\gamma \rightarrow 0} \Gamma^{B,\ell
	m}  = F(a_u)$.
Specifically, the autocorrelation responses for a single detector for the scalar breathing mode are
\begin{equation}
\Gamma_{i i}^{B,\ell m} = 2 \Gamma_{i i}^{T,\ell m}.
\end{equation}
For $a_u \gg 1$, the ORFs converge, but as $a_u$ increases, when $\gamma \rightarrow 0$, the ORFs tend to a constant.

Overall, the evolution of the ORFs for the scalar breathing mode is very similar to that of the tensor modes. This is because the ORF calculations for these two modes both contain the common term \eqref{eq:funcBT1} in Eqs.~\eqref{eq:RvuTsm} and \eqref{eq:RvuBLsm}, indicating that \eqref{eq:funcBT1} plays a dominant role in the ORF integrals for the tensor and scalar breathing modes.

\subsection{Scalar longitudinal mode}
\begin{figure*}[!t]
	\resizebox{230pt}{142.5pt}{\includegraphics{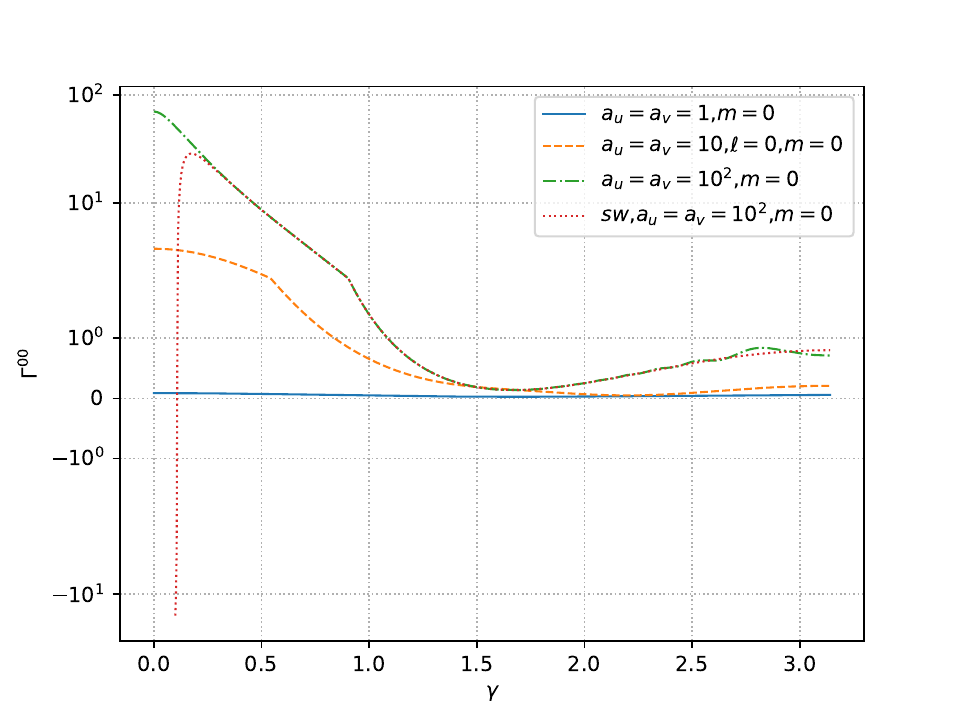}}\resizebox{230pt}{142.5pt}{\includegraphics{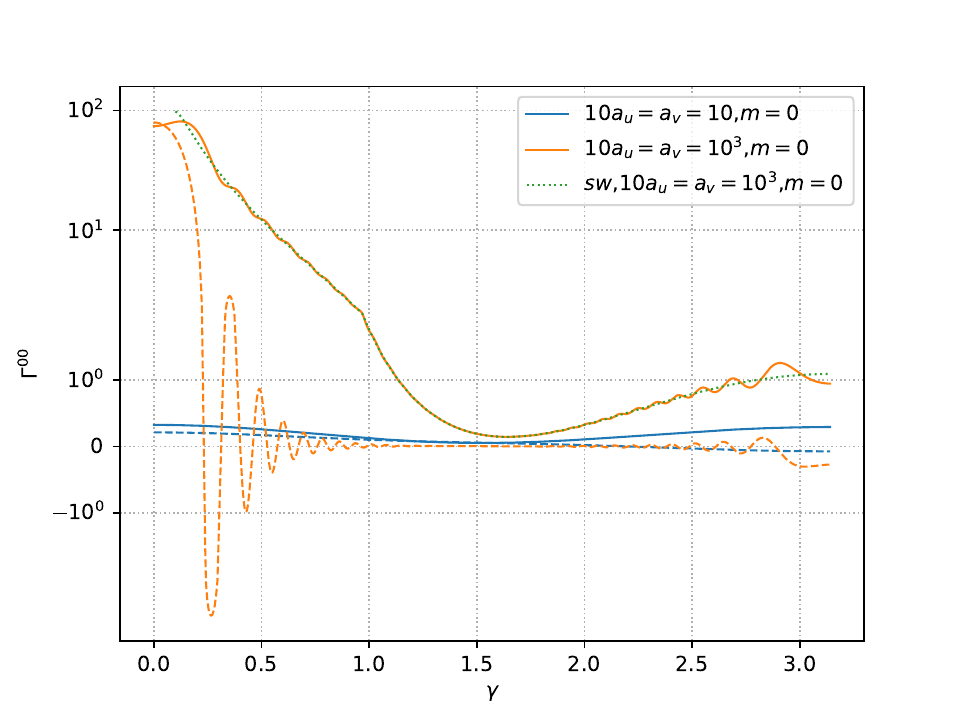}}
	
	\resizebox{230pt}{142.5pt}{\includegraphics{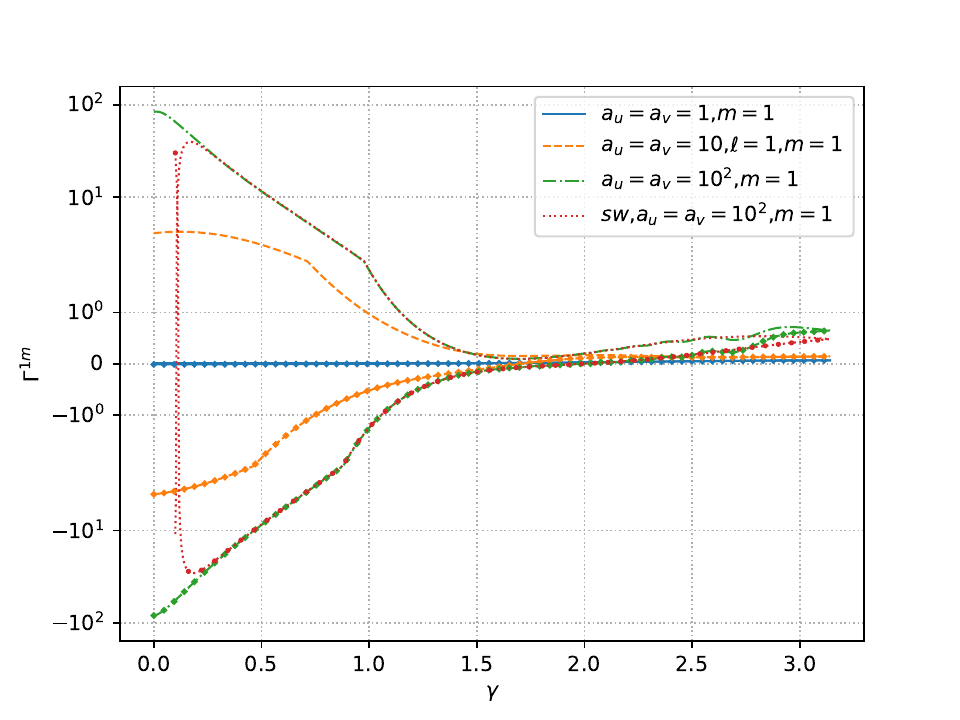}}\resizebox{230pt}{142.5pt}{\includegraphics{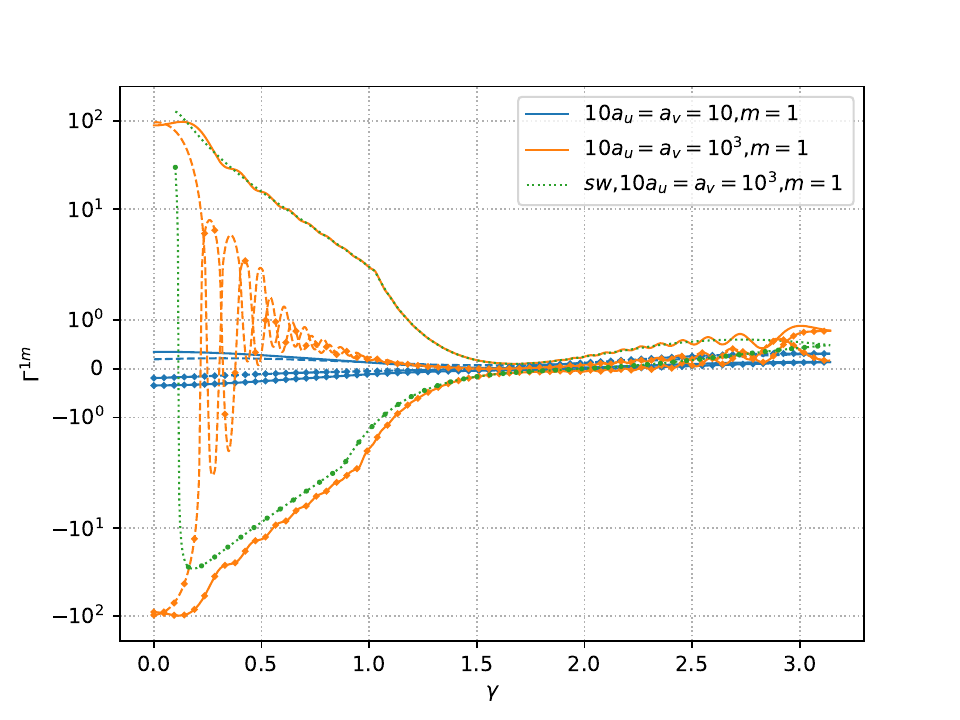}}
	
	\resizebox{230pt}{142.5pt}{\includegraphics{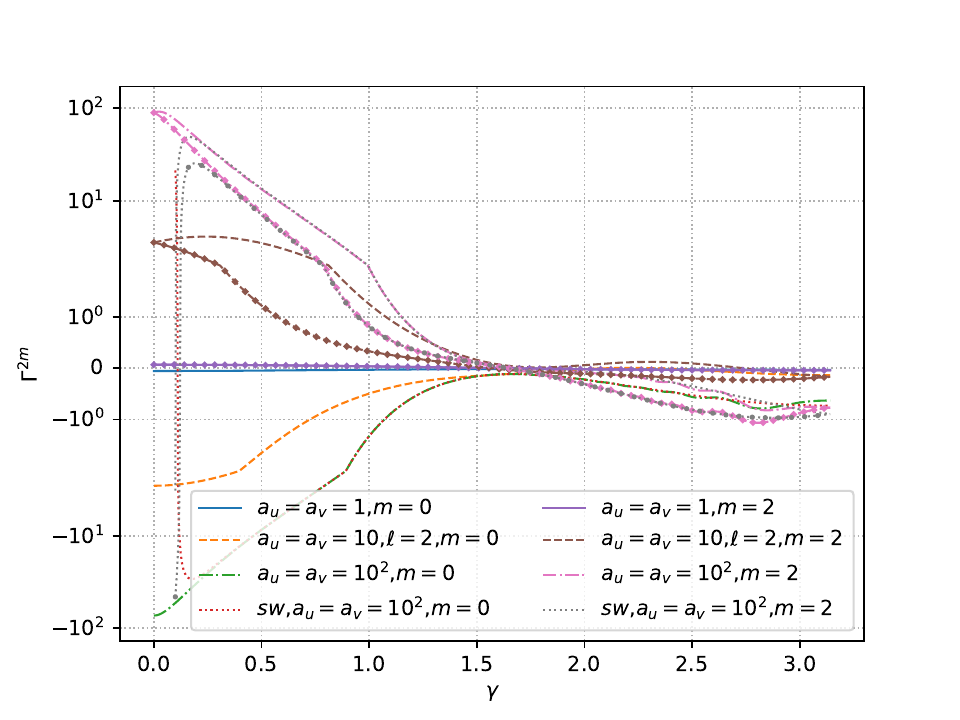}}\resizebox{230pt}{142.5pt}{\includegraphics{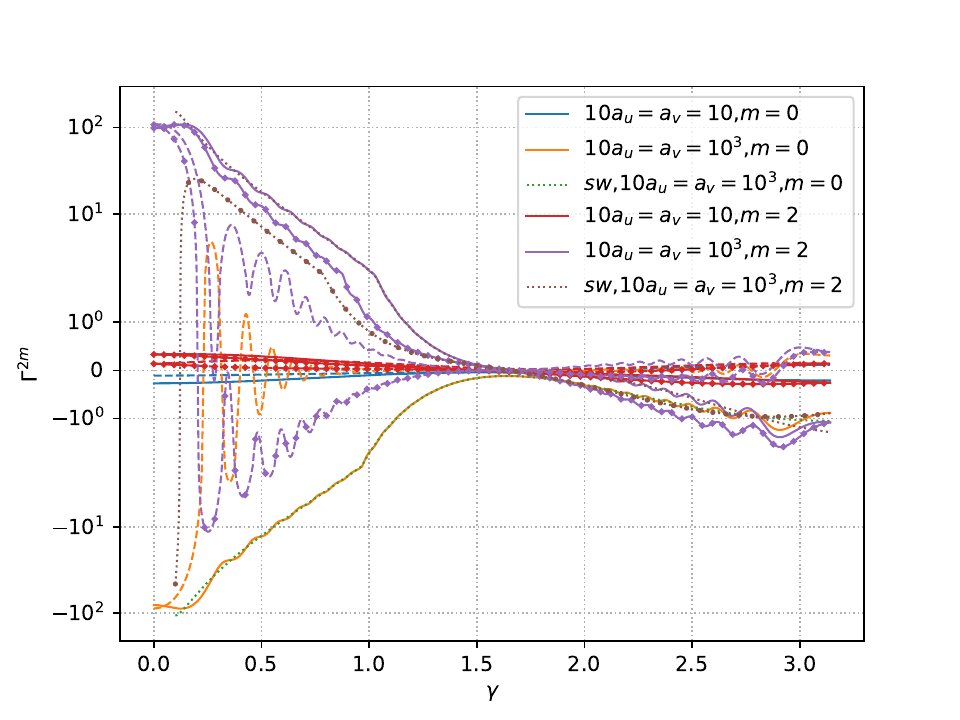}}
	\caption{\label{fig:orfL-lm-gamma} ORFs for scalar longitudinal mode with $\ell = 0, 1, 2$. The panel layout and graphical conventions follow Fig.~\ref{fig:orfT-lm-gamma}.}
\end{figure*}
\begin{figure*}[!t] 
	\resizebox{230pt}{142.5pt}{\includegraphics{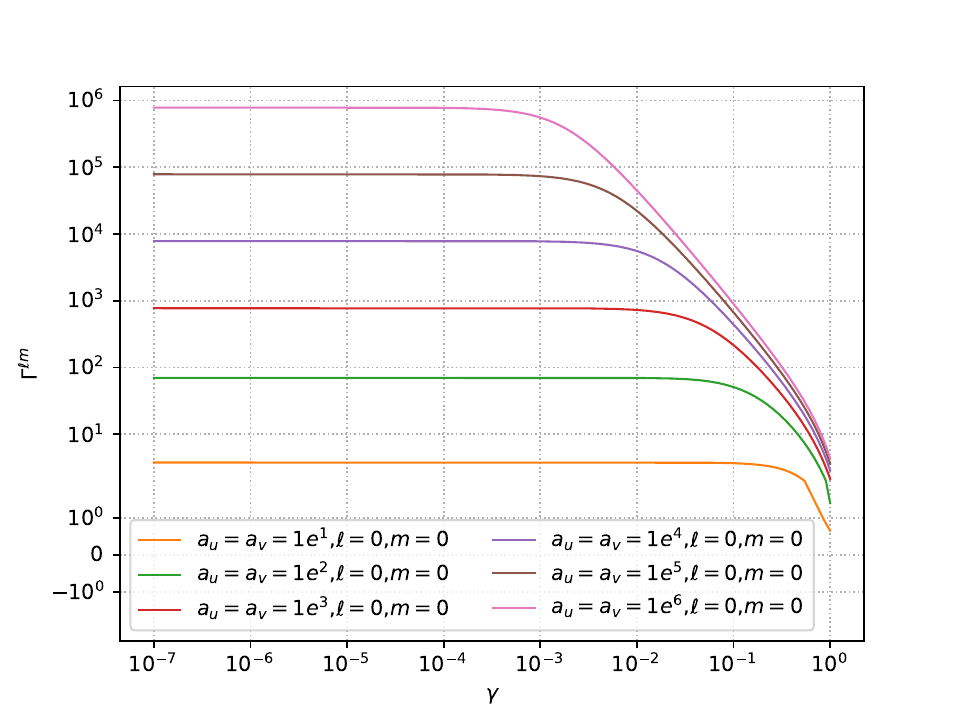}}\resizebox{230pt}{142.5pt}{\includegraphics{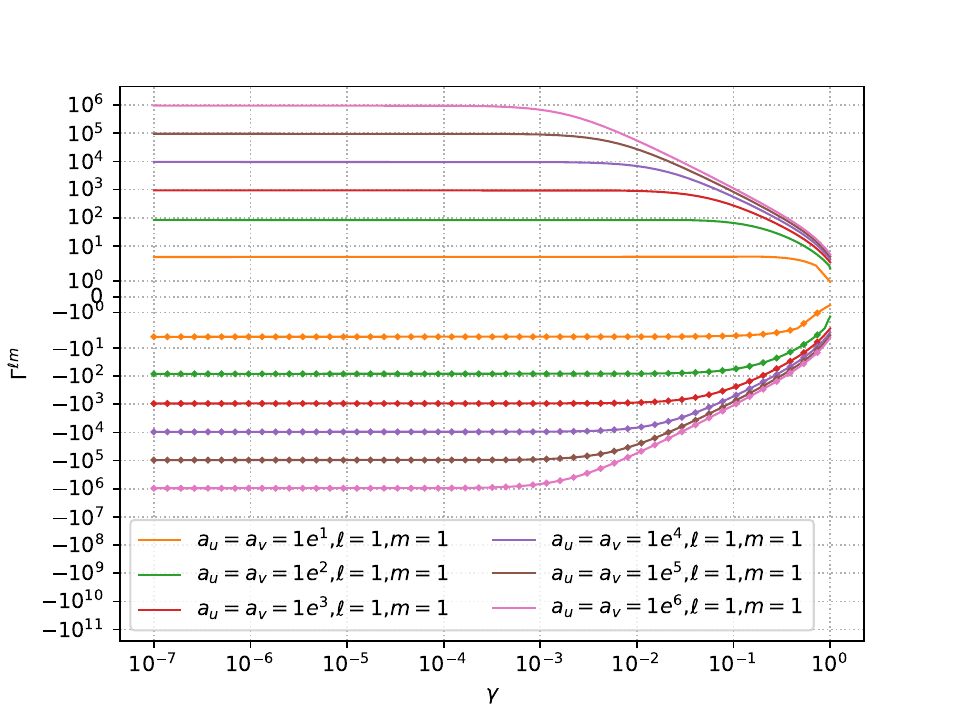}}
	\resizebox{230pt}{142.5pt}{\includegraphics{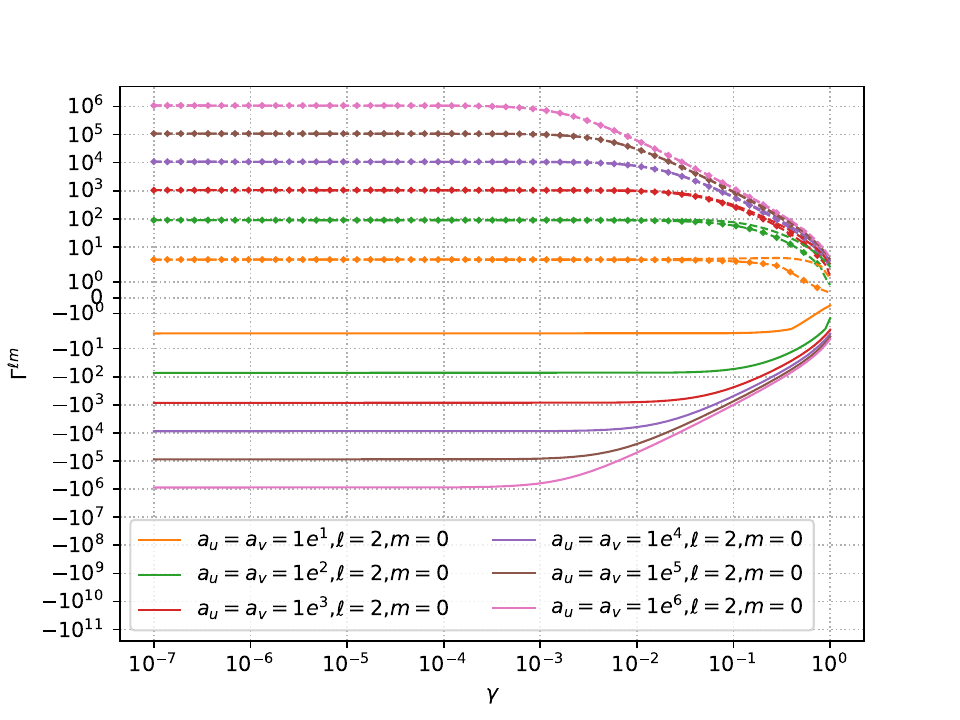}}\resizebox{230pt}{142.5pt}{\includegraphics{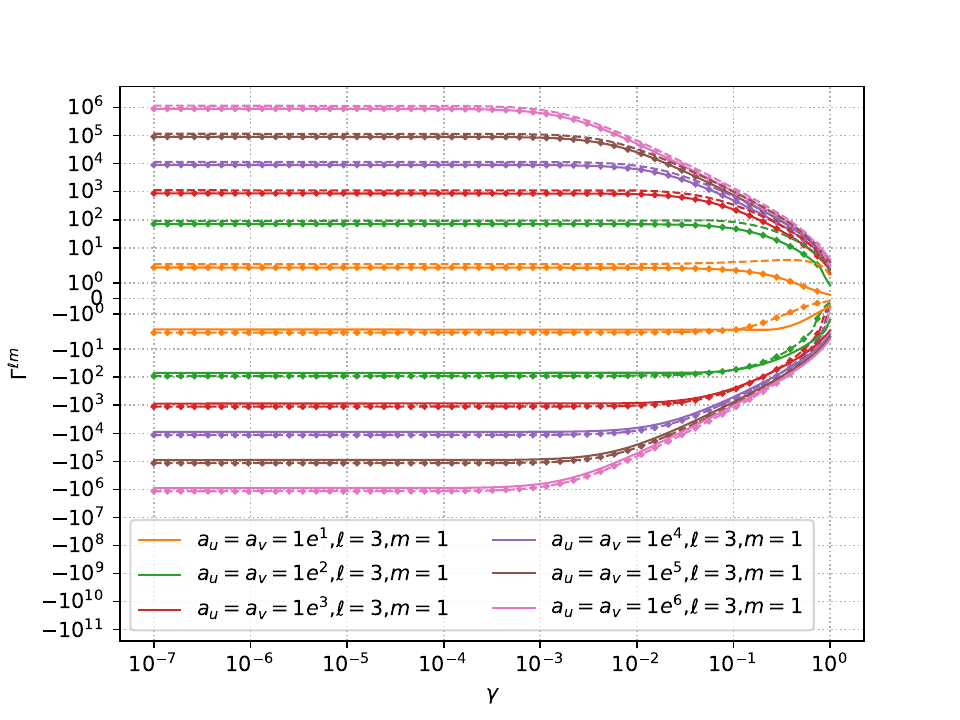}}
	\caption{\label{fig:orfL-lm-litgamma} Behavior of the scalar longitudinal mode at small angular separations for $L_u = L_v$. Line styles and markers as in Fig.~\ref{fig:orfT-lm-litgamma}.}
\end{figure*}

For the scalar longitudinal mode, a simple analytical expression for the ORFs in the isotropic case does not exist under the traditional short-wavelength approximation. This is because directly computing the integral by neglecting the exponential terms leads to divergence \cite{Romano2017}. The exponential terms must be included to overcome the divergence. However, we can obtain a short-wavelength expression from the full analytical form. For the scalar longitudinal mode, the analytical results in the short-wavelength approximation $a_v, a_u \gg 1$ simplify to the following forms.
\begin{eqnarray}
\Gamma_{sw}^{L,0, 0} & = & \frac{1}{4}   \gamma_E \cos (\gamma)^2
\csc (\tilde{\gamma})^2 + \frac{1}{8}   \cos (\gamma)^2 \csc
(\tilde{\gamma})^2 \log \left( 4  a_2 a_3 \right) \nonumber\\
&  & + \frac{1}{2}   \csc (\tilde{\gamma})^2 \log (\sin
(\tilde{\gamma})) + \frac{7}{6}   \cos (\gamma) + \frac{1}{2},
\end{eqnarray}
\begin{eqnarray}
\Gamma_{sw}^{L,1, 1} & = & \frac{1}{48}   \sqrt{6}  [ 3
\cos (\gamma)^2 \csc (\tilde{\gamma})^2 e^{\left( i
	\tilde{\gamma} \right)} \log \left( 4  a_2
a_3 \right) + \nonumber\\
&  & 6  \gamma_E \cos (\gamma)^2 \cot (\tilde{\gamma}) \csc
(\tilde{\gamma}) - 6 i  \cos (\gamma)^2 \csc (\tilde{\gamma})
 \nonumber\\&  & \log \left( 2  a_2 \right) + 24  \csc (\gamma) \csc (\tilde{\gamma}) \log (\sin
(\tilde{\gamma})) +  \nonumber\\
&  &  3  \pi \csc (\tilde{\gamma}) - 3
\pi \sin \left( 3  \tilde{\gamma} \right)
 - 3  \pi \sin (\tilde{\gamma}) + \nonumber\\
 &  & 18 
\cos \left( 3  \tilde{\gamma} \right) + 38 
\cos (\tilde{\gamma})]  ,
\end{eqnarray}
\begin{eqnarray}
\Gamma_{sw}^{L,2, 0} & = & - \frac{1}{240}   \sqrt{5} 
[30  \gamma_E \cos (\gamma)^2 \csc
(\tilde{\gamma})^2 + 15  \cos (\gamma)^2  \nonumber\\
&  & \csc
(\tilde{\gamma})^2  \log \left( 4  a_2 a_3 \right)- 240  \cos (\gamma) \csc (\gamma)^2  \nonumber\\
&  & \log (\sin(\tilde{\gamma})) +  480  \csc (\gamma)^2 \log (\sin
(\tilde{\gamma}))  \nonumber\\
&  & + 188  \cos (\gamma) + 180] ,
\end{eqnarray}
\begin{eqnarray}
\Gamma_{sw}^{L,2, 2} & = & \frac{1}{480}   \sqrt{30}  [\nobracket
30  \gamma_E \cos (\gamma)^3 \csc (\tilde{\gamma})^2 + 15
\cos (\gamma)^2 \nonumber\\
&  & \csc (\tilde{\gamma})^2 e^{\left( 2 i
	\tilde{\gamma} \right)} \log \left( 4  a_2
a_3 \right) - 60 i  \cos (\gamma)^2 \cot (\tilde{\gamma}) \nonumber\\
&  & \log
\left( 2  a_2 \right) + 240  \csc (\gamma)^2
\log (\sin (\tilde{\gamma})) + 30  \pi \cot (\tilde{\gamma})
\nonumber\\
&  & - 15  \pi \sin \left( 4  \tilde{\gamma}
\right) - 30  \pi \sin (\gamma) + 110  \cos
\left( 4  \tilde{\gamma} \right) + \nonumber\\
&  & 180  \cos
(\gamma) + 198] .
\end{eqnarray}

The terms with $m < 0$ can be derived by first taking the complex conjugate of the logarithmic parts containing $a_u$ and $a_v$ and then using the symmetry relation \eqref{eq:R-prop-menom-app-old} from the above expressions.

Moreover, as $\gamma \rightarrow 0$ (i.e., for pulsars with vanishing angular separation), the full analytical form of the ORFs converges, while the short-wavelength approximation diverges. Since no approximation is made, the analytical expressions are in excellent agreement with numerical integration. When the distances of the two pulsars are equal, i.e., $a_v = a_u$, the ORFs are functions of a real variable. As shown in Fig.~\ref{fig:orfL-lm-gamma}, as $a_u$ increases, the analytical expressions agree with the short-wavelength approximation for sufficiently large $\gamma$, but for $\gamma$ close to $0$ and $\pi$, the agreement between the short-wavelength approximation and the full analytical expression is poor. Furthermore, as $\ell$ increases, larger $a_v$ is needed for the full expression to agree well with the short-wavelength approximation. For long wavelengths, $\lim_{\gamma \rightarrow 0} \Gamma^{L, \ell m}$ is a fixed value that depends only on $a_u$ and $a_v$. In realistic scenarios characterized by differing pulsar distances, the ORFs form a set of complex-valued functions that obey the symmetries outlined in Sec.~\ref{sec:ORFssymmetry}. As shown in Fig.~\ref{fig:orfL-lm-gamma}, when $a_v \neq a_u$, i.e., when the two pulsars are at unequal distances, the imaginary part of the ORFs depends on the ratio of $a_v$ to $a_u$. The larger this ratio, the larger the imaginary part. However, when both $a_v$ and $a_u$ are large, the real part agrees with the short-wavelength approximation, while the imaginary part is comparable to the real part for $\gamma$ close to $0$ and $\pi$ and cannot be ignored.

As shown in Fig.~\ref{fig:orfL-lm-litgamma}, for $a_v = a_u$, the behavior when the angular separation is small is illustrated. As $a_u$ increases, the ORFs increase, and their evolution is smooth without significant decay, indicating that the exponential term plays a dominant role; as $\ell$ increases, the ORF values do not show a clear decreasing trend as in the tensor modes. For fixed $\ell$, there is no significant difference in ORF values between $m > 0$ and $m < 0$ cases, and as $|m|$ increases, the ORF values increase. From these highly similar curves, we see that the ORFs for two nearby pulsars depend only on their distance. For these analytical ORF formulas, direct substitution of $\gamma = 0$ may lead to divergence. Similar to the tensor modes, we have
\begin{eqnarray}
K_{sw}^{L,0, 0} & = & \frac{1}{8}   \left( - \frac{4 i}{a_0}
+ \frac{4 i}{a_0^3} + 1 \right) e^{\left( - 2 i  a_0 \right)}
+ \frac{1}{4}   \left( i  a_0 + 4 \right) \nonumber\\
&  & 
E_1  \left( 2 i  a_0 \right)+ \frac{1}{4}   \left( i  a_0 + 4
\right) \log \left( - 2  a_0 \right) - \frac{3 i}{2
	a_0}  \nonumber\\
&  & + \frac{1}{a_0^2} + \frac{i}{2  a_0^3} ,
\end{eqnarray}
\begin{eqnarray}
K_{sw}^{L,1, 1} & = & \frac{1}{16}   \sqrt{6}  \biggl[\nobracket
\left( - \frac{4 i}{a_0} + \frac{4}{a_0^2} - \frac{4 i}{a_0^3} -
\frac{12}{a_0^4} + 1 \right) e^{\left( - 2 i  a_0 \right)}  \nonumber\\
&  & - 2  \left(  i  a_0 + 5 \right) E_1  \left( 2 i
a_0 \right) - 2  \left( i  a_0 + 5 \right)  \nonumber\\
&  & \log (- a_0) + \frac{20 i}{a_0} - \frac{20}{a_0^2} - \frac{20 i}{a_0^3} +
\frac{12}{a_0^4} \biggl] \nobracket ,
\end{eqnarray}
\begin{eqnarray}
K_{sw}^{L,2, 0} & = & - \frac{1}{16}   \sqrt{5}  \biggl[\nobracket
\left( \frac{7 i}{a_0} - \frac{16 i}{a_0^3} - \frac{36}{a_0^4} + \frac{72
	i}{a_0^5} - 1 \right) e^{\left( - 2 i  a_0 \right)} \nonumber\\
&  & + 2
\left( i  a_0 + 7 \right) E_1  \left( 2 i
a_0 \right) + 2  \left( i  a_0 + 7 \right) \log (-
a_0) \nonumber\\
&  & - \frac{39 i}{a_0} + \frac{56}{a_0^2} + \frac{88 i}{a_0^3} -
\frac{108}{a_0^4} - \frac{72 i}{a_0^5} \biggl] \nobracket,
\end{eqnarray}
other terms can also be obtained from Eqs.~\eqref{eq:KAIJ-lm-rela22} to \eqref{eq:KAIJ-lm-rela31}.

At $\gamma = 0$, the ORFs are dominated by linear terms in $a_u$, while at $\gamma = \pi$, they are dominated by logarithmic terms in $a_u$ and are constant only at $\gamma = \pi$.

When $a_u$ is finite, $\lim_{\gamma \rightarrow 0} \Gamma^{L,\ell m} = F(a_u)$. For the longitudinal mode, considering the autocorrelation response for a single detector, when $a_u \gg 1$, the ORFs converge, but as $a_u$ increases and $\gamma \rightarrow 0$, the ORFs tend to a constant.
\begin{eqnarray}
\Gamma_{i i}^{L,0, 0} & = & - 2  \gamma_E - \frac{1}{2} 
\left( \pi - \mathrm{Si} \left( 2  a_v
\right) \right) a_v - \frac{2}{a_v^2} + \frac{1}{4}   \cos
\left( 2  a_v \right) \nonumber\\
&  & + 2  \mathrm{Ci} \left( 2  a_v \right) -
2  \log \left( 2  a_v \right) + \frac{37}{12} ,
\end{eqnarray}
\begin{eqnarray}
\Gamma_{i i}^{L,1, 1} & = & - \frac{1}{12}   \sqrt{6} 
\biggl[\nobracket 30  \gamma_E + 6  \left( \pi -
\mathrm{Si} \left( 2  a_v \right) \right) a_v -\nonumber\\
&  & \frac{12
	\left( \cos \left( 2  a_v \right) - 5
	\right)}{a_v^2} + \frac{36  \left( \cos \left( 2  a_v
	\right) - 1 \right)}{a_v^4} - \nonumber\\
&  & 3  \cos \left( 2
a_v \right) - 30  \mathrm{Ci} \left( 2
a_v \right) + 30  \log \left( 2
a_v \right)\nonumber\\
&  & - 53 \biggl] \nobracket ,
\end{eqnarray}
\begin{eqnarray}
\Gamma_{i i}^{L,2, 0} & = & \frac{1}{60}   \sqrt{5}
\biggl[\nobracket 210  \gamma_E + 30  \left( \pi -
\mathrm{Si} \left( 2  a_v \right) \right) a_v +
\frac{840}{a_v^2} -\nonumber\\
&  & \frac{540  \left( \cos \left( 2
	a_v \right) + 3 \right)}{a_v^4} - 15  \cos \left( 2  a_v \right) - 210
\mathrm{Ci} \left( 2  a_v \right) \nonumber\\
&  & + 210
\log \left( 2  a_v \right) - 443 \biggl] \nobracket ,
\end{eqnarray}
other terms satisfy Eqs.~\eqref{eq:KAIJ-lm-rela22} to \eqref{eq:KAIJ-lm-rela31}, with the letter $K$ replaced by $\Gamma$.

Similar to the vector modes, the ORFs for the scalar longitudinal mode cannot be normalized. The $(0, 0)$ term corresponds to the autocorrelation response for a single detector for the scalar longitudinal mode. For short wavelengths, $\Gamma^{L, 0, 0}_{ii} \propto \frac{\pi}{4} a_u$, and it grows faster with $a_u$ than the logarithmically growing vector modes. This agrees with Eq.~(A36) in \cite{Lee2008} and Eq.~(41) in \cite{2012prd_ORF_nonGR}.

In general, for fixed $(\ell, m)$, the real part of the ORF increases with growing $a_v$ and $a_u$, while the imaginary part decreases and approaches zero. As $\ell$ increases, the ORF values do not show a clear change in magnitude. For fixed $\ell$, the ORFs for $m > 0$ are larger than those for $m < 0$, and they increase with $|m|$.

\section{Discussion and Conclusions}\label{sec4}

We have derived the symmetries of the anisotropic ORFs based on the chosen reference frame and the inherent symmetries of the PTA response function and spherical harmonics. Specifically, when $\ell + m$ is odd, the corresponding ORFs vanish; therefore, only cases with $\ell + m$ even require computation. The ORFs also satisfy an exchange symmetry relation with respect to the dimensionless parameters $a_u$ and $a_v$. Furthermore, we have established relationships among different polarization modes in ORF calculations, as given by \Cref{eq:funcTT2,eq:funcV,eq:funcL}, and the limit relations between subintegrals. Utilizing these relations within our selected reference frame enables the computation of exact, approximation-free, analytical expressions for the anisotropic ORFs of a PTA with significantly reduced integral complexity, permitting further simplification of the final results. Their $(0, 0)$ components correspond to the isotropic ORFs provided in \cite{2019prd_analytic_response_function_TDI,2021prd_sensitivity_TDI,2021prd_sensitivity_TDI_nonGR}. In practical analyses involving multiple pulsar pairs where the reference frame may be arbitrary, the ORFs can be obtained via the symmetry relation Eq.~\eqref{rot-R}.

\begin{figure*}[!t] 
	\resizebox{230pt}{142.5pt}{\includegraphics{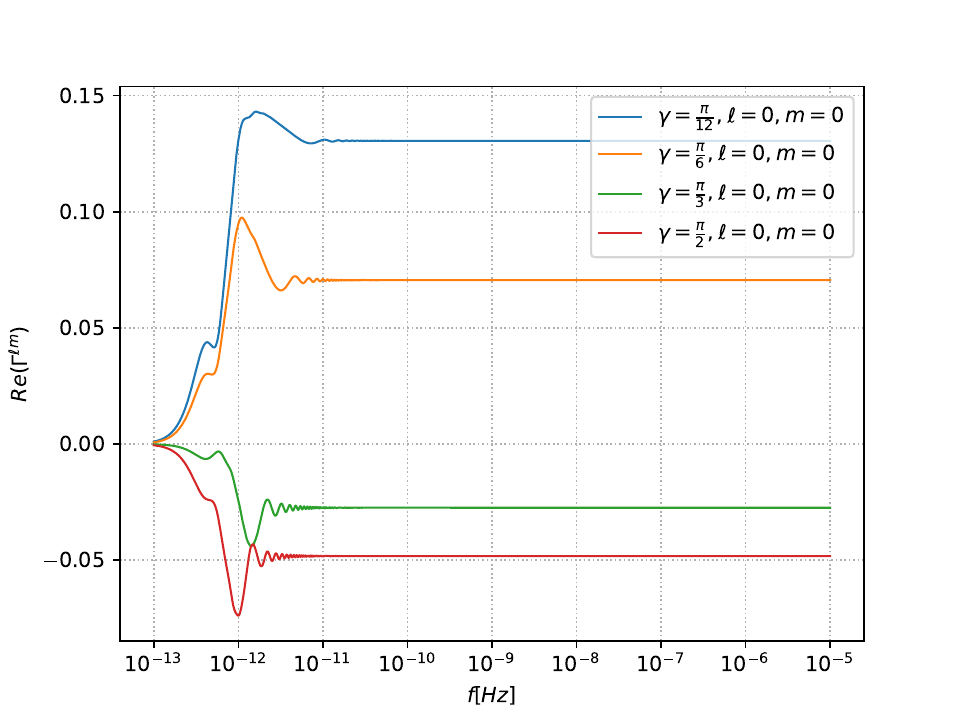}}\resizebox{230pt}{142.5pt}{\includegraphics{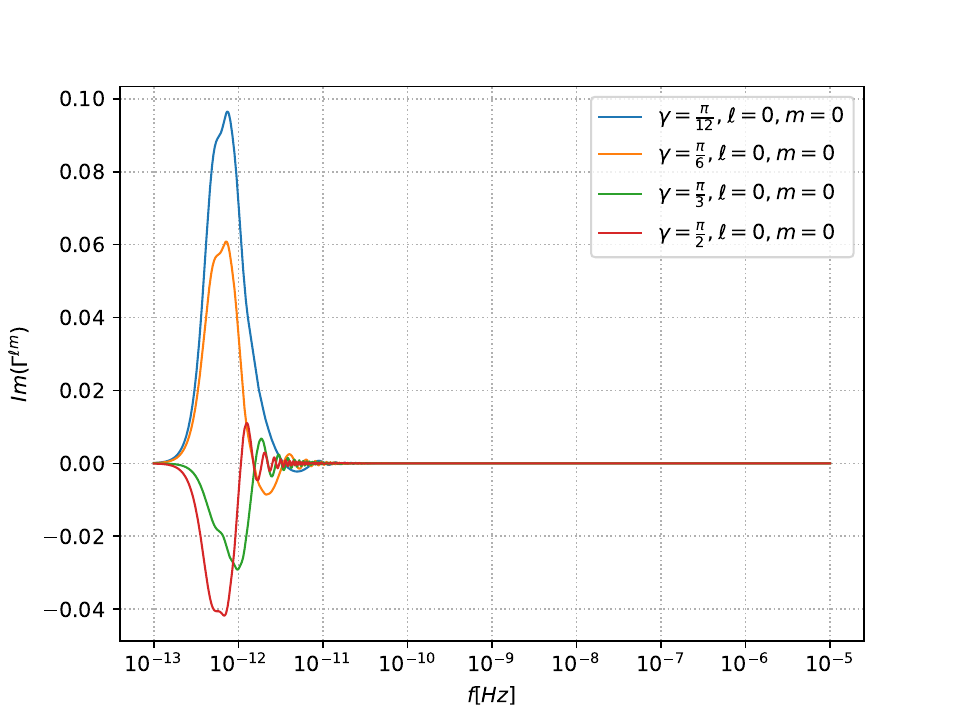}}
	\resizebox{230pt}{142.5pt}{\includegraphics{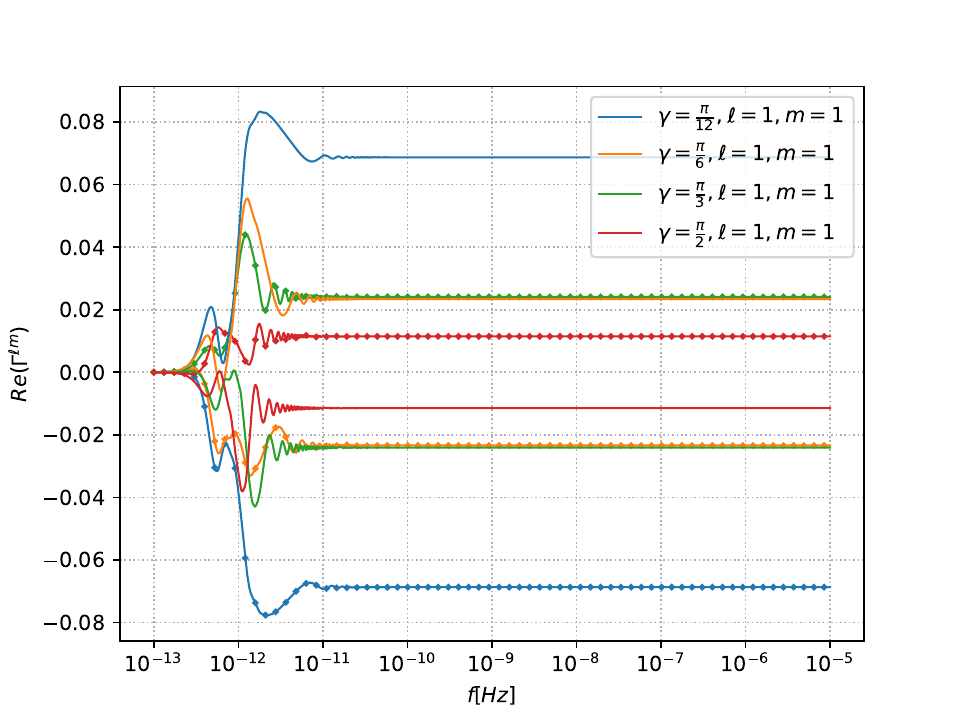}}\resizebox{230pt}{142.5pt}{\includegraphics{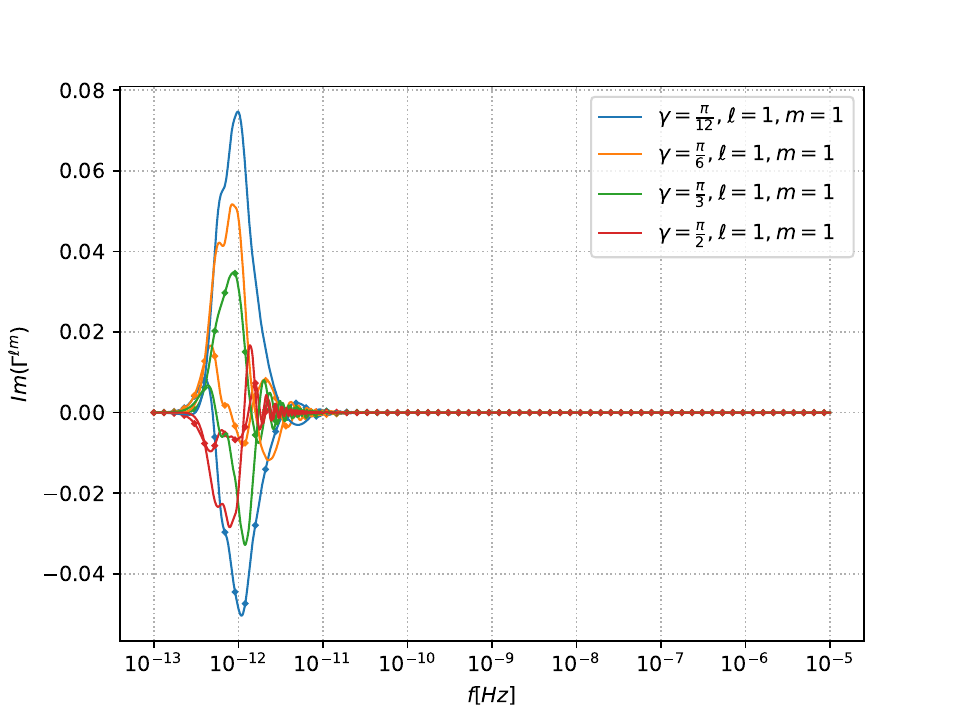}}
	\resizebox{230pt}{142.5pt}{\includegraphics{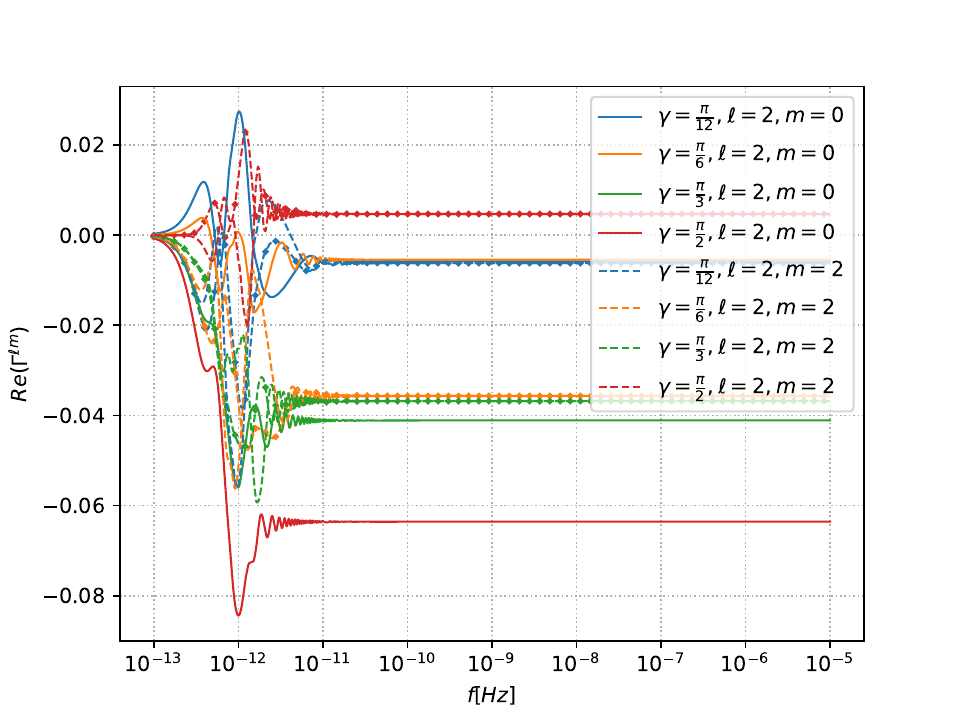}}\resizebox{230pt}{142.5pt}{\includegraphics{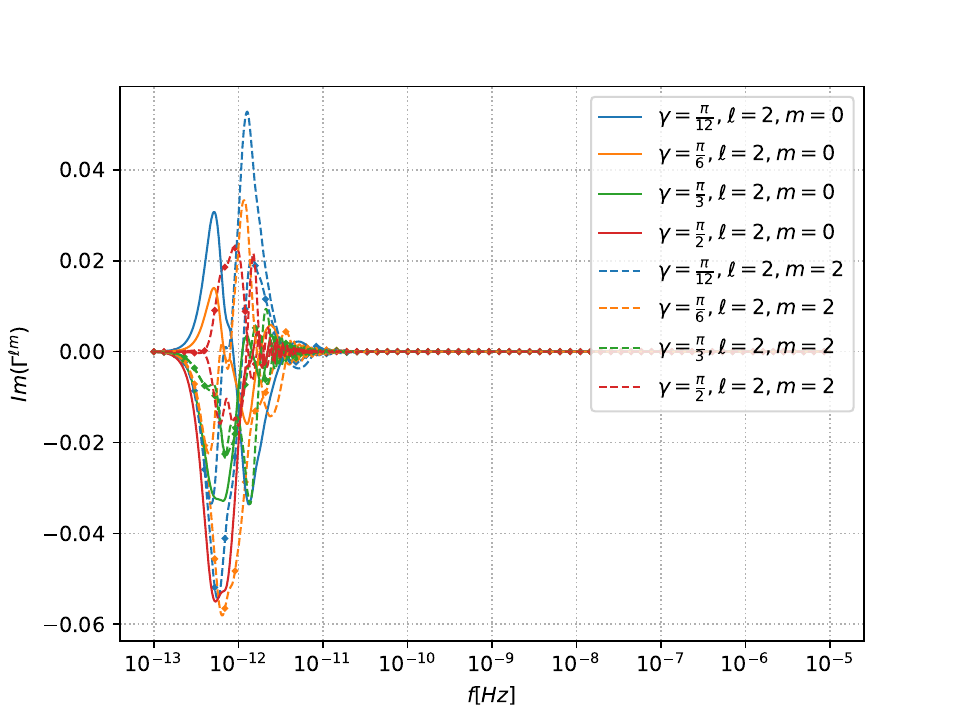}}
	\caption{\label{fig:orfT-lm-litfre} ORFs for tensor modes, $\ell = 0, 1, 2$, with $L_u = L_v/4 = 1$ kpc. Left panels: real part; right panels: imaginary part. Markers correspond to $m<0$ cases: diamonds for the general form and circles for the short-wavelength approximation.}
\end{figure*}

\begin{figure*}[!t] 
	\resizebox{230pt}{142.5pt}{\includegraphics{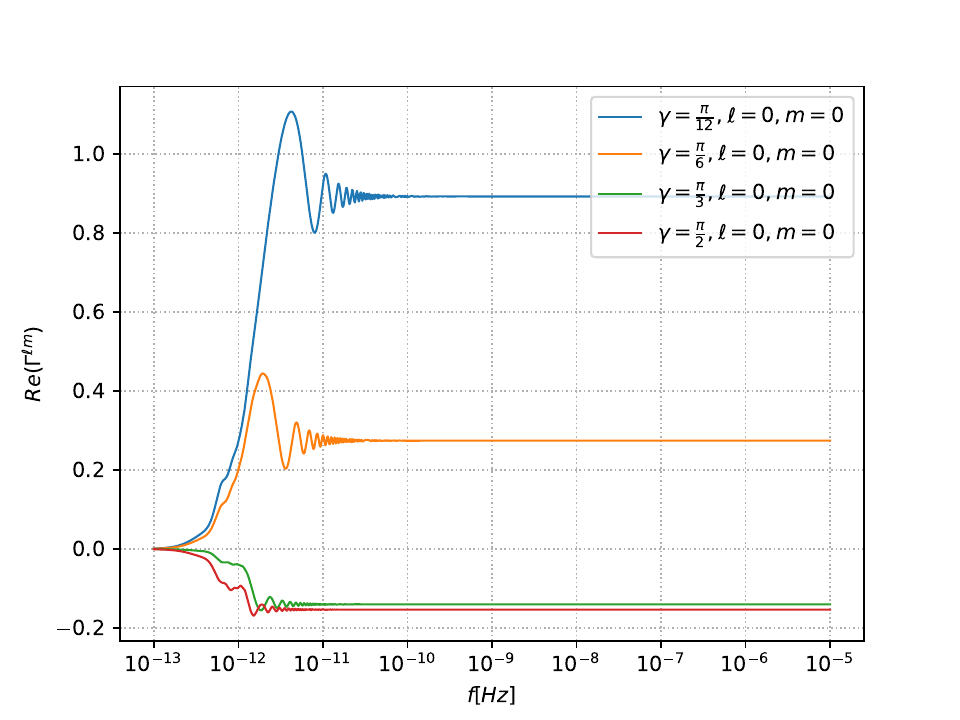}}\resizebox{230pt}{142.5pt}{\includegraphics{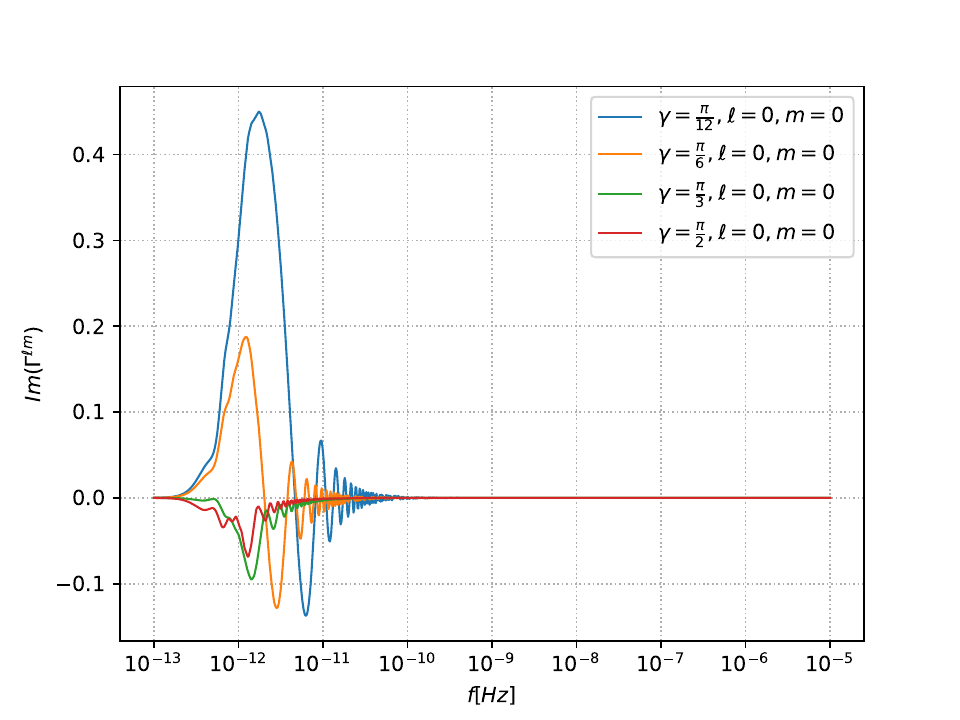}}
	\resizebox{230pt}{142.5pt}{\includegraphics{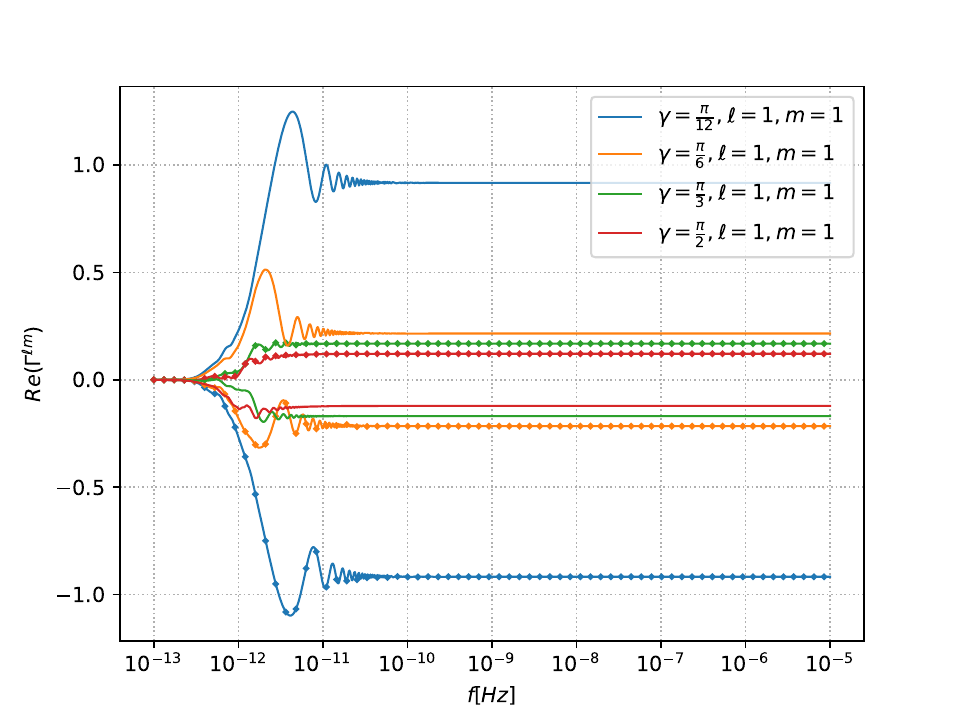}}\resizebox{230pt}{142.5pt}{\includegraphics{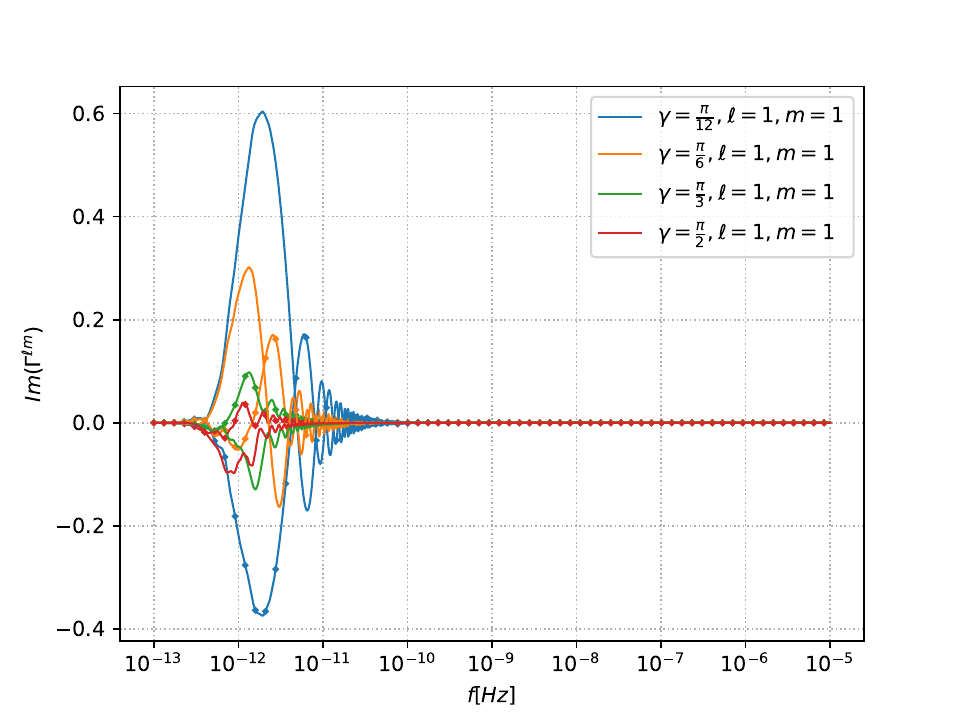}}
	\resizebox{230pt}{142.5pt}{\includegraphics{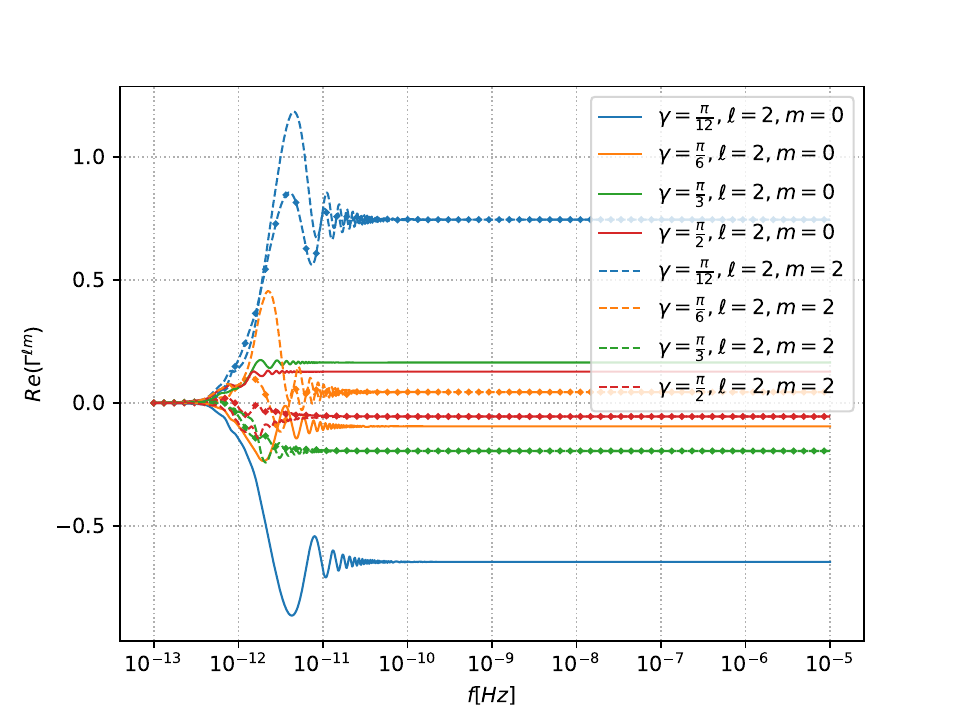}}\resizebox{230pt}{142.5pt}{\includegraphics{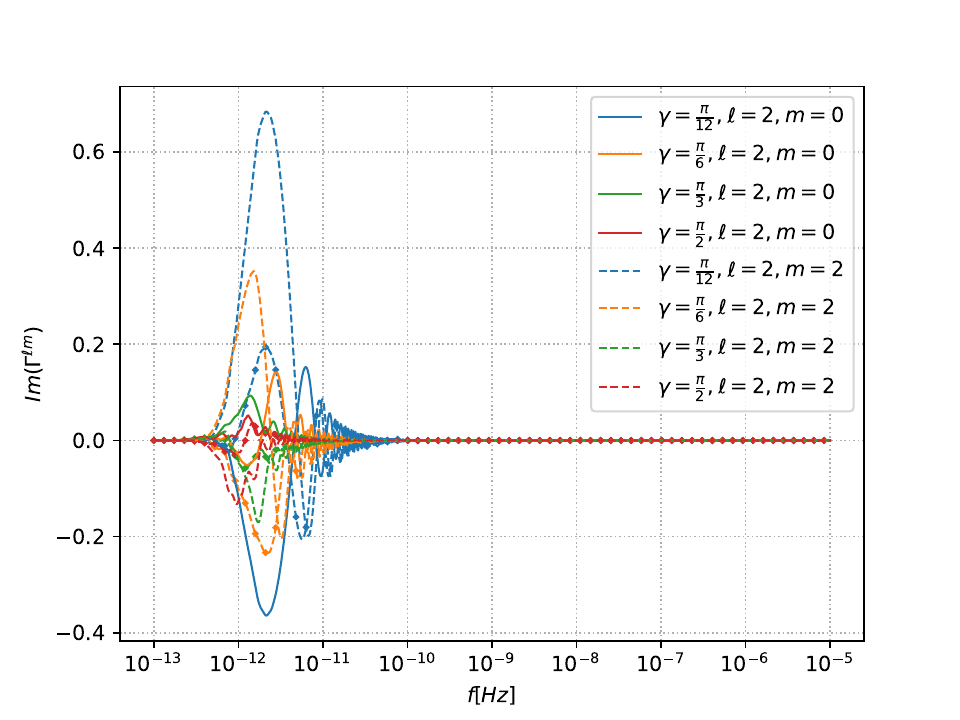}}
	\caption{\label{fig:orfV-lm-litfre} ORFs for vector modes, $\ell = 0, 1, 2$. Arm length and graphical conventions are as defined in Fig.~\ref{fig:orfT-lm-litfre}.}
\end{figure*}

\begin{figure*}[!t] 
	\resizebox{230pt}{142.5pt}{\includegraphics{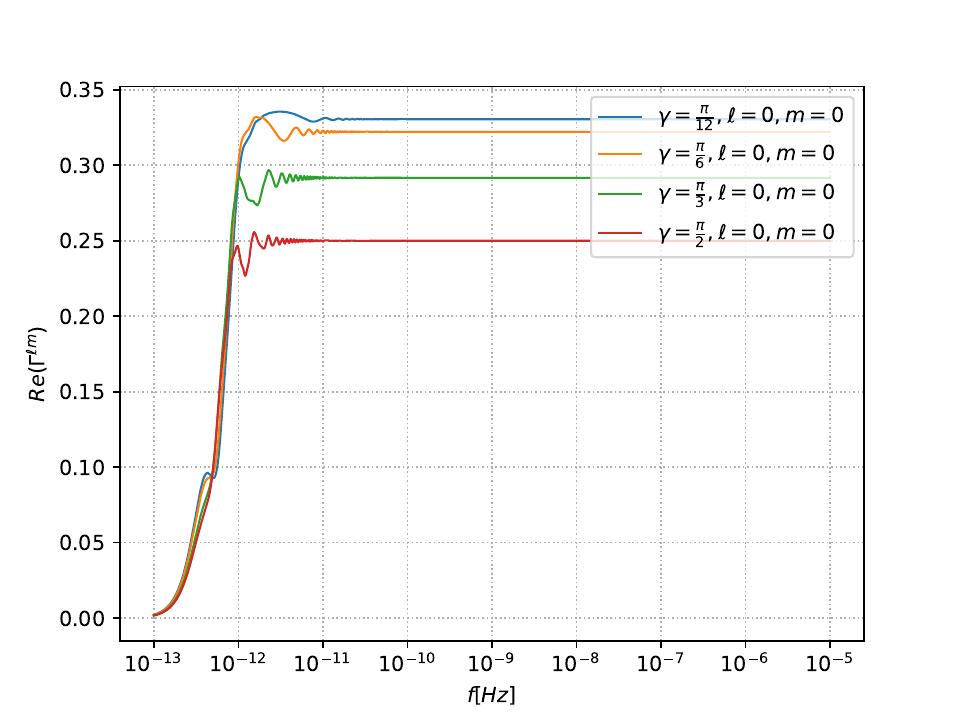}}\resizebox{230pt}{142.5pt}{\includegraphics{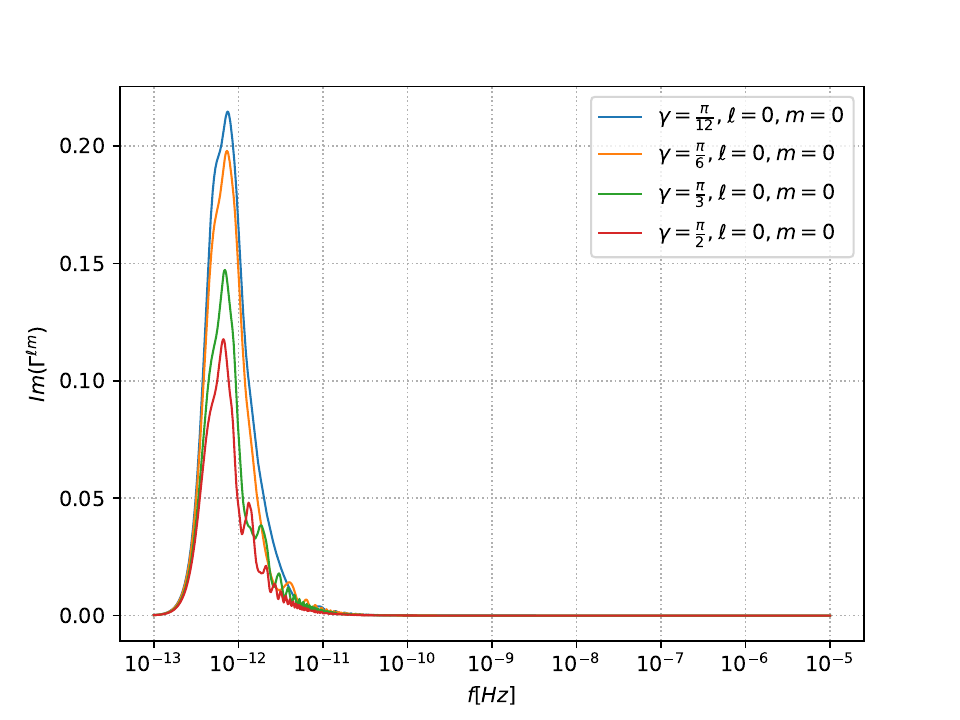}}
	\resizebox{230pt}{142.5pt}{\includegraphics{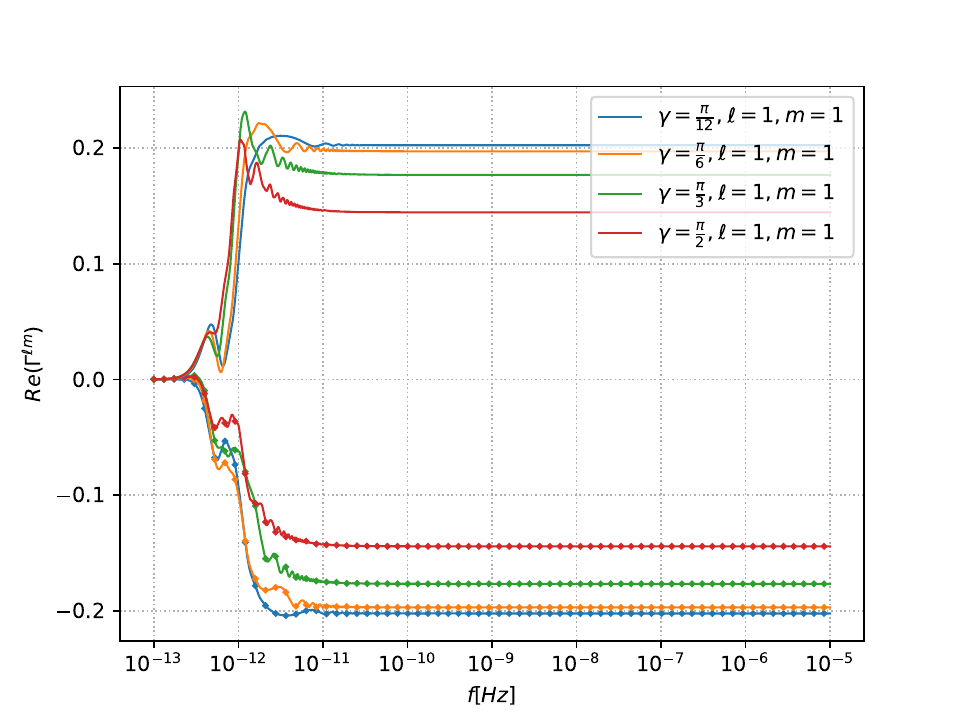}}\resizebox{230pt}{142.5pt}{\includegraphics{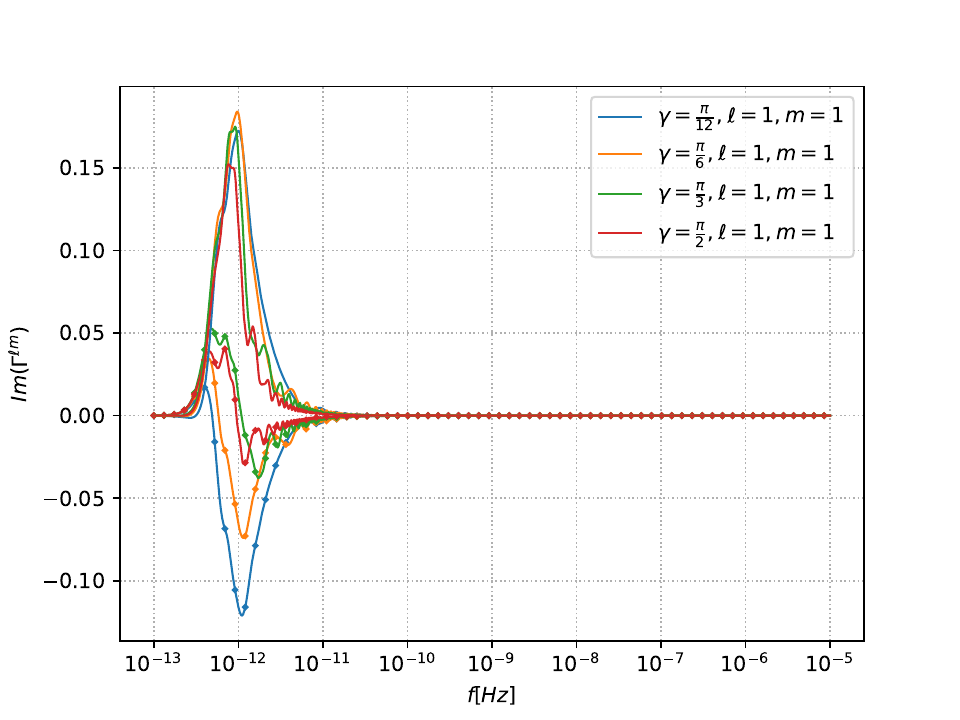}}
	\resizebox{230pt}{142.5pt}{\includegraphics{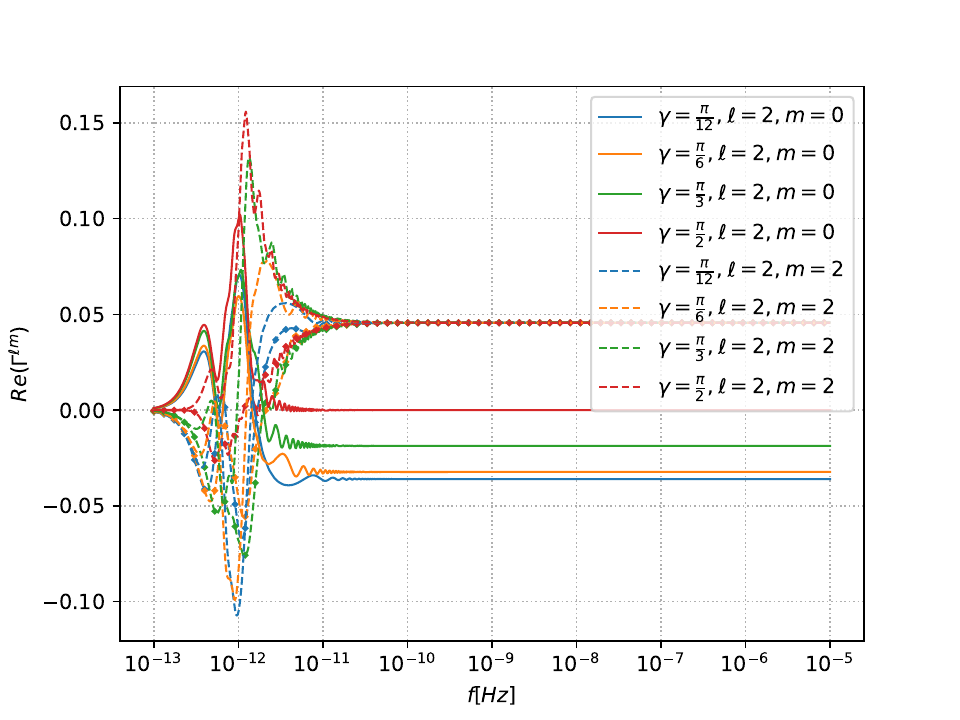}}\resizebox{230pt}{142.5pt}{\includegraphics{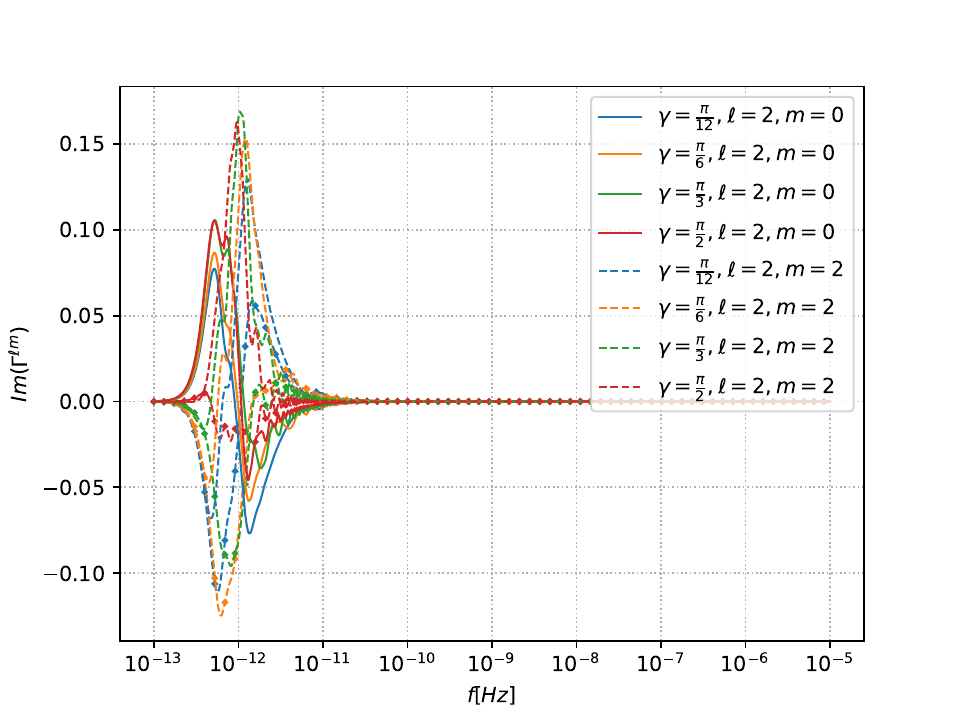}}
	\caption{\label{fig:orfB-lm-litfre} ORFs for scalar breathing mode, $\ell = 0, 1, 2$. Arm length and graphical conventions are as defined in Fig.~\ref{fig:orfT-lm-litfre}.}
\end{figure*}

\begin{figure*}[!t] 
	\resizebox{230pt}{142.5pt}{\includegraphics{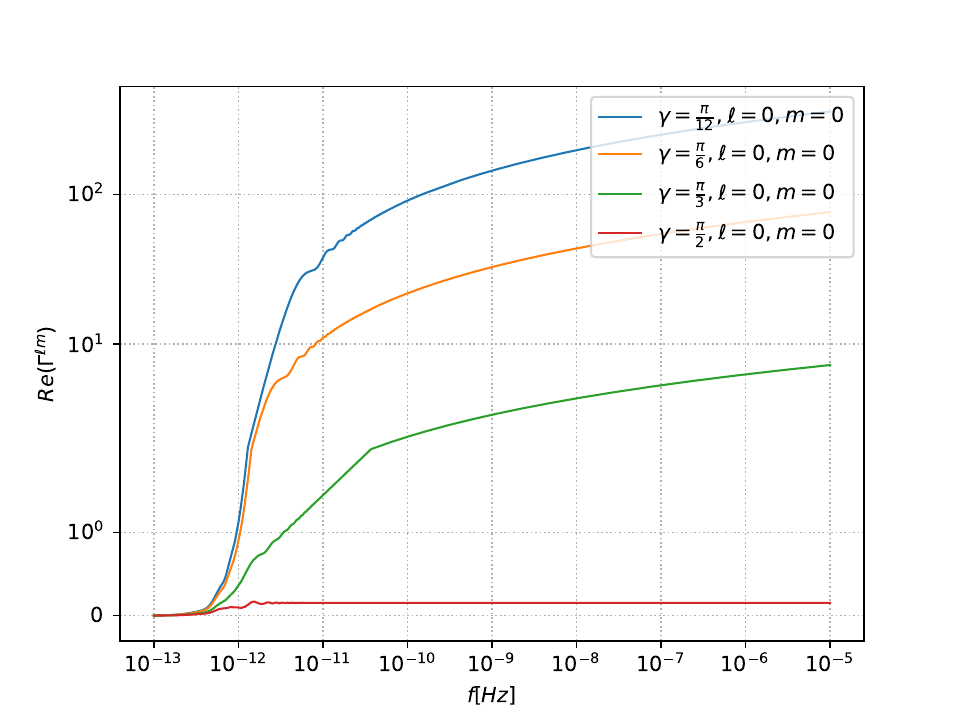}}\resizebox{230pt}{142.5pt}{\includegraphics{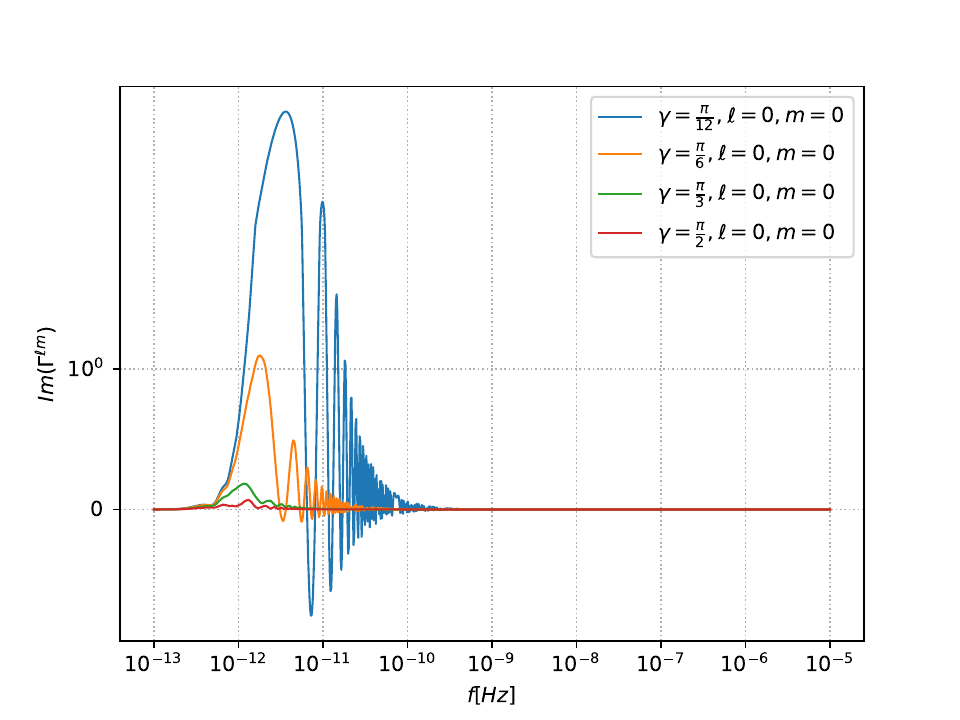}}
	\resizebox{230pt}{142.5pt}{\includegraphics{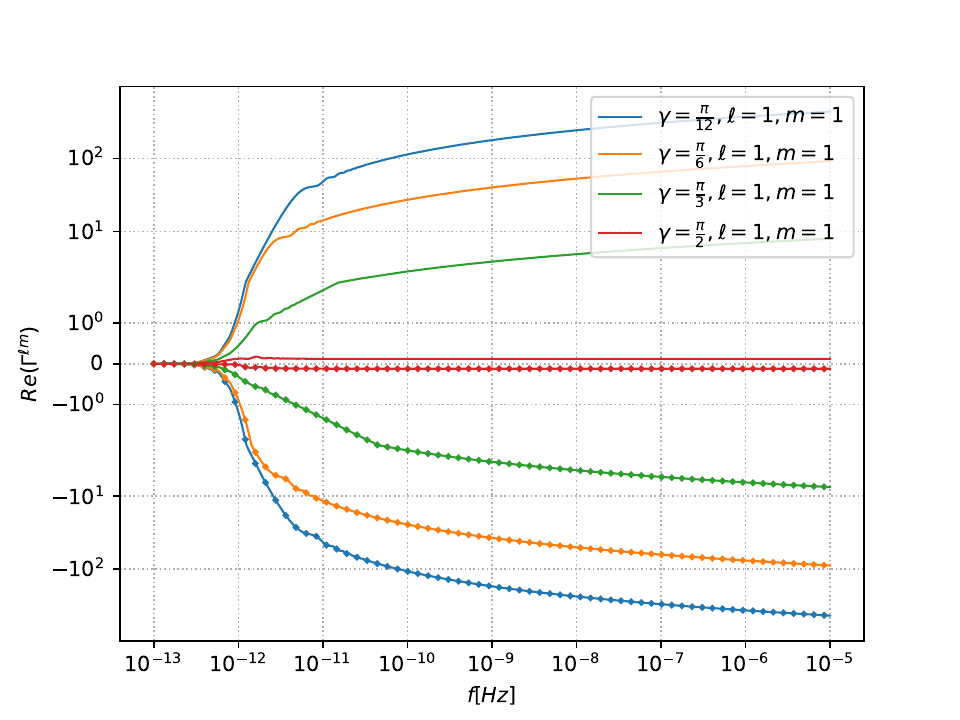}}\resizebox{230pt}{142.5pt}{\includegraphics{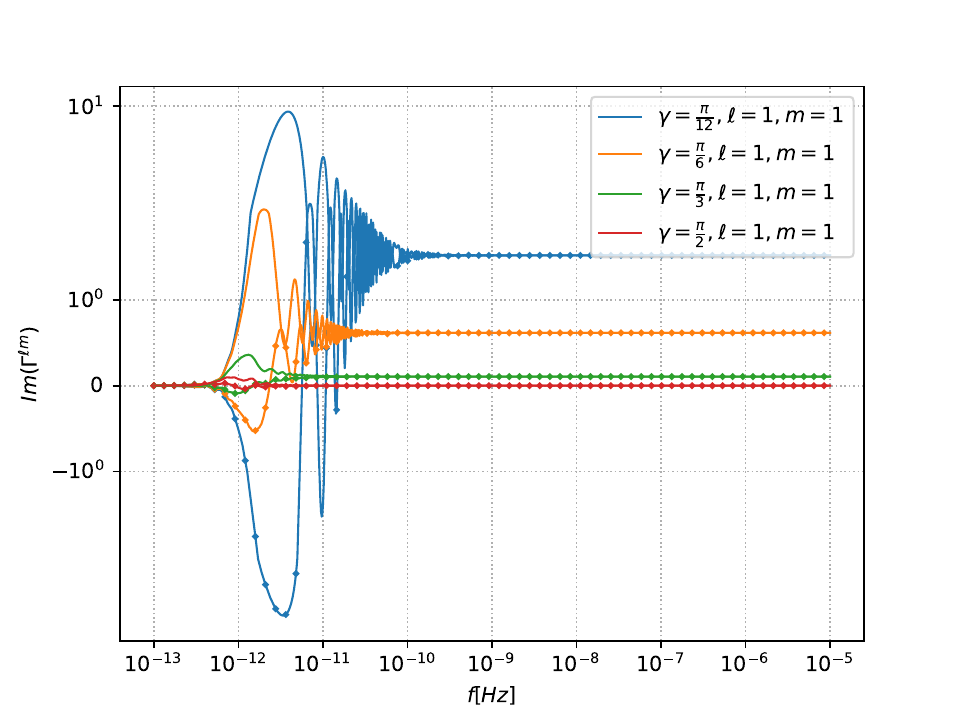}}
	\resizebox{230pt}{142.5pt}{\includegraphics{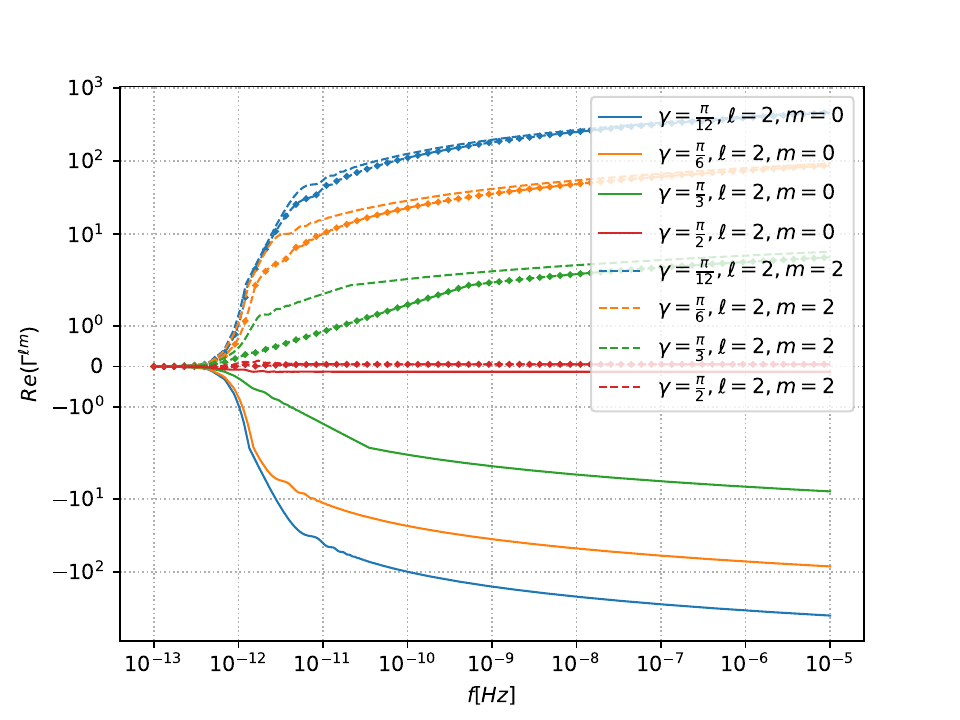}}\resizebox{230pt}{142.5pt}{\includegraphics{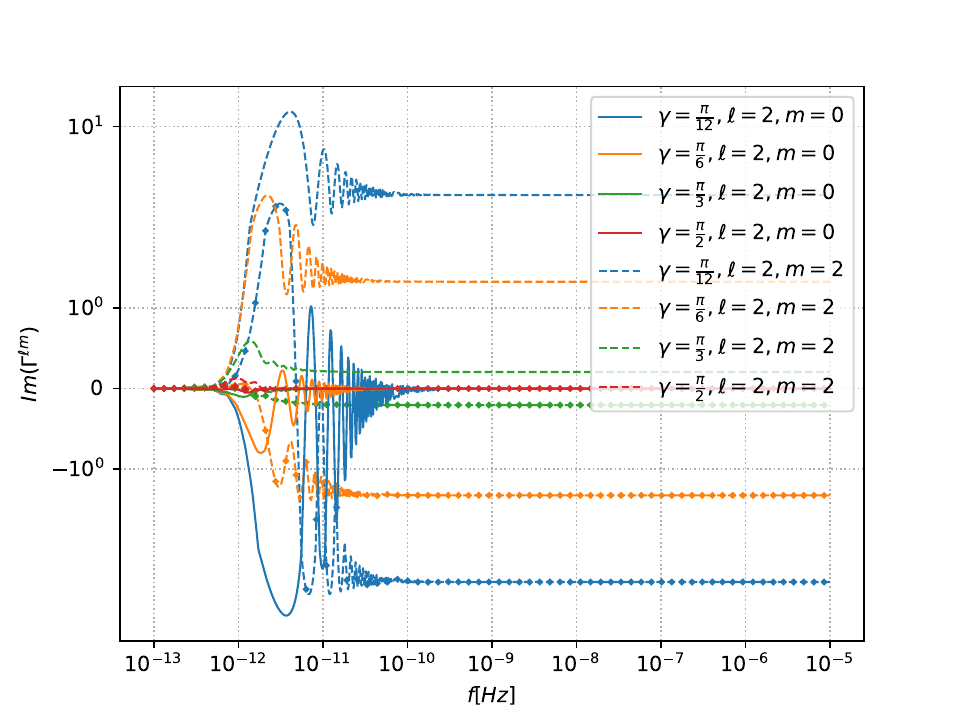}}
	\caption{\label{fig:orfL-lm-litfre} ORFs for scalar longitudinal mode, $\ell = 0, 1, 2$. Arm length and graphical conventions are as defined in Fig.~\ref{fig:orfT-lm-litfre}.}
\end{figure*}

Since no approximations are employed, our expressions allow for the rapid calculation of ORFs for any pair of pulsars at any multipole order. As illustrated in \Cref{fig:orfT-lm-litfre,fig:orfV-lm-litfre,fig:orfB-lm-litfre,fig:orfL-lm-litfre}, which display ORFs for different polarization modes (with frequency $f$ on the horizontal axis), the curves are plotted for various angular separations assuming pulsar distances of $L_u = 1$ kpc and $L_v = 4$ kpc. Key observations include: for tensor and scalar breathing modes, the $(\ell, m)$-order ORFs can be treated as constants depending only on the angular separation $\gamma$ within a certain detection band. This constant-behavior range shifts to higher frequencies as the ORF order increases. For instance, tensor and scalar breathing modes with $\ell \leqslant 2$ can be considered constant across the current PTA detection band ($10^{-9}$ to $10^{-5}$ Hz). Moreover, for the scalar breathing mode with $\ell \geqslant 3$, the ORFs are effectively zero. In contrast, the ORFs for vector modes may exhibit frequency dependence within the observational band for small $\gamma$, while the scalar longitudinal mode ORFs grow logarithmically with frequency for $\gamma < \pi/2$.

The tensor and scalar breathing modes exhibit similar behavior because both are transverse, and the dominant term in their integrals is Eq.~\eqref{eq:funcBT1}, which corresponds to an exponential function. Their ORFs generally decrease with increasing $\ell$. In the case of $\gamma = 0$ (co-aligned pulsars), the ORFs for transverse modes remain finite regardless of $f L$, whereas the ORFs for vector modes grow logarithmically with $f L$, and the scalar longitudinal mode ORF grows linearly with $f L$. Additionally, the phases of these modes are not consistent. In actual observations, pulsar distances from Earth are generally unequal, making the ORFs a set of complex-valued functions. For sufficiently large $f L$, the imaginary part of the transverse mode ORFs can always be neglected; for the longitudinal mode, this is possible only when $\gamma$ is near $\pi/2$.

The short-wavelength approximation is concise and effective in many cases. For the isotropic scenario \cite{Hellings1983,Lee2008}, it corresponds to the short-wavelength limit of our $(0, 0)$ term. However, it is crucial to note that for the scalar longitudinal mode---except for the special case of $\gamma=0$---no valid short-wavelength approximation exists, whereas our $(\ell, m)$-order analytical expressions are valid for arbitrary $\gamma$. The geometric characteristics of current PTA pulsar samples further highlight the limitations of the traditional short-wavelength approximation and impose higher demands on theory: pulsars are highly concentrated on the sky (e.g., the FAST GPPS survey covers regions with Galactic latitude $|b| < 10^\circ$ \cite{Han2025}), leading to naturally small angular separations $\gamma$ for a large number of pulsar pairs; distances are widely distributed (typical distances $0.1$--$10$ kpc \cite{Reardon2021, DOrazio:2020tne, Smits2011}), meaning the key parameter $a = 2\pi f L/c$ often lies in the range $10^1$--$10^4$, not always satisfying the $a \gg 1$ condition required for the short-wavelength limit; unequal arm lengths are the norm, and over 100 pulsar binary systems have been observed (FAST GPPS has discovered 157 binary pulsars \cite{Wang_2025}). The parameter space corresponding to these real observational features is precisely the ``blind spot'' of the traditional short-wavelength approximation. Compared to numerical integration, the analytical formulas we provide significantly enhance computational speed and accuracy. This improvement becomes more pronounced as $f L$ increases, with potential speedups exceeding a factor of $10^4$.

In summary, for current PTA detection within the framework of general relativity, these results provide a general formalism for anisotropic ORFs, whose $(0, 0)$ term corresponds to the isotropic case. The derived short-wavelength approximation is accurate and effective when $f L$ is sufficiently large. Our results are precisely valid for calculating the ORFs of pulsars relatively close to Earth and can also be applied to nearly collocated pulsars, such as those in binary systems. Notably, our results are particularly significant for non-Einsteinian polarization modes in modified gravity theories, especially the scalar longitudinal mode. For the vector modes, the short-wavelength approximation is not applicable when the angular separation between pulsars is small. For the scalar longitudinal mode, the short-wavelength approximation almost completely fails. In principle, the method presented in this paper can yield complete analytical expressions for the ORFs at any order, which are precisely valid for these two modes. This will be highly beneficial for future searches for anisotropic and polarized stochastic gravitational-wave backgrounds within modified gravity scenarios. Analytical expressions for anisotropic ORFs are highly valuable for advanced SGWB data analysis with PTAs.

\section{Acknowledgments}

This work was supported by the National Natural Science
Foundation of China (Grant No. 12575072,
No. 12547101 and No. 125B2102), the Fundamental Research
Funds for the Central Universities Project (Grant No.
2024IAIS-ZD009), and the Natural Science Foundation
of Chongqing (Grant No. CSTB2023NSCQ-MSX0103).

\section{Data availability}

The data that support the findings of this article are not publicly available. The data are available from the authors upon reasonable request.


\appendix
\phantomsection
\section*{APPENDIX: PROOF OF ORF SYMMETRIES}
\addcontentsline{toc}{section}{APPENDIX: PROOF OF ORF SYMMETRIES}
\markboth{APPENDIX: PROOF OF ORF SYMMETRIES}{}
\label{sec:ORFsymmetryproof}

The symmetry relations in Sec.~\ref{sec:ORFssymmetry} are extremely helpful for simplifying the calculation of ORFs. To prove these relations, we first prepare the following. Equation
\eqref{e:pulsarresponse-earthonly} is rewritten as
\begin{eqnarray}
  R^A_{\mathrm{doppler}}  (f, \hat{n}) & = & \frac{1}{2}  \frac{u^a u^b}{1 +
  \hat{u} \cdot \hat{n}}  (1 - e^{- i u (1 + \eta_i)}) e_{ab}^A (\hat{n}),
  \nonumber\\
  &  &  \label{eq:mich-frenq-A-New}
\end{eqnarray}
Define the polarization operator:
\begin{equation}
  \xi_k^A =\mathcal{G}^A (\hat{k}, \hat{l}_{ij}) \equiv \hat{l}_{ij}^a
  \hat{l}_{ij}^b e_{ab}^A (\hat{k}) . \label{eq:GAij-def-new}
\end{equation}
For any two unit vectors $\hat{l}_{ij}$ and
$\hat{l}_{ik}$, its property under rigid rotation is
\begin{eqnarray}
  \sum_A \mathcal{G}^A (\mathscr{R}  \hat{k}, \mathscr{R}  \hat{l}_{ij})
  \mathcal{G}^{A \ast} (\mathscr{R}  \hat{k}, \mathscr{R}  \hat{l}_{ik}) & = &
  \sum_{\lambda} \mathcal{G}^A (\hat{k}, \hat{l}_{ij}) \nonumber\\
  &  & \mathcal{G}^{A \ast} (\hat{k}, \hat{l}_{ik}).  \label{eq:GAGA-trace}
\end{eqnarray}
This property follows directly from the fact that it represents the contraction of the tensor
$\mathcal{G}^A$
with its own complex conjugate, reducing to the simplest form of a vector's modulus squared, which is naturally rotationally invariant. A detailed proof can be found in
\cite{Bartolo:2018qqn}.

We first prove \eqref{rot-R}. Under a rigid rotation, the pulsar positions transform as
$\vec{x}_i \to R \vec{x}_i$. Applying a similar rotation to the integration variables in Eq.~\eqref{eq:response}, we have
\begin{eqnarray}
  \Gamma_{\mathscr{R} i \mathscr{R} j}^{\ell m} (f) & \equiv & \frac{1}{8 \pi}
  \int d^2  \hat{k}  \tilde{Y}^{\ell m}  (\mathscr{R}  \hat{k})  \sum_A R_i^A
  (f, \mathscr{R}  \hat{k}, \mathscr{R}  \vec{l}_{p i}) \nonumber\\
  &  & R_j^{A \ast}  (f, \mathscr{R}  \hat{k}, \mathscr{R}  \vec{l}_{p j}) .
\end{eqnarray}
Using Eq.~\eqref{eq:GAGA-trace}, we obtain
\begin{equation}
  \Gamma_{\mathscr{R} i \mathscr{R} j}^{\ell m} (f) \equiv \frac{1}{4 \pi}
  \int d^2  \hat{k}  \tilde{Y}^{\ell m}  (\mathscr{R}  \hat{k})  \sum_A R_i^A
  (f, \hat{k}, \vec{l}_{p i}) R_j^{A \ast}  (f, \hat{k}, \vec{l}_{p j}),
\end{equation}
where for spherical harmonics we have
\begin{equation}
  \tilde{Y}^{\ell m}(\mathscr{R}\hat{k}) = \sum_{m'=-\ell}^{\ell} D_{mm'}^{(\ell)}(R) \, \tilde{Y}^{\ell m'}(\hat{k}).
\end{equation}
Substituting this into the above calculation yields Eq.~\eqref{rot-R}. Equation \eqref{rotz-R} is a special case of \eqref{rot-R}.

To prove \eqref{eq:R-prop-lm}, it is easy to show
\begin{equation}
  \begin{aligned}
    R_{ij}  (a_u, a_v, \tilde{x}, \tilde{y}) \odot \tilde{Y}^{\ell m}
    (\tilde{x}, \tilde{y}) & = R_{ij}  (a_u, a_v, - \tilde{x}, - \tilde{y})
    \odot\\
    & \tilde{Y}^{\ell m} (- \tilde{x}, - \tilde{y}) .
  \end{aligned}
\end{equation}
Then we have
\begin{eqnarray}
\Gamma_{i j}^{\ell m} (a_v, a_u, \gamma) & = & R_{i j} (a_v, a_u, -
\tilde{x}, - \tilde{y}) \odot \tilde{Y}^{\ell m}(- \tilde{x}, - \tilde{y})\nonumber\\
& = & (- 1)^{\ell} R_{i j} (a_v, a_u, - \tilde{x}, - \tilde{y}) \odot
\tilde{Y}^{\ell m}(\tilde{x}, \tilde{y})\nonumber\\
& = & (- 1)^{\ell + m} R_{i j} (a_v, a_u, \tilde{x}, \tilde{y}) \odot
\tilde{Y}^{\ell m}(\tilde{x}, \tilde{y})\nonumber\\
& = & (- 1)^{\ell + m} \Gamma_{ij}^{\ell m} (a_u, a_v, \gamma) 
\end{eqnarray}
Here we used the symmetry of spherical harmonics:
\begin{eqnarray}
  \tilde{Y}^{\ell m} (\hat{k}) \to \tilde{Y}^{\ell m}  (- \hat{k}) & = & (-
  1)^{\ell}  \tilde{Y}^{\ell m} (\hat{k}) \to \nonumber\\
  (- 1)^{\ell}  \tilde{Y}^{\ell m}  (\mathscr{R}_{z, \pi}  \hat{k}) & = & (-
  1)^{\ell + m}  \tilde{Y}^{\ell m} (\hat{k}) \hspace{0.27em} .
  \label{eq:sphareharmon}
\end{eqnarray}
That is, for any $(\ell, m)$,
\begin{equation}
  \Gamma_{ij}^{\ell m} (a_u, a_v, \gamma) = (- 1)^{(\ell + m)}
  \Gamma_{ij}^{\ell m} (a_u, a_v, \gamma) .
\end{equation}
When $\ell + m = 2 n + 1, n \in
\mathbb{N}$, we obtain Eq.~\eqref{eq:R-prop-lm}.

To prove Eq.~\eqref{eq:R-rop-lm4}, we use the property of spherical harmonics $\tilde{Y}^{\ell
m} (\tilde{y}, \tilde{x}) = \tilde{Y}^{\ell  m \ast} (\tilde{x},
\tilde{y})$. Thus, for these integrals, we have
\begin{eqnarray}
  \Gamma_{ij}^{\ell m} (a_u, a_v, \gamma) & = & R_{ij}  (a_u, a_v, \tilde{x},
  \tilde{y}) \odot \tilde{Y}^{\ell m} (\tilde{x}, \tilde{y}) \nonumber\\
  & = & R^{\ast}_{ji}  (a_v, a_u, \tilde{y}, \tilde{x}) \odot \tilde{Y}^{\ell  m
  \ast} (\tilde{y}, \tilde{x}) \nonumber\\
  & = & (R_{ji} (a_v, a_u, \tilde{y}, \tilde{x}) \odot \tilde{Y}^{\ell m} (\tilde{y}, \tilde{x}))^{\ast} \nonumber\\
  & = & \Gamma_{ji}^{\ell m \ast} (a_v, a_u, \gamma)
\end{eqnarray}
To prove \eqref{eq:R-prop-menom-app-old}, from Eq.~\eqref{eq:response}, we have
\begin{eqnarray}
  \Gamma_{ij}^{\ell, - m} (a_u, a_v, \gamma) & = & \tilde{Y}^{\ell, - m}
  (\tilde{x}, \tilde{y}) \odot R_{ij}  (a_u, a_v, \tilde{x}, \tilde{y})
  \nonumber\\
  & = & \tilde{Y}^{\ell m \ast} (\tilde{x}, \tilde{y}) \odot R_{ij}  (a_u,
  a_v, \tilde{x}, \tilde{y}),
\end{eqnarray}
because when $\gamma = 0$, $\tilde{x} = \tilde{y} = x$, or when
$a_u, a_v \rightarrow + \infty$,
\begin{equation}
R_{i j} (a_u, a_v, \theta, - \phi, \gamma) = R_{i j} (a_u, a_v,
 \theta, \phi, \gamma).
\end{equation}
Therefore,
\begin{eqnarray}
\Gamma_{ij}^{\ell, - m} (a_u, a_v, \gamma) & = & (- 1)^m \tilde{Y}_{\ell m}^{\ast} (\theta, - \phi) \odot
R_{i j} (a_v, a_u, \theta, - \phi, \gamma)
\nonumber\\ 
& = & (- 1)^m \tilde{Y}_{\ell m} (\theta, \phi) \odot
R_{i j} (a_v, a_u, \theta, \phi, \gamma)
\nonumber\\
& = & (- 1)^m \Gamma_{ij}^{\ell m} (a_u, a_v, \gamma),
\end{eqnarray}
Using Eq.~\eqref{eq:R-rop-lm4} we obtain Eq.~\eqref{eq:R-prop-menom-app-old}. Here, from the second to the third step, we performed a rotation of the entire detector constellation by $\pi$ about the $z$ axis, $\mathscr{R}_z (\pi)$, under which $R_{ij}  (f,
\hat{k}) \rightarrow R_{ji}  (f,
\hat{k})$, and we used the spherical harmonic symmetry Eq.~\eqref{eq:sphareharmon}.

\bibliography{reference}
\bibliographystyle{apsrev}
\end{document}